\documentclass[ejsv2,preprint,noshowframe]{imsart}
\RequirePackage[numbers]{natbib}
\RequirePackage[colorlinks,citecolor=blue,urlcolor=blue]{hyperref}
\RequirePackage{graphicx}

\startlocaldefs
\theoremstyle{plain}
\newtheorem{axiom}{Axiom}

\newtheorem{theorem}{Theorem}[section]
\newtheorem{lemma}[theorem]{Lemma}
\newtheorem{proposition}[theorem]{Proposition}
\newtheorem{corollary}[theorem]{Corollary}

\theoremstyle{definition}
\newtheorem{definition}[theorem]{Definition}
\newtheorem{assumption}[theorem]{Assumption}

\theoremstyle{remark}
\newtheorem{remark}[theorem]{Remark}

\usepackage{algorithm}
\usepackage{algorithmic}
\usepackage{booktabs}

\theoremstyle{definition}
\theoremstyle{remark}
\endlocaldefs

\begin{document}
\begin{frontmatter}
\title{Order-Constrained Spectral Causality for Multivariate Time Series}
\runtitle{Order-Constrained Spectral Causality for Multivariate Time Series}

\begin{aug}
\orcid{0000-0002-2400-1097}               
\author[A]{\fnms{Alejandro}~\snm{Rodriguez Dominguez}\ead[label=e1]{arodriguez@miraltabank.com}}
\address[A]{Head of Quantitative Analysis and Artificial Intelligence,
Miralta Finance Bank S.A.\printead[presep={,\ }]{e1}}

\end{aug}

\begin{abstract}
We introduce an operator-theoretic framework for analyzing directional dependence
in multivariate time series based on order-constrained spectral non-invariance.
Directional influence is defined as the sensitivity of second-order dependence
operators to admissible, order-preserving temporal deformations of a designated
source component, summarized through orthogonally invariant spectral functionals.
We show that the resulting supremum--infimum dispersion functional is the unique
diagnostic within this class satisfying order consistency, orthogonal invariance,
Loewner monotonicity, second-order sufficiency, and continuity, and that classical
Granger causality, directed coherence, and Geweke frequency-domain causality arise
as special cases under appropriate restrictions. An information-theoretic
impossibility result establishes that entrywise-stable edge-based tests require
quadratic sample size scaling in distributed (non-sparse) regimes, whereas spectral
tests detect at the optimal linear scale. We establish uniform consistency and
valid shift-based randomization inference under weak dependence. Simulations
confirm correct size and strong power across distributed and nonlinear
alternatives, and an empirical application illustrates system-level directional
causal structure in financial markets.
\end{abstract}



\end{frontmatter}
\tableofcontents

\section{Introduction}

Causal analysis of time series has traditionally been framed through a small number of
dominant paradigms. Predictive approaches, most notably Granger causality and its
extensions, define causality through improvements in conditional prediction based on
temporal ordering \cite{Granger1969,Geweke1982,Eichler2007}. Interventional frameworks
define causality via counterfactual responses to external manipulations within structural
causal models \cite{Pearl2009,PetersJanzingScholkopf2017}, while information-theoretic approaches quantify
directional dependence using asymmetric measures of information flow
\cite{Massey1990,Schreiber2000,Amblard2012}. Each paradigm is mathematically rigorous and
widely used, but each formalizes a distinct causal primitive (predictive improvement,
manipulability, or information transfer) and is primarily designed to detect localized,
edge-level effects.

In high-dimensional economic, financial, networked, and biological systems, influence often manifests through changes in dominant modes, correlation geometry, or collective dependence structure, without producing substantial gains in any individual linear prediction, and may remain stable at the system level while appearing unstable when represented through individual components. Such phenomena are well documented in multivariate analysis and random matrix
theory, where dependence is frequently driven by distributed or low-rank structure rather
than isolated coefficients \cite{Anderson2003,BaiSilverstein2010}. In these settings,
purely predictive or edge-based notions of causality may be statistically underpowered,
unstable, or conceptually misaligned with the underlying mechanism.

This paper introduces a causal framework designed to capture this form of interaction.
Rather than targeting pairwise links or isolated predictive improvements, the proposed
approach treats causality as a property of collective dependence geometry. Specifically,
we define causality as \emph{non-invariance of a family of second-order dependence operators
under admissible, order-preserving temporal deformations of a designated source
component}. Temporal ordering enters the definition structurally, through constrained
deformation, rather than through conditioning, regression, or hypothetical intervention.
Causal influence is assessed via variation across an entire order-indexed operator family,
yielding an axiomatic, order-based causal primitive that is distinct from predictive,
interventional, and information-theoretic formulations. Throughout, causality is understood
in the sense of directional predictive content encoded in second-order dependence structure,
rather than interventional causality.

From a mathematical perspective, the framework treats causality as a property of an
operator family rather than of a single regression, projection, or transfer function.
Dependence is summarized through orthogonally invariant spectral functionals, leading to
causal statistics defined as supremum--infimum dispersions over admissible temporal
deformations. This construction avoids reliance on collections of pairwise edges and
instead captures joint directional effects acting on subspaces of the system. Such a
representation is particularly relevant in financial and economic systems, where causal
influence often manifests through coordinated group behavior and low-dimensional modes
rather than isolated bilateral interactions.

A key aspect of the framework is that it operates at the level of second-order structure
in a chosen feature representation. The underlying notion of causality is invariant to the
specific parametrization of the feature space, and nonlinearity is introduced entirely
through the choice of feature maps applied to the observed processes. In particular, the
framework is second-order in the feature space, but not necessarily linear in the original
variables: nonlinear dependence structures can be captured through appropriate embeddings
without altering the causal definition, the order constraint, or the inferential procedure.

An operator interpretation clarifies the relationship to classical notions of causality.
Directed influence can be represented through a whitened cross-covariance (directed
coherence) operator whose spectrum characterizes maximal correlations over linear
projections. Under restrictive linear Gaussian assumptions, the deformation-based
criterion coincides exactly with linear Granger causality. Outside this regime, admissible
temporal deformation can alter the geometry of second-order dependence without inducing
any change in linear predictability, particularly when directional influence is mediated
through nonlinear, high-rank, or distributed transformations. This distinction is
empirically relevant in high-dimensional systems, as demonstrated by both simulations and
financial applications in this paper.

We establish that this supremum--infimum dispersion functional is not an arbitrary
design choice but the unique diagnostic within the proposed class compatible with a minimal set of
structural axioms: order consistency, orthogonal invariance, Loewner monotonicity,
second-order sufficiency, and continuity. Several classical directionality
notions---including linear Granger causality, directed coherence, and Geweke
frequency-domain causality---arise as special cases under explicit restrictions
on features and deformation sets. Conversely, an information-theoretic impossibility
theorem shows that entrywise-stable edge-based procedures cannot achieve nontrivial
power in distributed dependence regimes unless the sample size scales quadratically
in the feature dimension, whereas the proposed spectral tests detect at the optimal
linear scale in this regime. These results provide a formal justification for the shift from
edge-level to operator-level causal analysis in high-dimensional systems.

To address regimes in which causal influence is distributed rather than concentrated, we
extend the framework from scalar spectral summaries to the full spectral distribution of
the dependence operators. This spectral-measure extension detects causal structure driven
by bulk redistribution of dependence and strictly dominates any fixed collection of
scalar or edge-based criteria, while preserving invariance and interpretability.

From an inferential standpoint, the proposed causal statistics are non-smooth, involving
suprema and infima over admissible deformations. We establish existence and uniform
consistency of the causal functionals and develop randomization-based inference procedures
that exploit order-induced group invariance under the null of causal invariance. These
procedures yield exact or asymptotically valid inference under weak dependence without
requiring parametric assumptions or functional central limit theorems.

The framework is deliberately minimal and model free. It does not aim to replace
interventional notions of causality or to identify counterfactual effects of the form
$\mathbb{E}[Y \mid \mathrm{do}(X=x)]$. Instead, it introduces a complementary causal
primitive tailored to high-dimensional systems where interventions are infeasible and
causal influence is inherently collective: sensitivity of dependence geometry to
admissible, order-preserving temporal deformation. The resulting methodology is intended
for system-level analysis and monitoring rather than structural parameter identification.

The empirical analysis illustrates the practical implications of this perspective in a large financial system. Directional causal organization is episodic rather than persistent, intensifying during periods of market stress. These episodes are characterized by increases in low-dimensional spectral concentration, while the effective dimensionality of dependence remains broadly stable. At the same time, coordinate-level representations of influence exhibit substantial instability: the identities of dominant causal hubs vary considerably across episodes, even when aggregate spectral structure is robust. Transmission at the edge level remains sparse and heterogeneous, with only a small subset of statistically significant channels exhibiting amplification and nontrivial propagation delays. Together, these findings highlight a central distinction between stable system-level directional structure and unstable coordinate-level representations, and illustrate the value of operator-based causal diagnostics in high-dimensional settings.

The remainder of the paper is organized as follows. Section~\ref{sec:related} reviews
related work and positions the proposed approach within predictive, interventional,
invariance-based, information-theoretic, and spectral frameworks for causal analysis in
time series. Section~\ref{sec:framework} develops the order-constrained spectral
framework: Part~I presents the constructive theory, including the causal definition,
operator interpretation, relationship to linear Granger causality, spectral distribution
extension, and inferential procedures; Part~II establishes the theoretical foundations,
including the axiomatic characterization, systematic comparison with existing paradigms,
exact unification with classical directionality notions, and impossibility results for
edge-based methods in distributed regimes. Section~\ref{sec:implementation} presents the
operator construction and practical implementation in asymmetric and fully multivariate
settings. Section~\ref{sec:simulations} reports simulation studies assessing finite-sample
behavior across edge-dominated, bulk-dominated, and nonlinear regimes, and presents a
large-scale empirical study of global financial markets illustrating system-level causal
monitoring in high dimensions. Section~\ref{sec:conclusion} concludes with a discussion
of implications, limitations, and directions for future research. All technical proofs
are collected in the appendices and supplementary material.

\section{Related Work}
\label{sec:related}

The study of directionality and causality in time series spans econometrics,
statistics, information theory, and signal processing. The present work is related
to, but distinct from, four major strands: (i) predictive notions of causality
based on Granger-type criteria; (ii) interventional and structural causal models;
(iii) invariance- and constraint-based approaches to causal structure; and (iv)
spectral and operator-theoretic summaries of dependence. Our framework connects
most directly to (iii) and (iv), while coinciding with (i) only under restrictive
linear and Gaussian assumptions.

\subsection{Predictive Causality and Granger-Type Criteria}

Predictive causality is most commonly formalized through Granger causality, which
declares a component directional if its past improves prediction of another
component beyond what is achievable using the latter's own past
\cite{Granger1969}. This notion has been extensively developed in multivariate
settings, including measures of linear feedback \cite{Geweke1982}, graphical
representations for vector autoregressive processes \cite{Eichler2007}, and
frequency-domain decompositions \cite{Geweke1984}. Frequency-domain formulations
express Granger-type directionality through spectral factorization and transfer
functions, leading to directed transfer functions and partial directed coherence
\cite{Kaminski1991,Baccala2001}. These approaches provide reliable inference for
linear predictive dependence under parametric assumptions
\cite{Barnett2014}.

By construction, predictive criteria target incremental improvements in
conditional prediction error at the level of individual components. When
directional structure manifests through collective, distributed, or high-rank
changes in dependence geometry, predictive gains for any single component may be
weak or absent, motivating causal formulations that operate beyond localized
predictive effects. Section~\ref{subsec:impossibility-quantitative} formalizes this
limitation as an information-theoretic impossibility.

\subsection{Interventional and Structural Causal Models}

A complementary paradigm defines causality through interventions and
counterfactual reasoning, most prominently via structural causal models (SCMs)
\cite{Pearl2009}. Causal effects are defined through stability of structural
mechanisms under manipulation and are typically represented by directed graphs,
with statistical treatments emphasizing identifiability under explicit
assumptions \cite{AngristPischke2009,ImbensRubin2015}. Extensions to time series
include structural vector autoregressions and state-space representations
\cite{HannanDeistler2012LinearSystems}, often assuming known or partially known
temporal ordering.

The present work does not invoke interventions and does not claim equivalence to
interventional causality. Instead, it defines directionality through
order-constrained invariance of second-order dependence structure, which we view
as complementary to intervention-based notions, particularly in high-dimensional
systems where interventions are infeasible and causal influence is inherently
collective.

\subsection{Invariance-, Constraint-, and Order-Based Perspectives}

Beyond predictive and interventional paradigms, causal structure has long been
studied through invariance and stability properties rather than explicit
manipulation. In econometrics, this perspective traces back to
\cite{Haavelmo1944}, while more recent work formalizes causality as persistence
across admissible environments or transformations
\cite{Peters2016,PetersJanzingScholkopf2017,Pfister2019ICPSequential}. Related ideas appear in
anchor regression and domain adaptation, where causal structure is characterized
through robustness to perturbations \cite{Rothenhausler2021}.

A closely related line of work adopts a constraint-based view of causality,
recovering directed structure via conditional independence testing and graphical
models. Classical PC- and FCI-type algorithms are known to suffer from order
dependence in high-dimensional settings, a limitation addressed by
order-independent variants \cite{JMLR:v15:colombo14a}. These ideas have been
extended explicitly to time series through methods such as PCMCI and related
algorithms, which exploit temporal ordering and conditional independence tests to
recover causal graphs under faithfulness assumptions
\cite{Runge2019}. Related approaches formulate causal
discovery as a constrained optimization or constraint satisfaction problem,
enforcing temporal precedence and independence constraints to identify admissible
graphical structures \cite{Malinsky2018OptimizationCausal}.

While these methods impose temporal or logical constraints on admissible causal
graphs, they remain focused on recovering edge-level structure through
conditional independence logic. In contrast, the present framework treats
temporal order as a constraint on admissible deformations of second-order
dependence structure and defines causal directionality directly through
non-invariance under order-preserving temporal realignment, without aiming to
recover a causal graph.

Order constraints have also played a central role in statistical inference more
broadly, including order-restricted estimation
\cite{Robertson1988OrderRestricted} and quantile rearrangement inequalities
\cite{Chernozhukov2009Rearrangement}.
These works formalize shape or order restrictions as admissible transformations
but do not address directionality or causal asymmetry in multivariate time series.

\subsection{Spectral and Operator-Theoretic Approaches}

Spectral summaries are central in multivariate analysis, where eigenvalues and
eigenvectors of covariance operators provide canonical descriptions of dependence
\cite{Anderson2003}. In time series analysis, spectral methods are foundational,
particularly in the frequency domain \cite{Brillinger1981}, and canonical
correlation analysis motivates the use of whitened cross-covariance operators
\cite{Hotelling1936}. Operator-valued measures of dependence and conditional
structure have also been developed in Hilbert-space settings
\cite{Bosq2000,Gretton2005,Fukumizu2007}.

Random matrix theory further clarifies how collective dependence structure
manifests through spectral bulk behavior and low-rank perturbations
\cite{Baik2005,BaiSilverstein2010}. In financial applications, eigenvalue dynamics
of correlation matrices have been used to detect structural change and systemic
stress \cite{Bouchaud2005,BouchaudPotters2009}. These approaches focus on symmetric
dependence and do not encode temporal ordering, admissible directional asymmetry,
or causal invariance under deformation.

A closely related contribution is \cite{RodriguezDominguezYadav2024}, which uses
extreme variations of the leading eigenvalue of lagged correlation matrices to
detect directional interaction when predictive measures are weak, but is
inherently pairwise, scalar, and tied to fixed lag structures. The present work
generalizes these spectral perspectives by defining causal directionality through
order-based spectral deformation of families of second-order operators. Unlike
dependence-only diagnostics, the proposed framework explicitly encodes temporal
ordering and admissible directional deformation, yielding a causal primitive that
is sensitive to collective directional structure while remaining agnostic to
specific parametric or interventional assumptions.

We use the term \emph{directional causality} exclusively to denote
order-constrained spectral non-invariance as defined in
Section~\ref{sec:framework}. Predictive improvement, linear Granger causality, and
interventional causal effects are referred to explicitly when intended and are not
used interchangeably.

\section{Order-Constrained Spectral Framework}
\label{sec:framework}

This section develops the theoretical framework underlying the proposed notion of
order-constrained spectral causality. The framework proceeds in two parts.
Part~I (Sections~\ref{subsec:setup}--\ref{subsec:inference}) develops the constructive
framework: the causal definition, its operator interpretation, its relationship to
classical notions, the spectral distribution extension, and the associated inferential
theory. Part~II (Sections~\ref{subsec:axiomatic}--\ref{subsec:impossibility-quantitative}) provides
the theoretical foundations: an axiomatic characterization establishing uniqueness,
a systematic comparison with existing paradigms, exact unification with classical
directionality notions, and impossibility results for edge-based methods in
distributed regimes.

\subsection{Basic Setup and Admissible Temporal Deformations}
\label{subsec:setup}

Let $\{X_t\}_{t\in\mathbb{Z}}$ be a strictly stationary stochastic process in $\mathbb{R}^K$
with $\mathbb{E}X_t=0$ and finite second moments. No parametric or distributional
assumptions are imposed unless stated otherwise. The time index induces a fixed total
order, which is assumed to be meaningful and invariant throughout the analysis.

Directionality is introduced through restrictions on admissible temporal deformations.
Let $\mathcal P \subset \mathbb{R}^m$ denote a nonempty collection of lag configurations
such that each $\tau \in \mathcal P$ preserves temporal order. Typical examples include
finite sets of nonnegative integer lags or compact sets defined by linear constraints such
as $\tau \ge 0$ and $\|\tau\|_1 \le \tau_{\max}$. The set $\mathcal P$ is fixed by the
analyst and encodes which order-preserving temporal displacements are regarded as
meaningful in a given application.

Fix distinct components $i \neq j$. For each $\tau \in \mathcal P$, we consider an
asymmetric deformation of the system obtained by temporally displacing the source
component $X^{(i)}$ according to $\tau$, while leaving all other components unchanged.
This deformation protocol encodes directionality structurally and does not rely on
predictive modeling, conditioning on future information, or hypothetical interventions.

\subsection{Order-Constrained Spectral Causality}
\label{subsec:ocsc}

Let $\{X_t\}_{t\in\mathbb{Z}}\subset\mathbb{R}^d$ be a stochastic process, and fix
components $i$ and $j$.
Let $\mathcal{P}\subset\mathbb{R}_+$ be a set of admissible temporal deformations
(e.g., lags), and for each $\tau\in\mathcal{P}$ let $C_{i\to j}(\tau)\in\mathbb{S}_+^d$
denote a dependence operator constructed from the deformed source $X^{(i)}_{t-\tau}$
and the target $X^{(j)}_t$.

The operator $C_{i\to j}(\tau)$ is assumed to be a second-order object encoding
dependence between suitably defined feature representations of source and target.
Although $C_{i\to j}(\tau)$ is symmetric positive semidefinite, directional
content does not arise from asymmetry of the operator itself, but from the
asymmetric construction of source and target under temporal deformation.
In particular, when the operator is formed from stacked feature vectors
$Z_t(\tau)=(V_t^\top,U_t(\tau)^\top)^\top$, its block structure takes the form
\[
C_{i\to j}(\tau)
=
\begin{pmatrix}
\Sigma_{VV} & \Sigma_{VU}(\tau) \\
\Sigma_{UV}(\tau) & \Sigma_{UU}(\tau)
\end{pmatrix},
\]
where $\Sigma_{VU}(\tau)=\mathrm{Cov}(V_t,U_t(\tau))$ captures cross-dependence
between target and lagged source features. Directional dependence is therefore
encoded through these cross-components and their interaction with the marginal
blocks, rather than through any intrinsic asymmetry of the operator.

\begin{definition}[Order-Constrained Spectral Causality]
\label{def:ocsc}
We say that $i$ does not Granger-cause $j$ under order-constrained spectral causality
if
\[
C_{i\to j}(\tau)=C_{i\to j}(\tau')
\quad \text{for all } \tau,\tau'\in\mathcal{P}.
\]
Equivalently, causality is defined as non-invariance of the dependence operator
under admissible temporal deformations.
\end{definition}

This definition formalizes directional dependence as a property of the family
$\{C_{i\to j}(\tau)\}_{\tau\in\mathcal{P}}$, rather than of any single operator.
Under the null hypothesis, the spectral structure of $C_{i\to j}(\tau)$ remains
invariant across $\tau$, while under the alternative, temporal deformation induces
systematic changes in the operator and its spectrum. To obtain scalar test statistics, we consider spectral functionals of
$C_{i\to j}(\tau)$. Let $\lambda_1(\tau)\ge\cdots\ge\lambda_d(\tau)\ge0$
denote the eigenvalues of $C_{i\to j}(\tau)$.
For a measurable function $f:\mathbb{R}_+\to\mathbb{R}$, define
\[
L_f(\tau)
=
\frac{1}{d}\sum_{r=1}^d f\!\big(\lambda_r(\tau)\big).
\]
The order-constrained dispersion functional is then given by
\[
T_f
=
\sup_{\tau\in\mathcal{P}} L_f(\tau)
-
\inf_{\tau\in\mathcal{P}} L_f(\tau).
\]

Under the null hypothesis of invariance, $T_f=0$; under the alternative, $T_f>0$
captures variation of spectral summaries across admissible deformations.
This formulation is invariant under orthogonal transformations of the feature
space and does not depend on a particular coordinate representation. More generally, the full spectral distribution may be used by defining the empirical
spectral measure
\[
\mu_\tau=\frac{1}{d}\sum_{r=1}^d \delta_{\lambda_r(\tau)},
\]
and considering dispersion in the space of probability measures,
\[
T_{\mathrm{spec}}
=
\sup_{\tau_1,\tau_2\in\mathcal{P}}
d\!\left(\mu_{\tau_1},\mu_{\tau_2}\right),
\]
for a suitable metric $d(\cdot,\cdot)$.
This extension captures changes in the entire spectral distribution rather than
only scalar summaries and provides sensitivity to both concentrated and diffuse
forms of directional dependence.

\subsection{Operator Interpretation and Variational Characterization}
\label{subsec:operator}

For a given admissible deformation $\tau\in\mathcal P$, let $U_\tau$ denote a vector of
features derived from the lagged source component $X^{(i)}$, and let $V$ denote features
derived from the target component $X^{(j)}$.
Conditional analysis may be incorporated via residualization, but is not required for the
definition. Define covariance blocks
\[
\Sigma_{UU}(\tau)=\mathrm{Cov}(U_\tau),\qquad
\Sigma_{VV}=\mathrm{Cov}(V),\qquad
\Sigma_{VU}(\tau)=\mathrm{Cov}(V,U_\tau),
\]
with inverse square roots interpreted as Moore--Penrose pseudoinverses when necessary. Define the directed coherence operator
\[
A(\tau)
=
\Sigma_{VV}^{-1/2}\Sigma_{VU}(\tau)\Sigma_{UU}(\tau)^{-1/2}.
\]
Its operator norm
\[
\kappa(\tau)=\|A(\tau)\|_2
\]
admits the variational characterization
\[
\kappa(\tau)
=
\sup_{\|a\|=\|b\|=1}
\mathrm{Corr}\!\left(a^\top V,\; b^\top U_\tau\right),
\]
identifying $\kappa(\tau)$ as the largest canonical correlation between $V$ and $U_\tau$
\citep{Anderson2003}.

Within this formulation, causality corresponds to non-invariance of $\kappa(\tau)$, or of
more general spectral characteristics of $A(\tau)$, over $\mathcal P$.
This operator viewpoint clarifies the distinction between edge-dominated and distributed
directional structure and motivates the computational procedures developed later. It also highlights a key representation effect: while spectral properties of the operator may remain stable under admissible deformations, their coordinate-level projections (e.g., leading directions or loadings) need not be stable, particularly in high-dimensional or near-degenerate regimes.

\subsection{Relation to Linear Granger Causality}
\label{subsec:granger}

To situate the proposed criterion, we focus on linear Granger causality, the canonical
order-based predictive notion for time series
\citep{Granger1969,Geweke1982,Eichler2007}.
Linear Granger causality is defined entirely in terms of temporal ordering and linear
predictability and admits a precise Hilbert-space formulation via orthogonal projections.

\subsubsection{Projection Invariance Formulation}

Let $\mathcal H_{t-1}^{(p)}=\mathrm{span}\{X_{t-1},\dots,X_{t-p}\}$ and
$\mathcal H_{t-1}^{(-i,p)}$ denote the corresponding information sets with and without the
$i$th component.

\begin{definition}[Linear Granger noncausality]
Component $i$ is linearly Granger-noncausal for component $j$ at order $p$ if
\[
\Pi_{\mathcal H_{t-1}^{(p)}} X_t^{(j)}
=
\Pi_{\mathcal H_{t-1}^{(-i,p)}} X_t^{(j)}.
\]
\end{definition}

Thus, linear Granger causality is fundamentally a statement about invariance of linear
projections under removal of lagged information.

\subsubsection{Coincidence under Linear Gaussian Dynamics}

\begin{theorem}[Coincidence under Gaussian VAR($p$)]
\label{thm:granger_coincidence_main}
Suppose $\{X_t\}$ follows a stable Gaussian VAR($p$) process with correctly specified lag
order, nonsingular innovation covariance, and no omitted variables. Let $\mathcal{P}
= \{1,\ldots,p\}$ and define order-constrained spectral causality using the residualized
directed coherence operator. Then, for any $i \neq j$, linear Granger noncausality,
vanishing directed coherence, and zero VAR coefficients are equivalent.
\end{theorem}

Thus, linear Granger causality arises as a special case of order-constrained spectral
causality when dependence geometry is fully determined by linear second-order structure. Outside this restricted regime, the notions generally diverge. In particular, order-constrained
spectral causality may detect directional deformation of dependence geometry even when
linear predictive relationships are absent or cancel under projection.

\subsubsection{Distinctness beyond Linear Predictability}

Outside the linear Gaussian class, admissible temporal deformation may alter second-order
dependence geometry without inducing any change in linear predictability.

\begin{theorem}[Distinctness under nonlinear dependence]
There exist stationary processes for which linear Granger causality fails at all finite
orders, while order-constrained spectral causality holds.
\end{theorem}

Here, deformation changes the alignment of nonlinear or distributed features, producing
spectral variation without affecting linear projections.
This reflects genuine structural directionality rather than a pathological exception. Order-constrained spectral causality defines a structural, order-based notion of directional
dependence. It is neither necessary nor sufficient for interventional (SCM-style) causal effects
without additional assumptions, such as absence of latent confounding.
These limitations are stated explicitly to avoid over-interpretation.

\subsection{Spectral Distribution Extension and Collective Effects}
\label{subsec:spectral_extension_core}

The causal definition introduced above is formulated in terms of non-invariance of scalar
spectral summaries of the lag-indexed dependence operators
$\{C(\tau)\}_{\tau\in\mathcal P}$.
Such summaries are effective when directional influence manifests through amplification or
attenuation of a dominant mode.
However, in high-dimensional or weak-signal regimes, admissible temporal deformation may
redistribute dependence across multiple modes without substantially affecting any single
eigenvalue.

To capture such effects, we consider an extension based on the \emph{entire spectral
distribution} of the dependence operator.
Importantly, this extension does not redefine causality.
It strengthens operational sensitivity while preserving the same order-based invariance
criterion defined in Section~\ref{subsec:ocsc}. For each admissible deformation $\tau\in\mathcal P$, let
$C(\tau)\in\mathbb{S}_+^d$ denote the dependence operator introduced earlier, with ordered
eigenvalues
\[
\lambda_1(\tau)\ge \cdots \ge \lambda_d(\tau)\ge 0.
\]
Define the empirical spectral distribution
\[
\mu_\tau := \frac{1}{d}\sum_{r=1}^d \delta_{\lambda_r(\tau)}.
\]

The family $\{\mu_\tau:\tau\in\mathcal P\}$ provides an orthogonally invariant
representation of how second-order dependence geometry changes under admissible,
order-preserving temporal deformation. Stability at the level of spectral distributions
does not imply stability of the associated eigenvectors or coordinate representations,
which may vary substantially even when global dependence geometry is preserved. Scalar criteria arise as projections of $\mu_\tau$.
For any integrable function $f:\mathbb{R}_+\to\mathbb{R}$, define the associated linear
spectral statistic
\[
L_f(\tau)
=
\int f(\lambda)\,d\mu_\tau(\lambda)
=
\frac{1}{d}\sum_{r=1}^d f\!\big(\lambda_r(\tau)\big).
\]
Convex functions $f$ emphasize edge behavior, while smooth bounded functions emphasize
bulk structure. All scalar summaries used in the core framework correspond to specific
choices of $f$. Causality is assessed via dispersion of either scalar or measure-level quantities over
$\mathcal P$. For a fixed $f$, define
\[
T_f := \sup_{\tau\in\mathcal P} L_f(\tau) - \inf_{\tau\in\mathcal P} L_f(\tau).
\]
More generally, let $d(\cdot,\cdot)$ be a metric on probability measures on $\mathbb{R}_+$,
and define
\[
T_{\mathrm{spec}}
:=
\sup_{\tau_1,\tau_2\in\mathcal P}
d\!\left(\mu_{\tau_1},\mu_{\tau_2}\right).
\]

Spectral-measure dispersion detects directional dependence whenever admissible temporal
deformation induces any change in second-order dependence geometry, regardless of whether
that change is concentrated in a single mode or distributed across many. It therefore
provides a maximally sensitive second-order criterion within the order-constrained
framework.

The relationship between scalar and spectral-measure criteria is exact.
If $\mu_{\tau_1}=\mu_{\tau_2}$, then $L_f(\tau_1)=L_f(\tau_2)$ for all integrable $f$.
Conversely, equality of $L_f$ over a separating class of functions implies equality of
spectral measures. Thus, scalar criteria are complete if and only if they span a separating
class. When the bounded-Lipschitz metric is used, spectral-measure dispersion admits the
representation
\[
T_{\mathrm{spec}}
=
\sup_{\tau_1,\tau_2\in\mathcal P}
\sup_{\|f\|_{\mathrm{BL}}\le1}
\big|L_f(\tau_1)-L_f(\tau_2)\big|,
\]
showing that measure-based causality can be equivalently expressed as a uniform supremum
over normalized linear spectral statistics.

In practice, spectral-measure dispersion is primarily useful in high-dimensional or
distributed regimes. In low-dimensional or strongly rank-one settings, scalar criteria may
suffice. All inferential results developed in the next subsection apply to both
formulations without modification.

\subsection{Asymptotic Theory and Inference}
\label{subsec:inference}

We establish existence, consistency, and valid inference for the proposed
order-constrained spectral causal functionals.
The primary objects of interest are linear spectral statistics of the
lag-indexed dependence operators $\{C(\tau)\}_{\tau\in\mathcal P}$
evaluated uniformly over the admissible deformation set.
Edge-based summaries, such as the largest eigenvalue or directed coherence norm,
arise as non-smooth special cases and are treated as optional refinements rather
than core components. Throughout this subsection, the feature dimension $d$ is
treated as fixed. Extensions to regimes in which $d$ grows with the sample size
are beyond the scope of the present work.

For each admissible deformation $\tau\in\mathcal P$, let $Z_t(\tau)\in\mathbb{R}^d$ denote
the (possibly residualized) feature vector used to construct the dependence operator.
Define the population operator
\[
C(\tau) := \mathbb{E}\!\left[ Z_t(\tau) Z_t(\tau)^\top \right],
\]
and its empirical estimator
\[
\widehat C_T(\tau) := \frac{1}{T}\sum_{t=1}^T Z_t(\tau) Z_t(\tau)^\top.
\]

Let $\lambda_1(\tau)\ge\cdots\ge\lambda_d(\tau)$ and
$\widehat\lambda_1(\tau)\ge\cdots\ge\widehat\lambda_d(\tau)$ denote the eigenvalues of
$C(\tau)$ and $\widehat C_T(\tau)$, respectively.
For a measurable function $f:\mathbb{R}_+\to\mathbb{R}$, define the linear spectral
statistics
\[
L_f(\tau)=\frac{1}{d}\sum_{r=1}^d f\!\left(\lambda_r(\tau)\right),
\qquad
\widehat L_f(\tau)=\frac{1}{d}\sum_{r=1}^d f\!\left(\widehat\lambda_r(\tau)\right),
\]
and the associated dispersion functionals
\[
T_f=\sup_{\tau\in\mathcal P} L_f(\tau)-\inf_{\tau\in\mathcal P} L_f(\tau),
\qquad
\widehat T_f=\sup_{\tau\in\mathcal P} \widehat L_f(\tau)
-\inf_{\tau\in\mathcal P} \widehat L_f(\tau).
\]

\subsubsection{Causal Null Hypothesis}

Inference is formulated relative to a null hypothesis defined in terms of
order-constrained spectral invariance of the lag-indexed operator family.

\begin{definition}[Null of causal invariance]
\label{def:null}
The null hypothesis of absence of causal influence from component $i$ to component $j$ is
\[
H_0:\quad L_f(\tau)\ \text{is constant over }\tau \in \mathcal P,
\]
equivalently $T_f = 0$.
\end{definition}

Under $H_0$, admissible temporal deformation of the source component leaves the
second-order dependence geometry invariant, as summarized by the chosen spectral
functional $f$. The alternative corresponds to \emph{order-constrained spectral
non-invariance} of the operator family, manifested through variation of $L_f(\tau)$ over
$\mathcal P$.

The null hypothesis $H_0 : T_f = 0$ corresponds to \emph{global invariance} of the chosen
second-order dependence summary across all admissible temporal deformations in
$\mathcal{P}$ and is therefore stronger than the absence of a localized causal effect at a
single lag. Failure to reject $H_0$ does not imply the absence of directional dependence,
but rather the absence of deformation-sensitive directional structure over the admissible
set.

\paragraph{Relation to aggregated operators.}
The inferential framework is defined entirely in terms of dispersion over the
lag-indexed family $\{C(\tau)\}_{\tau\in\mathcal P}$.
In later sections, aggregated operators of the form
\[
C = \sum_{\tau\in\mathcal P} w_\tau C(\tau)
\]
are used for monitoring and representation.
Such aggregation removes the index over $\tau$ and therefore does not preserve
order-dependent variation. Consequently, inference based on $\widehat T_f$
targets invariance across $\tau$, while aggregated operators provide
complementary summaries of directional dependence once invariance is violated.

\subsubsection{Assumptions}

Let $\|\cdot\|$ denote the operator norm.
The following assumptions are standard for covariance operators and linear spectral
statistics of weakly dependent time series
\citep{Bosq2000,Bradley2005,BaiSilverstein2010}.

\begin{enumerate}
\item[(A1)] \emph{Weak dependence.}
For each $\tau\in\mathcal P$, the process $\{Z_t(\tau)\}$ is strictly stationary and
$\alpha$-mixing with
\[
\sum_{h=1}^\infty \alpha(h)^{\delta/(2+\delta)}<\infty
\quad\text{for some }\delta>0.
\]

\item[(A2)] \emph{Uniform moments.}
\[
\sup_{\tau\in\mathcal P}\mathbb{E}\|Z_t(\tau)\|^{4+\delta}<\infty.
\]

\item[(A3)] \emph{Admissible deformation set.}
Either (i) $\mathcal P$ is finite, or (ii) $\mathcal P$ is compact and
$\tau\mapsto Z_t(\tau)$ is continuous in $L^{4+\delta}$.

\item[(A4)] \emph{Spectral boundedness.}
There exist constants $0<m<M<\infty$ such that
\[
\mathrm{spec}\big(C(\tau)\big)\subset[m,M]
\quad\text{for all }\tau\in\mathcal P.
\]
\end{enumerate}

\subsubsection{Consistency}

Under Assumptions (A1)--(A4), the dependence operators, their spectral summaries, and the
associated dispersion functionals are uniformly consistent.

\begin{theorem}[Uniform consistency]
\label{thm:uniform_consistency}
\[
\sup_{\tau\in\mathcal P}
\big\|\widehat C_T(\tau)-C(\tau)\big\|
\xrightarrow{p}0,
\qquad
\sup_{\tau\in\mathcal P}
\big|\widehat L_f(\tau)-L_f(\tau)\big|
\xrightarrow{p}0,
\]
and consequently $\widehat T_f\xrightarrow{p}T_f$.
\end{theorem}

\subsubsection{Asymptotic Normality}

For any fixed $\tau\in\mathcal P$ and Lipschitz function $f$,
\[
\sqrt{T}\big(\widehat L_f(\tau)-L_f(\tau)\big)
\xrightarrow{d}
\mathcal N\!\big(0,\sigma_f^2(\tau)\big),
\]
where $\sigma_f^2(\tau)$ is a finite long-run variance. Although $T_f$ involves a supremum over $\mathcal P$, inference does not rely on a
functional central limit theorem for the process
$\{\widehat L_f(\tau):\tau\in\mathcal P\}$.
Instead, resampling and randomization procedures approximate the distribution of
$\widehat T_f$ directly under the null hypothesis.
Pointwise asymptotic normality therefore suffices for the inferential development.

\subsubsection{Inference Procedures and Interpretation}

Block bootstrap and stationary bootstrap procedures that preserve temporal dependence yield
consistent approximations to the distribution of $\widehat T_f$ under standard conditions
\citep{Lahiri2003,PolitisRomano1994}.
Alternatively, shift-based randomization exploits invariance of the joint distribution
under circular shifts of the source component when the null holds.
Appendix~\ref{app:rand} formalizes the required invariance condition and establishes
finite-sample exactness under exact invariance and asymptotic validity otherwise.

Rejection of the null implies that admissible temporal deformation induces nontrivial
deformation of the system's second-order dependence geometry, providing evidence of causal
influence in the order-constrained spectral sense.
Such rejection does not, in general, imply an interventional causal effect without
additional assumptions; see Appendix~\ref{app:scope}.

Non-smooth spectral summaries, such as the largest eigenvalue or directed coherence norm,
may be employed as optional refinements.
While consistency follows from the above results, their limiting distributions typically
require stronger assumptions and specialized techniques.
They are therefore not required for the core inferential framework.

\subsection{Axiomatic Characterization}
\label{subsec:axiomatic}

The constructive framework developed above defines order-constrained spectral
causality through a supremum--infimum dispersion functional over the
lag-indexed operator family $\{C(\tau)\}_{\tau\in\mathcal P}$.
This subsection shows that this choice is not arbitrary, but the unique causal
diagnostic compatible with a minimal set of structural requirements.
The result formalizes the sense in which causality, when defined as a property of
collective second-order dependence under temporal ordering, must appear as
extremal spectral variation across admissible deformations.

Let $\{X_t\}_{t\in\mathbb{Z}}$ be a strictly stationary stochastic process in
$\mathbb{R}^K$ with $\mathbb{E}X_t=0$ and $\mathbb{E}\|X_t\|^2<\infty$.
Let $\mathcal{P}\subset\mathbb{R}^m$ be a nonempty compact set of admissible,
order-preserving temporal deformations. For each $\tau\in\mathcal{P}$, let
$C(\tau)\in\mathcal{S}_d^+$ denote a symmetric positive semidefinite operator
summarizing second-order dependence between designated source and target
components under deformation $\tau$.

A \emph{causal criterion} is a functional
\[
\mathcal{C}:\{C(\tau):\tau\in\mathcal{P}\}\longrightarrow\mathbb{R}.
\]

\begin{axiom}[Order consistency]
\label{a1}
If $\pi:\mathcal{P}\to\mathcal{P}$ is a bijection preserving temporal order, then
\[
\mathcal{C}\big(\{C(\tau)\}_{\tau\in\mathcal{P}}\big)
=
\mathcal{C}\big(\{C(\pi(\tau))\}_{\tau\in\mathcal{P}}\big).
\]
\end{axiom}

\begin{axiom}[Orthogonal invariance]
\label{a2}
For any orthogonal matrix $Q\in O(d)$,
\[
\mathcal{C}\big(\{C(\tau)\}_{\tau\in\mathcal{P}}\big)
=
\mathcal{C}\big(\{Q C(\tau) Q^\top\}_{\tau\in\mathcal{P}}\big).
\]
\end{axiom}

\begin{axiom}[Monotonicity under Loewner strengthening]
\label{a3}
If two operator families $\{C_1(\tau)\}$ and $\{C_2(\tau)\}$ satisfy
$C_1(\tau)\preceq C_2(\tau)$ for all $\tau\in\mathcal{P}$ in the Loewner order,
then
\[
\mathcal{C}\big(\{C_1(\tau)\}_{\tau\in\mathcal{P}}\big)
\le
\mathcal{C}\big(\{C_2(\tau)\}_{\tau\in\mathcal{P}}\big).
\]
\end{axiom}

\begin{axiom}[Second-order sufficiency]
\label{a4}
The value of $\mathcal{C}$ depends on the process $\{X_t\}$ only through the
operator family $\{C(\tau)\}_{\tau\in\mathcal{P}}$.
\end{axiom}

\begin{axiom}[Continuity in admissible deformations]
\label{a5}
The mapping $\tau\mapsto C(\tau)$ is continuous in operator norm, and
$\mathcal{C}$ is continuous with respect to the induced uniform topology on
$\{C(\tau)\}_{\tau\in\mathcal{P}}$.
\end{axiom}

\begin{theorem}[Uniqueness up to spectral functional]
\label{thm:axiomatic-uniqueness}
Let $\mathcal{C}$ be a causal criterion satisfying Axioms~1--5. Then there exists
a continuous, orthogonally invariant spectral functional
$\varphi:\mathcal{S}_d^+\to\mathbb{R}$ such that
\[
\mathcal{C}\big(\{C(\tau)\}_{\tau\in\mathcal{P}}\big)
=
\sup_{\tau\in\mathcal{P}}\varphi\big(C(\tau)\big)
-
\inf_{\tau\in\mathcal{P}}\varphi\big(C(\tau)\big).
\]
Conversely, any functional of the above form satisfies Axioms~1--5.
\end{theorem}

The proof, given in Appendix~\ref{app:axiomatic-proof}, relies on representation
results for orthogonally invariant operator functionals and extremal aggregation
under monotonicity.
Theorem~\ref{thm:axiomatic-uniqueness} shows that order-constrained spectral
dispersion is the unique causal diagnostic compatible with order consistency,
orthogonal invariance, monotonicity under Loewner strengthening, second-order
sufficiency, and continuity.
Under these minimal commitments, causality must appear as extremal variation of
a spectral functional across admissible temporal deformations.

\subsection{Relationship to Existing Causality Notions via Axiom Failures}
\label{subsec:axiom-map}

The axioms above characterize causal diagnostics based on the operator family
$\{C(\tau)\}_{\tau\in\mathcal P}$.
We now relate common causality notions to the axioms they violate.
This comparison clarifies that different paradigms correspond to different causal
primitives and are therefore not interchangeable.

\paragraph{Predictive Granger causality and VAR-based tests.}
Linear Granger causality is defined via improvements in conditional prediction.
As a functional of second-order operators, it is not invariant under general
orthogonal transformations mixing source and target subspaces, and therefore
violates Axiom~\ref{a2} unless restricted to partition-preserving transforms.
It may also violate Axiom~\ref{a3} in high-dimensional settings where dependence is
distributed across many directions.

\paragraph{Transfer entropy and directed information.}
Information-theoretic notions depend on the full joint distribution and are not
functions of $\{C(\tau)\}$ alone.
They therefore violate Axiom~\ref{a4}.
They are also generally not invariant under arbitrary linear reparameterizations,
so Axiom~\ref{a2} may fail.

\paragraph{Structural causal models and interventional causality.}
Interventional quantities such as $\mathbb{E}[Y\mid do(X=x)]$ are not determined by
the second-order operator family without additional assumptions.
Thus, they violate Axiom~\ref{a4}.
This is consistent with the interpretation limits discussed in
Appendix~\ref{app:scope}.

\paragraph{Graphical VAR and edge-based methods.}
Edge-based procedures are not invariant under orthogonal mixing of coordinates,
violating Axiom~\ref{a2}.
They may also violate Axiom~\ref{a3} when dependence is distributed across many
directions, reducing detectability of individual edges despite stronger global
dependence.

\paragraph{Frequency-domain causality.}
Frequency-domain measures depend on spectral factorization and model structure,
and are not purely functions of $\{C(\tau)\}$ without additional assumptions.
They therefore violate Axiom~\ref{a4} in the model-free sense and may fail
Axiom~\ref{a1} when representation choices affect the outcome.

\paragraph{Invariant causal prediction.}
ICP methods define causality via invariance across environments rather than
temporal deformations.
They therefore violate Axiom~\ref{a1} unless environments coincide with
$\mathcal P$, and typically violate Axiom~\ref{a4}.

\medskip
In summary, most existing paradigms fail at least one axiom because they either
(i) depend on distributional or interventional structure beyond second-order
operators (violating Axiom~\ref{a4}), (ii) are not invariant under orthogonal
reparameterization (violating Axiom~\ref{a2}), or (iii) do not respect monotonicity
under Loewner strengthening (violating Axiom~\ref{a3}).
The present framework is tailored to settings in which causal influence is
collective, order-constrained, and captured through second-order geometry.

\begin{table}[t]
\caption{Compatibility of common causality paradigms with Axioms~\ref{a1}--\ref{a5}.}
\label{tab:axiom-failures}
\begin{tabular}{@{}lccccc@{}}
\hline
Method / notion
& A1 Order
& A2 Ortho.\ inv.
& A3 Monotone
& A4 2nd-order
& A5 Cont. \\
\hline
Linear Granger
& $\checkmark$
& $\times$
& $\times$
& $\checkmark$
& $\checkmark$ \\
Graphical VAR
& $\times$
& $\times$
& $\times$
& $\checkmark$
& $\times$ \\
Transfer entropy / DI
& $\checkmark$
& $\times$
& $\checkmark$
& $\times$
& $\checkmark$ \\
SCM / interventional
& $\checkmark$
& $\checkmark$
& $\checkmark$
& $\times$
& $\checkmark$ \\
ICP
& $\times$
& $\checkmark$
& $\times$
& $\times$
& $\checkmark$ \\
\hline
Order-constrained spectral (this work)
& $\checkmark$
& $\checkmark$
& $\checkmark$
& $\checkmark$
& $\checkmark$ \\
\hline
\end{tabular}
\end{table}

\subsection{Unification with Classical Notions and Edge-Based Impossibility}
\label{subsec:unification-and-impossibility}

\subsubsection{Exact Unification: Granger, Directed Coherence, and Lead--Lag Asymmetry}
\label{subsubsec:unification}

This subsection establishes that several classical notions of directional
dependence arise as special cases of the order-constrained spectral framework
under specific choices of (i) feature maps, (ii) admissible deformation sets,
and (iii) spectral functionals.

The construction proceeds from the lag-indexed operator family
$\{C(\tau)\}_{\tau\in\mathcal P}$ introduced in Section~\ref{subsec:ocsc}.
As discussed there, $C(\tau)$ is a symmetric positive semidefinite operator
whose directional content is encoded in its cross-block components.

\paragraph{Lag-embedded feature specialization.}
Fix components $i\neq j$. Let $p,q\in\mathbb{N}$ and define lag-embedded
vectors
\[
u_t(\tau)=\big(X^{(i)}_{t-\tau},\,X^{(i)}_{t-\tau-1},\ldots,X^{(i)}_{t-\tau-p+1}\big)^\top\in\mathbb{R}^p,
\qquad
v_t=\big(X^{(j)}_{t},\,X^{(j)}_{t-1},\ldots,X^{(j)}_{t-q+1}\big)^\top\in\mathbb{R}^q.
\]
Let $\Psi,\Phi$ be identity maps and define
\[
Z_t(\tau)=(v_t^\top,u_t(\tau)^\top)^\top\in\mathbb{R}^{q+p}.
\]

The corresponding second-order operator
\[
C(\tau)=\mathbb{E}[Z_t(\tau)Z_t(\tau)^\top]
=
\begin{pmatrix}
\Sigma_{VV} & \Sigma_{VU}(\tau)\\
\Sigma_{UV}(\tau) & \Sigma_{UU}(\tau)
\end{pmatrix}
\]
has directional content entirely encoded in the cross-block
$\Sigma_{VU}(\tau)=\mathbb{E}[v_t u_t(\tau)^\top]$. To obtain an orthogonally invariant representation of this directional
dependence, we consider the whitened cross-operator
\begin{equation}
\label{eq:whitened-cross}
A(\tau)
=
\Sigma_{VV}^{-1/2}\,\Sigma_{VU}(\tau)\,\Sigma_{UU}(\tau)^{-1/2},
\qquad
D(\tau)=A(\tau)A(\tau)^\top\in\mathcal{S}_q^+.
\end{equation}

The operator $D(\tau)$ is a reparameterization of the cross-block dependence
in $C(\tau)$ and is invariant to invertible linear transformations of the
feature spaces. Its eigenvalues are the squared canonical correlations
between the lag spaces spanned by $v_t$ and $u_t(\tau)$.

\paragraph{Order-constrained spectral statistic.}
For any continuous orthogonally invariant spectral functional
$\varphi:\mathcal{S}_q^+\to\mathbb{R}$, define
\[
T_\varphi(\mathcal{P})
=
\sup_{\tau\in\mathcal{P}}\varphi\big(D(\tau)\big)
-
\inf_{\tau\in\mathcal{P}}\varphi\big(D(\tau)\big).
\]
This is a specialization of the general dispersion functional applied to
the lag-indexed operator family through the cross-dependence structure.

\begin{proposition}[Exact unification]
\label{prop:unification}
Under the above specialization, the following classical notions arise as
special cases.

\begin{enumerate}

\item \textbf{Linear Granger causality (coincidence regime).}
Under a correctly specified Gaussian VAR model, let $v_t^\perp$ denote
the residual after projection onto past information excluding component $i$.
Then Granger noncausality is equivalent to
\[
\forall \tau\in\mathcal{P}:\quad D^\perp(\tau)=0,
\]
which implies $T_\varphi(\mathcal{P})=0$ for all admissible $\varphi$.

\item \textbf{Directed coherence / canonical correlation.}
For $\varphi(M)=\lambda_1(M)$,
\[
\varphi\big(D(\tau)\big)=\|A(\tau)\|_2^2,
\]
which corresponds to squared directed coherence.

\item \textbf{Lead--lag correlation asymmetry (scalar case).}
For $p=q=1$,
\[
D(\tau)=\rho_{ji}(\tau)^2.
\]
With $\mathcal{P}=\{\ell,-\ell\}$ and $\varphi(M)=\sqrt{M}$,
\[
T_\varphi(\mathcal{P})
=
|\rho_{ji}(\ell)|-|\rho_{ji}(-\ell)|.
\]

\end{enumerate}
\end{proposition}

\begin{proof}
See Appendix~\ref{app:proof-unification}.
\end{proof}

\subsubsection{Qualitative Impossibility: Edge Selection under Orthogonal Invariance}
\label{subsubsec:impossibility-qualitative}

In high-dimensional systems, directional dependence may be distributed across
many modes rather than concentrated on coordinate-wise edges.
The following result shows that edge-based selection is fundamentally incompatible
with orthogonal invariance.

\paragraph{Edge-based criteria.}
A criterion $\mathcal{E}$ is edge-based if it depends only on finitely many
entries of $M\in\mathcal{S}_d^+$ in a fixed coordinate system:
\[
\mathcal{E}(M)=H\big(\{M_{ab}:(a,b)\in\mathcal{I}\}\big).
\]

\begin{corollary}[Impossibility of invariant edge selection]
\label{cor:edge-impossibility}
If $\mathcal{E}$ is orthogonally invariant, then for all $M$,
\[
\mathcal{E}(M)=\mathcal{E}(\lambda I_d),
\qquad
\lambda=\tfrac{1}{d}\mathrm{tr}(M).
\]
\end{corollary}

Thus, no nontrivial edge-based representation can be invariant to orthogonal
transformations.

\begin{corollary}[Failure of monotonicity under distributed dependence]
\label{cor:monotonicity-failure}
There exist $M_1\preceq M_2$ such that entrywise edge selection is not monotone.
\end{corollary}

This reflects the fundamental limitation of coordinate-based methods in
distributed regimes: strengthening global dependence can reduce detectable
edges.

Proofs are given in Appendix~\ref{app:proof-edge-impossibility-strong} and
Appendix~\ref{app:proof-monotonicity-failure}.

\subsubsection{Frequency-Domain Specialization: Geweke/Brillinger as a Circle-Deformation Case}
\label{subsubsec:freq-domain}

This subsection establishes a frequency-domain specialization of the
order-constrained spectral framework. Classical Brillinger coherence and
Geweke frequency-domain causality arise as spectral functionals of the
lag-indexed operator family when (i) feature maps are chosen as Fourier
projections and (ii) admissible deformations correspond to phase shifts
on the unit circle.

\paragraph{Fourier feature class and circle deformations.}
Let $\{X_t\}_{t\in\mathbb{Z}}$ be a zero-mean, $K$-variate, second-order stationary
process with absolutely summable autocovariances, so that the spectral density
matrix $f_X(\omega)$ exists and is continuous on $\omega\in[-\pi,\pi]$.
Fix index sets $I,J\subset\{1,\dots,K\}$ and define the corresponding spectral
blocks
\[
f_{JJ}(\omega),\quad f_{II}(\omega),\quad f_{JI}(\omega),\quad f_{IJ}(\omega).
\]

Define Fourier feature maps (on a window of size $T$) by
\[
U_T(\omega)
=
\frac{1}{\sqrt{T}}\sum_{t=1}^T X^{(I)}_t\,e^{-i\omega t}\in\mathbb{C}^{|I|},
\qquad
V_T(\omega)
=
\frac{1}{\sqrt{T}}\sum_{t=1}^T X^{(J)}_t\,e^{-i\omega t}\in\mathbb{C}^{|J|}.
\]

A temporal shift by $\tau$ induces the transformation
\[
U_T(\omega)\mapsto e^{-i\omega\tau}U_T(\omega),
\]
so admissible deformations correspond to the circle group
\[
\mathcal{P}_\omega=\{e^{-i\theta}:\theta\in[0,2\pi)\}.
\]

Define the stacked feature
\[
Z_T(\omega)=(V_T(\omega)^\top,U_T(\omega)^\top)^\top,
\]
and the associated second-order operator
\[
C_\omega
=
\lim_{T\to\infty}\mathbb{E}[Z_T(\omega)Z_T(\omega)^*]
=
\begin{pmatrix}
f_{JJ}(\omega) & f_{JI}(\omega)\\
f_{IJ}(\omega) & f_{II}(\omega)
\end{pmatrix}.
\]

As in Section~\ref{subsec:ocsc}, directional dependence is encoded in the
cross-block $f_{JI}(\omega)$.

Define the whitened cross-spectrum operator
\begin{equation}
\label{eq:freq-whitened-cross}
A(\omega)
=
f_{JJ}(\omega)^{-1/2}\,f_{JI}(\omega)\,f_{II}(\omega)^{-1/2},
\qquad
D(\omega)=A(\omega)A(\omega)^*\in\mathcal{S}_{|J|}^+,
\end{equation}
which is an orthogonally invariant reparameterization of the cross-block
dependence.

\begin{corollary}[Frequency-domain spectral specializations]
\label{cor:geweke-brillinger}
Assume $f_{JJ}(\omega)\succ 0$ and $f_{II}(\omega)\succ 0$ for a fixed
$\omega\in(-\pi,\pi)$. Then:

\begin{enumerate}

\item \textbf{Brillinger coherence.}
The maximal squared coherence is
\[
\gamma_{\max}^2(\omega)=\|A(\omega)\|_2^2=\lambda_1(D(\omega)).
\]

\item \textbf{Geweke causality.}
Let
\[
f_{J\mid I}(\omega)
=
f_{JJ}(\omega)-f_{JI}(\omega)f_{II}(\omega)^{-1}f_{IJ}(\omega).
\]
Then
\[
\mathcal{G}_{I\to J}(\omega)
=
\log\frac{\det f_{JJ}(\omega)}{\det f_{J\mid I}(\omega)}
=
-\log\det(I-D(\omega)),
\]
provided $\|D(\omega)\|_2<1$.

\end{enumerate}
\end{corollary}

\begin{proof}
See Appendix~\ref{app:proof-geweke-brillinger}.
\end{proof}

\subsection{Quantitative Impossibility: Edge-Based Causality Under Distributed Dependence}
\label{subsec:impossibility-quantitative}

The qualitative impossibility result established that edge-based criteria are
incompatible with orthogonal invariance. We now show a stronger statement:
even without axiomatic constraints, edge-based procedures are
information-theoretically underpowered in distributed regimes.

\subsubsection{Gaussian Matrix Model}
\label{subsubsec:gaussian-matrix}

Let $d\in\mathbb{N}$ and consider observations
\begin{equation}
\label{eq:gaussian-matrix-observation}
\widehat{M}
=
M
+
\frac{1}{\sqrt{T}}Z,
\qquad
Z_{ab}\stackrel{\mathrm{iid}}{\sim}\mathcal{N}(0,1).
\end{equation}

\[
H_0:\quad M=0,
\qquad
H_1(\delta):\quad M=\delta uv^\top,\quad \|u\|=\|v\|=1.
\]

Under $H_1$, entries satisfy $|M_{ab}|\asymp \delta/d$.

\paragraph{Edge-based procedures.}
We formalize edge-based methods as entrywise-stable tests.

\begin{definition}[Entrywise-stable test]
A test $\psi$ is entrywise-stable if
\[
|\psi(A)-\psi(B)| \le L\|A-B\|_\infty,
\quad
\|A\|_\infty=\max_{a,b}|A_{ab}|.
\]
\end{definition}

\paragraph{Impossibility result.}

\begin{theorem}[Edge-based detection barrier]
\label{thm:edge-impossibility-strong}
Let $\psi$ satisfy
\[
\mathbb{P}_0(\psi=1)\le\alpha,
\]
and be entrywise-stable and permutation invariant.
Then for some constant $c>0$, if
\begin{equation}
\label{eq:d2logd-barrier}
T \le c\,\frac{d^2}{\delta^2}\log d,
\end{equation}
\[
\sup_\psi\inf_{u,v}\mathbb{P}_{\delta uv^\top}(\psi=1)
\le \alpha + o(1).
\]
\end{theorem}

\paragraph{Spectral escape.}

\begin{theorem}[Spectral detection rate]
\label{thm:spectral-detection}
Define
\[
\chi(\widehat{M})
=
\mathbf{1}\{\|\widehat{M}\|_2 \ge \tau_{d,T}(\alpha)\}.
\]
Then for some $C>0$, if
\begin{equation}
\label{eq:d-barrier}
T \ge C\,\frac{d}{\delta^2},
\end{equation}
\[
\inf_{u,v}\mathbb{P}_{\delta uv^\top}(\chi=1)\to 1.
\]
\end{theorem}

Thus, spectral methods detect at scale $T\asymp d$, while edge-based methods
require $T\asymp d^2\log d$.

\medskip

These results provide a formal explanation for the instability of pairwise
Granger-network procedures in high-dimensional distributed regimes:
entrywise signals decay as $1/d$, while spectral aggregation preserves
signal strength at the operator level.

Proofs are given in Appendix~\ref{app:proof-edge-impossibility-strong}.

\subsubsection{VAR($L$) Generalization}
\label{subsubsec:var-generalization}

This subsection lifts the Gaussian matrix impossibility result from
Section~\ref{subsubsec:gaussian-matrix} to a stable Gaussian VAR($L$)
model with lag-embedded operators consistent with the construction of
Section~\ref{sec:implementation}.

\paragraph{Stable VAR($L$) with lag-embedded operators.}
Let $\{X_t\}_{t\in\mathbb{Z}}$ be a $K$-variate, zero-mean, stable Gaussian VAR($L$):
\begin{equation}
\label{eq:varL}
X_t
=
\sum_{\ell=1}^L A_\ell X_{t-\ell} + \varepsilon_t,
\qquad
\varepsilon_t\stackrel{\mathrm{iid}}{\sim}\mathcal{N}(0,\Sigma_\varepsilon),
\end{equation}
with $\Sigma_\varepsilon\succ 0$ and standard stability conditions. Fix $i\neq j$ and consider the directional pair $(i\to j)$. Within a rolling window $W_t=\{t-T+1,\ldots,t\}$, define lag embeddings
\[
u_s^{(i)}
=
\big(X^{(i)}_{s-1},\ldots,X^{(i)}_{s-L}\big)^\top,
\qquad
v_s^{(j)}
=
\big(X^{(j)}_{s},\ldots,X^{(j)}_{s-L+1}\big)^\top.
\]

Let $v_{s,\perp}^{(j)}$ denote the residual after projection onto the
conditioning space $\mathcal{H}_Y$, consistent with the residualized
construction of Section~\ref{subsec:operator}. Define the stacked feature
\[
Z_s(\tau)
=
\big(v_{s,\perp}^{(j)\top},u_{s-\tau}^{(i)\top}\big)^\top,
\]
and the corresponding covariance operator
\[
C(\tau)
=
\mathbb{E}[Z_s(\tau)Z_s(\tau)^\top]
=
\begin{pmatrix}
\Sigma_{VV} & \Sigma_{VU}(\tau) \\
\Sigma_{UV}(\tau) & \Sigma_{UU}(\tau)
\end{pmatrix}.
\]

Directional dependence is encoded in the cross-block
$\Sigma_{VU}(\tau)$. The whitened cross-operator is the reparameterization
\begin{equation}
\label{eq:Ahat}
\widehat{A}(\tau)
=
\widehat{\Sigma}_{VV}^{-1/2}\,
\widehat{\Sigma}_{VU}(\tau)\,
\widehat{\Sigma}_{UU}(\tau)^{-1/2},
\end{equation}
where covariance matrices are computed over $W_t$ and regularized if necessary
to ensure invertibility.

\paragraph{Null and distributed alternatives.}
The pairwise Granger-null is
\begin{equation}
\label{eq:granger-null}
H_0(i\to j):
\quad
\forall\tau\in\mathcal{P},\quad \Sigma_{VU}(\tau)=0
\quad\Longleftrightarrow\quad A(\tau)=0.
\end{equation}

Under a distributed alternative at fixed $\tau$,
\begin{equation}
\label{eq:distributed-alt}
H_1(i\to j;\delta):
\quad
A(\tau)=\delta ab^\top,
\qquad
\|a\|=\|b\|=1,
\end{equation}
so $\|A(\tau)\|_2=\delta$ but entries satisfy $|A_{rs}|\asymp \delta/L$.

\paragraph{Gaussian approximation.}
Under stability and mixing conditions (Section~\ref{subsec:inference}),
\begin{equation}
\label{eq:Ahat-gauss-approx}
\widehat{A}(\tau)
=
A(\tau)
+
\frac{1}{\sqrt{T}}Z_T(\tau)
+
R_T(\tau),
\end{equation}
where $Z_T(\tau)$ is asymptotically Gaussian with bounded variances and
$\|R_T(\tau)\|_2=o_{\mathbb{P}}(T^{-1/2})$.

Thus $\widehat{A}(\tau)$ reduces to the Gaussian matrix model with dimension $L$.

\paragraph{Edge-based tests.}
An edge-based test is any entrywise-stable functional:
\begin{equation}
\label{eq:linfty-stability}
|\psi(B)-\psi(C)| \le L_\psi\|B-C\|_\infty.
\end{equation}

\begin{theorem}[VAR($L$) edge-based impossibility]
\label{thm:var-edge-impossibility}
Let $\psi$ control size $\alpha$ and satisfy \eqref{eq:linfty-stability}.
If
\begin{equation}
\label{eq:var-edge-barrier}
T \le c\,\frac{L^2}{\delta^2}\log L,
\end{equation}
then
\[
\inf_{a,b}
\mathbb{P}_{H_1}(\psi(\widehat{A}(\tau))=1)
\le \alpha + o(1).
\]
\end{theorem}

\begin{theorem}[VAR($L$) spectral detection]
\label{thm:var-spectral-detection}
Let
\[
\chi(\widehat{A}(\tau))
=
\mathbf{1}\{\|\widehat{A}(\tau)\|_2 \ge \tau_{L,T}(\alpha)\}.
\]
If
\begin{equation}
\label{eq:var-spectral-barrier}
T \ge C\,\frac{L}{\delta^2},
\end{equation}
then
\[
\inf_{a,b}
\mathbb{P}_{H_1}(\chi(\widehat{A}(\tau))=1)
\to 1.
\]
\end{theorem}

\begin{proof}
Follows from reduction to the Gaussian matrix model using
\eqref{eq:Ahat-gauss-approx}. See Supplement~\ref{app:var-proof-barriers-full}.
\end{proof}
\subsubsection{Sample-Size Barriers: Pairwise and Network Scaling}

In the implementation, $d_u=d_v=L$, so the operator dimension is governed by $L$.

\smallskip
\emph{Pairwise detection:}
\[
\text{edge-based:}\quad
T \gtrsim \frac{L^2}{\delta^2}\log L,
\qquad
\text{spectral:}\quad
T \gtrsim \frac{L}{\delta^2}.
\]

\smallskip
\emph{Network-level detection ($K$ variables):}
Multiple testing inflates thresholds to order
\[
\sqrt{\frac{\log(KL)}{T}},
\]
yielding
\[
T \gtrsim \frac{L^2}{\delta^2}\log(KL)
\]
for edge-based methods, while spectral/operator methods remain at
\[
T \gtrsim \frac{L}{\delta^2}.
\]

This separation explains the instability of pairwise Granger-network inference
in high-dimensional distributed regimes, where signal is diffuse across modes
but preserved at the operator level.

\section{Methodology}
\label{sec:implementation}

This section documents the implementation of order-constrained spectral causality through
a single operator-valued construction indexed by admissible temporal deformations.
All empirical procedures used in the paper are exact specializations of this construction.
No alternative algorithms or competing estimators are introduced.

The purpose of this section is documentation rather than methodological development.
It makes explicit how the theoretical objects defined in
Section~\ref{sec:framework} are instantiated in practice, and how inference is carried out
in finite samples.
A complete algorithmic description is given in
Algorithm~\ref{alg:ocsc}, and its theoretical validity is justified in
Supplement~\ref{app:implementation}.

\subsection{Unified Order-Indexed Operator Construction}

Let $\{X_t\}_{t=1}^T$ be a $K$-dimensional strictly stationary time series with
$\mathbb E X_t=0$ and $\mathbb E\|X_t\|^2<\infty$.
Fix nonempty index sets
$\mathcal I,\mathcal J\subset\{1,\dots,K\}$ corresponding to source and target components,
and an admissible deformation set $\mathcal P\subset\mathbb R_+$.

Let $\Psi$ and $\Phi$ be measurable feature maps applied to the source and target components,
respectively. These maps are assumed to be fixed \emph{a priori} or selected on an auxiliary
sample that is independent of the evaluation sample, as formalized in
Assumption~\ref{ass:embed}. In particular, the feature maps are not tuned on the data used
for inference.

Let $\tau_{\max}=\max\mathcal P$. For each $\tau\in\mathcal P$ and
$t=\tau_{\max}+1,\dots,T$, define
\[
U_t(\tau)=\Psi\!\left(X^{(\mathcal I)}_{t-\tau}\right)\in\mathbb R^{d_u},
\qquad
V_t=\Phi\!\left(X^{(\mathcal J)}_t\right)\in\mathbb R^{d_v},
\]
and stack
\[
Z_t(\tau)=
\begin{pmatrix}
V_t\\
U_t(\tau)
\end{pmatrix}
\in\mathbb R^{d},
\qquad d=d_v+d_u.
\]

The population dependence operator is
\[
C(\tau)=\mathbb E\!\left[Z_t(\tau)Z_t(\tau)^\top\right]\in\mathbb S_+^d,
\]
with empirical estimator
\[
\widehat C_T(\tau)=\frac{1}{T-\tau_{\max}}\sum_{t=\tau_{\max}+1}^T Z_t(\tau)Z_t(\tau)^\top.
\]

Although $C(\tau)$ is symmetric positive semidefinite, directional dependence is not encoded
through asymmetry of the operator itself but through its block structure induced by the
stacking of target and lagged source features. In particular,
\[
C(\tau)
=
\begin{pmatrix}
\Sigma_{VV} & \Sigma_{VU}(\tau) \\
\Sigma_{UV}(\tau) & \Sigma_{UU}(\tau)
\end{pmatrix},
\]
where $\Sigma_{VU}(\tau)=\mathrm{Cov}(V_t,U_t(\tau))$ captures cross-dependence between
target and lagged source components. Directionality is therefore encoded in the
cross-block structure and its interaction with the marginal covariances, rather than in
any intrinsic asymmetry of $C(\tau)$.

In subsequent constructions, whitened operators (e.g.\ based on
$\Sigma_{VV}^{-1/2}\Sigma_{VU}(\tau)\Sigma_{UU}^{-1/2}$) are used as
orthogonally invariant reparameterizations of this cross-block dependence.
These do not define alternative estimators, but rather equivalent representations of the
same underlying second-order operator. If conditional analysis is required, $Z_t(\tau)$ is replaced by residualized features
$Z_t^\perp(\tau)$ as defined in Section~\ref{sec:framework}.
All subsequent constructions remain unchanged.

The consistency and inferential validity of the operator estimates rely on the
mixing and moment conditions specified in Section~\ref{subsec:inference} and
Supplement~\ref{app:asymptotic_theory}. These assumptions are implicitly imposed
for all constructions in this section.

\subsection{Spectral Summaries and Dispersion Statistics}

Let
$\widehat\lambda_1(\tau)\ge\cdots\ge\widehat\lambda_d(\tau)\ge0$
denote the eigenvalues of the estimated operator $\widehat C_T(\tau)$ associated with
the lag-indexed operator family $\{C(\tau)\}_{\tau\in\mathcal P}$.
For a scalar spectral functional $f$, define the linear spectral statistic
\[
\widehat L_f(\tau)
=
\frac{1}{d}\sum_{r=1}^d f\!\big(\widehat\lambda_r(\tau)\big),
\]
and the associated dispersion statistic
\[
\widehat T_f
=
\sup_{\tau\in\mathcal P}\widehat L_f(\tau)
-
\inf_{\tau\in\mathcal P}\widehat L_f(\tau).
\]

Alternatively, define the empirical spectral measure
\[
\widehat\mu_\tau=\frac{1}{d}\sum_{r=1}^d \delta_{\widehat\lambda_r(\tau)},
\]
and the spectral-measure dispersion
\[
\widehat T_{\mathrm{spec}}
=
\sup_{\tau_1,\tau_2\in\mathcal P}
d\!\left(\widehat\mu_{\tau_1},\widehat\mu_{\tau_2}\right),
\]
where $d(\cdot,\cdot)$ is a metric on probability measures. Both statistics quantify deviations from invariance of the lag-indexed operator family
$\{C(\tau)\}_{\tau\in\mathcal P}$ and target the same null hypothesis of
order-constrained spectral invariance, as established in
Section~\ref{subsec:spectral_extension_core}. In particular, inference is conducted at the
level of variation across $\tau$. Any aggregation across lags (e.g.\ constructions of the
form $C(t)=\sum_{\tau} w_\tau C_\tau(t)$ in Section~\ref{sec:methodology}) is treated as a
post-estimation descriptive summary and does not replace the invariance-based test defined
by $\widehat T_f$ or $\widehat T_{\mathrm{spec}}$.

Algorithm~\ref{alg:ocsc} summarizes the complete computational procedure.
Each step corresponds directly to the operator-theoretic construction above and to the
inferential framework of Section~\ref{subsec:inference}.

\begin{algorithm}[t]
\caption{Unified Order-Constrained Spectral Causality Procedure}
\label{alg:ocsc}
\begin{algorithmic}[1]
\REQUIRE Time series $\{X_t\}_{t=1}^T$, source indices $\mathcal I$, target indices $\mathcal J$,
admissible deformation set $\mathcal P$, feature maps $\Psi,\Phi$, spectral summary $f$
or metric $d(\cdot,\cdot)$
\ENSURE Dispersion statistic $\widehat T$ and randomization $p$-value $\widehat p$

\STATE Let $\tau_{\max} = \max_{\tau \in \mathcal P} \tau$.

\STATE For each $\tau\in\mathcal P$ and $t=\tau_{\max}+1,\dots,T$, construct
$U_t(\tau)=\Psi(X^{(\mathcal I)}_{t-\tau})$ and $V_t=\Phi(X^{(\mathcal J)}_t)$,
and form $Z_t(\tau)=(V_t^\top,U_t(\tau)^\top)^\top$.

\STATE Estimate the dependence operator
\[
\widehat C_T(\tau)
=
(T-\tau_{\max})^{-1}
\sum_{t=\tau_{\max}+1}^T Z_t(\tau)Z_t(\tau)^\top
\]
for all $\tau\in\mathcal P$.

\STATE Compute eigenvalues $\{\widehat\lambda_r(\tau)\}_{r=1}^d$ and evaluate either
$\widehat L_f(\tau)$ or $\widehat\mu_\tau$.

\STATE Compute the dispersion statistic $\widehat T_f$ or $\widehat T_{\mathrm{spec}}$.

\STATE Generate circular shifts of the source component, recompute the statistic, and
compute the randomization $p$-value
\[
\widehat p
=
\frac{1+\sum_{b=1}^B \mathbf 1\{\widehat T^{(b)}\ge \widehat T^{\mathrm{obs}}\}}{B+1}.
\]

\RETURN $\widehat T$ and $\widehat p$
\end{algorithmic}
\end{algorithm}

Let $d=d_u+d_v$ denote the feature dimension.
For each $\tau\in\mathcal P$, operator estimation requires $O(Td^2)$ operations and spectral
decomposition requires $O(d^3)$ operations.
The total computational cost is therefore
$O(|\mathcal P|(Td^2+d^3))$, multiplied by the number of randomization replicates $B$.
All computations rely on standard linear algebra routines and involve no iterative
optimization procedures.

\begin{lemma}[Correctness of Algorithm~\ref{alg:ocsc}]
\label{lem:alg_correctness}
Under the assumptions of Section~\ref{subsec:inference},
Algorithm~\ref{alg:ocsc} computes a consistent estimator of the population
dispersion functional associated with the lag-indexed operator family
$\{C(\tau)\}_{\tau\in\mathcal P}$. Under the null hypothesis of
order-constrained spectral invariance, the randomization $p$-value is
asymptotically valid.
\end{lemma}

\begin{proof}
Consistency of $\widehat C_T(\tau)$, uniform convergence of spectral summaries, and
consistency of the dispersion functional follow from
Supplement~\ref{app:implementation} and Supplement~\ref{app:asymptotic_theory}.
Validity of the randomization procedure follows from the group-invariance arguments in
Appendix~\ref{app:rand}.
\end{proof}

All operators are symmetric and positive semidefinite by construction, and numerically
stable eigensolvers for symmetric matrices may be used.
In finite samples with moderately large feature dimension, centering of $Z_t(\tau)$ and,
if necessary, addition of a small ridge regularization
\[
\widehat C_T(\tau)\leftarrow\widehat C_T(\tau)+\epsilon I_d
\]
ensures numerical stability and invertibility without affecting the null hypothesis or
the theoretical guarantees.

For spectral-measure dispersion, bounded-Lipschitz or Wasserstein metrics computed from
finite spectra are numerically stable and insensitive to eigenvalue ordering.
Shift-based randomization preserves marginal dependence and avoids the instability of
block-resampling schemes in strongly dependent settings.

\subsection{Operator-Based Multivariate Causal Monitoring}
\label{sec:methodology}

We develop a methodology for monitoring time-varying directional causal
relationships between multivariate stochastic processes using a rolling
operator framework.
The central object of inference is a positive semidefinite operator whose
spectral structure provides a characterization of second-order directional
dependence across multiple lags and feature dimensions.

Let $\{X_t\}_{t\in\mathbb{Z}}\subset\mathbb{R}^d$ denote a multivariate driver
process and $\{Y_t\}_{t\in\mathbb{Z}}\subset\mathbb{R}^q$ a multivariate target
process.
In the notation of Section~\ref{sec:implementation}, $X$ and $Y$ correspond to
the source and target components $X^{(\mathcal{I})}$ and $X^{(\mathcal{J})}$,
respectively; the separate notation is adopted here to emphasize the
driver--target asymmetry of the monitoring framework.
Our objective is to assess whether, and how, past values of $X$ improve the
prediction of $Y$ beyond the information contained in the past of $Y$ itself,
and to monitor how this directional influence evolves over time.

While the theoretical construction in Section~\ref{sec:framework} assumes strict
stationarity, the rolling-window framework adopted here is interpreted under a
local stationarity regime: within each window $W_t$, the process is assumed to be
approximately stationary, so that the operator estimates are consistent for a
time-indexed dependence structure.
Unlike pairwise or edge-based approaches, we do not seek to identify isolated
causal links.
Instead, we characterize directional causality as a geometric object acting on
the target lag space, allowing simultaneous assessment of causal strength,
dimensionality, and affected subspaces.

\subsubsection{Directional causal operator}

For each window $W_t$ and lag $\tau \in \mathcal{T}$, the lag-specific operator
$C_\tau(t)$ is a windowed empirical counterpart of the population dependence
operator $C(\tau)$ introduced in Section~\ref{sec:framework}, expressed in a
whitened coordinate system.
Specifically, whitening constitutes an equivalent reparameterization of the
second-order dependence structure and does not introduce a distinct estimator.

Fix an embedding order $p\ge1$ and lag $\tau\ge1$.
Within a rolling window $W_t=\{t-W+1,\dots,t\}$, define lag-embedded vectors
\[
\mathbf{v}_s=(Y_s^\top,\dots,Y_{s-p+1}^\top)^\top\in\mathbb{R}^{pq},\qquad
\mathbf{u}_s(\tau)=(X_{s-\tau}^\top,\dots,X_{s-\tau-p+1}^\top)^\top\in\mathbb{R}^{pd}.
\]

Let $S_{VV}(t)$, $S_{UU}(t,\tau)$, and $S_{VU}(t,\tau)$ denote the corresponding
sample covariance blocks, regularized if necessary to ensure invertibility.
We define the whitened cross-covariance operator
\[
A_\tau(t)
=
S_{VV}(t)^{-1/2} S_{VU}(t,\tau) S_{UU}(t,\tau)^{-1/2},
\]
and the associated directional operator
\[
C_\tau(t)=A_\tau(t)A_\tau(t)^\top\succeq0.
\]

Aggregating over a finite lag set $\mathcal{T}$ with nonnegative weights
$\{w_\tau\}_{\tau\in\mathcal{T}}$ yields
\[
C(t)=\sum_{\tau\in\mathcal{T}} w_\tau\, C_\tau(t),
\]
which provides a time-indexed summary of directional dependence across admissible
lags within the window.

Importantly, $C(t)$ is a derived, aggregated operator used for monitoring and
interpretation. Inference on causal presence is conducted at the level of the
lag-indexed family $\{C_\tau(t)\}_{\tau\in\mathcal{T}}$ through invariance-based
statistics defined over $\tau$ (Section~\ref{subsec:inference}), rather than on
the aggregated operator itself.

\subsubsection{Why $C(t)$ encodes directional causality}

The operator $C(t)$ captures directional predictive structure rather than mere
contemporaneous dependence for three reasons. First, $C(t)$ is constructed from lagged values of $X$ and $Y$, ensuring temporal
ordering consistent with the admissible deformation framework.
Second, whitening by $S_{VV}(t)$ removes second-order structure internal to $Y$,
so that the resulting operator reflects predictive content attributable to $X$
relative to the target's own past.
Third, under standard linear prediction assumptions, $C_\tau(t)=0$ for all
$\tau\in\mathcal{T}$ if and only if past values of $X$ provide no linear
predictive improvement for $Y$ given its own history within $W_t$.

Consequently, $C(t)=0$ implies absence of directional predictive content across
all admissible lags, while $C(t)\neq0$ indicates the presence of at least one
direction in the target lag space along which past $X$ contributes predictive
information.
Precise equivalence statements and proofs are provided in
Supplement~\ref{app:operator_theory}.

\subsubsection{Multiscale causal decomposition}

The eigendecomposition
\[
C(t)=\sum_{j=1}^{pq}\lambda_j(t)v_j(t)v_j(t)^\top
\]
induces a hierarchy of directional resolutions. The leading eigenvalue
$\lambda_1(t)$ measures maximal directional dependence strength,
corresponding to the strongest achievable predictive gain.
The trace $\mathrm{tr}(C(t))$ captures total dependence energy, while the effective
rank
\[
r_{\mathrm{eff}}(t)=\frac{\mathrm{tr}(C(t))^2}{\mathrm{tr}(C(t)^2)}
\]
quantifies the dimensionality of directional transmission. The leading eigenspaces define subspaces of the target lag space most affected
by directional dependence. Projecting these subspaces onto coordinate axes yields
variable-level hub scores, measuring exposure to dominant transmission channels.

These summaries provide a descriptive decomposition of the aggregated operator
$C(t)$. They do not constitute primary inferential objects, but rather
interpretations of the directional structure once dependence has been detected
through the lag-indexed framework of Section~\ref{subsec:inference}.
All reported empirical indicators are functionals of $C(t)$ or its spectrum,
ensuring internal methodological coherence.

\subsubsection{Rolling monitoring and inference}

Applying the above construction in rolling windows produces an operator-valued
time series $\{C(t_k)\}$.
Monitoring the evolution of its spectral characteristics allows detection of
the emergence, persistence, and dissipation of multivariate directional
dependence structures.

Statistical significance is assessed via circular-shift nulls applied to the
driver process, preserving marginal temporal dependence while destroying
cross-dependence. Inference is conducted using statistics derived from the lag-indexed operator
family $\{C_\tau(t)\}_{\tau\in\mathcal{T}}$, in accordance with the
order-constrained spectral framework of Section~\ref{subsec:inference}.
In particular, testing is based on invariance (or lack thereof) across $\tau$,
rather than on the aggregated operator $C(t)$ itself.

The aggregated operator $C(t)$ serves as a descriptive summary within each
window. Under the circular-shift null, cross-dependence between source and target
is removed jointly across all lags, implying $C_\tau(t)=0$ for all
$\tau\in\mathcal{T}$ in population, and therefore $C(t)=0$.
Rejection of the null indicates that admissible temporal deformation induces
nontrivial variation in the second-order dependence structure across lags,
providing evidence of directional influence in the order-constrained spectral
sense.

Importantly, nonzero aggregated operators $C(t)$ indicate the presence of
second-order directional dependence, but do not by themselves establish
order-constrained spectral causality in the sense of Definition~\ref{def:ocsc}.
Causality is formally defined through non-invariance across admissible
deformations $\tau$, and is therefore assessed via dispersion statistics over
the lag-indexed operator family. Aggregated operators provide descriptive
summaries conditional on the presence of such non-invariance.

This separation ensures coherence between the inferential framework
(variation across $\tau$) and the monitoring representation (aggregation over
$\tau$), avoiding ambiguity between detection and summarization.
Inference is thus conducted without multiple testing across lags or pairs.

\subsubsection{Relation to alternative causal frameworks}

The proposed methodology admits a unified interpretation of several classical
notions of directional dependence through its operator-based formulation.
In particular, it generalizes linear Granger causality by replacing
coefficient-level hypothesis testing with analysis of second-order dependence
operators indexed by admissible temporal deformations.

Under the linear Gaussian VAR($p$) setting with correct model specification,
the absence of Granger causality from component $i$ to component $j$ is
equivalent to the vanishing of the whitened cross-covariance (directed coherence)
operator across all admissible lags:
\[
\forall \tau \in \mathcal{P}:\quad A(\tau)=0.
\]
In this regime, the order-constrained spectral framework coincides exactly with
linear Granger causality at the operator level, and the proposed dispersion-based
statistics reduce to tests of invariance of the corresponding operator family.
Thus, classical Granger causality arises as a special case of the present framework
under linear second-order structure.

Outside this restricted setting, the two notions diverge. In particular,
order-constrained spectral causality may detect directional structure in the
second-order geometry of the system even when all finite-dimensional linear
projections exhibit no predictive improvement. This occurs, for example, in
high-dimensional or distributed regimes where dependence is spread across many
weak directions, or when nonlinear transformations induce second-order structure
in an appropriate feature space.

More generally, the framework provides an exact unification of several classical
directionality notions through appropriate choices of feature maps, admissible
deformation sets, and spectral functionals, as established in
Section~\ref{sec:framework}. These include directed coherence, canonical
correlation-based measures, and frequency-domain causality statistics such as
Geweke's decomposition. In all cases, directional dependence is represented as
variation of an operator-valued object under admissible temporal transformations,
rather than through coordinate-wise or edge-level effects.

A key implication of this operator formulation is that causal influence is treated
as a collective geometric property of the system rather than as a collection of
pairwise links. As shown by the impossibility results in
Section~\ref{sec:framework}, entrywise or edge-based procedures are
information-theoretically underpowered in distributed dependence regimes:
stable edge-based tests require sample sizes scaling on the order of
$d^2\log d$ to achieve nontrivial power, whereas operator-based spectral
procedures detect directional dependence at the optimal linear scale $d$.
This gap reflects a fundamental limitation of coordinate-level representations
in high-dimensional systems, rather than a deficiency of specific estimation
techniques.

The framework therefore avoids reliance on collections of pairwise edges and
instead captures joint directional effects acting on subspaces of the system.
This perspective is particularly well suited to financial and economic systems,
where directional influence often manifests through coordinated group behavior
and low-dimensional modes rather than isolated bilateral interactions.

The resulting methodology enables scalable, rolling inference on multivariate
directional dependence without parametric model estimation or multiple hypothesis
testing across edges or lags. Inference is conducted either through dispersion
of operator families across admissible deformations or, in rolling settings,
through operator-level statistics that aggregate directional structure over lags.

Finally, it is important to clarify the scope of the framework. The methodology is
second-order in the chosen feature space: all causal statements are defined in
terms of second-order dependence operators constructed from feature maps applied
to the data. Consequently, nonlinearity is entirely governed by the embedding.
When nonlinear feature maps are used, the framework captures directional
dependence in the induced second-order geometry of the transformed processes,
without modifying the causal definition or inferential procedures. At the same
time, the framework does not aim to identify structural or interventional causal
effects, which require additional assumptions beyond second-order dependence.

These trade-offs are appropriate for high-dimensional applications where
interpretability, stability, and scalability are central, and where causal
influence is inherently collective rather than localized.

\subsection{Nonlinear extensions}
\label{subsec:nonlinear_extensions}

The notion developed in this paper is defined at the level of order-constrained
non-invariance of second-order dependence operators and does not rely on linearity
of the underlying processes. Linearity enters only through the choice of feature
maps used to construct the empirical operators. The framework is second-order in
the chosen feature space; nonlinearity is entirely governed by the embedding.

Formally, let $\Psi:\mathcal{X}\to\mathbb{R}^{d_U}$ and
$\Phi:\mathcal{Y}\to\mathbb{R}^{d_V}$ denote measurable feature maps applied to
the source and target components, respectively.
All definitions and results in Sections~\ref{sec:framework} and
\ref{sec:implementation} remain valid when $\Psi$ and $\Phi$ are nonlinear,
provided the resulting feature dimensions are fixed and the second moments of
the transformed processes exist.

Under such nonlinear embeddings, order-constrained spectral causality
characterizes directional dependence in the second-order geometry of the
feature-transformed processes.
In particular, detection reflects sensitivity of transformed dependence
structure to admissible temporal deformations, rather than linear predictability.
Examples include polynomial expansions, spline bases, interaction terms, or
random-feature approximations to reproducing kernel Hilbert space embeddings.

The operator-based formulation ensures that nonlinear extensions do not alter
the underlying dependence criterion, the order constraint, or the inferential
procedure. Consequently, nonlinear monitoring can be performed by replacing the
feature maps in Algorithm~\ref{alg:ocsc}, without introducing additional tuning
parameters or modifying the asymptotic framework.

\section{Experimental Results}
\label{sec:simulations}

This section presents two complementary sets of experiments designed to examine the
theoretical and practical implications of the proposed order-constrained spectral
framework, followed by a unified discussion of the empirical findings.

Section~\ref{subsec:simulations} presents a controlled simulation study that isolates
specific aspects of the causal functional, including calibration of the shift-based
randomization test, the distinction between edge-dominated and bulk-dominated causal
effects, sensitivity to the rank of cross-series dependence, detection of nonlinear
directional dependence beyond linear Granger causality, and robustness to conditioning
and latent confounding. These experiments provide direct empirical confirmation of the
theoretical properties established in Section~\ref{sec:framework}, particularly the
spectral distribution extension (Section~\ref{subsec:spectral_extension_core}), the
impossibility results for edge-based methods
(Section~\ref{subsec:impossibility-quantitative}), and the distinctness beyond linear
predictability (Section~\ref{subsec:granger}).

Section~\ref{subsec:empirical} presents a large-scale empirical study of global financial
markets that validates lag-level directional structure, identifies statistically significant
episodes of system-level causal organization, characterizes the joint phase structure of
causal strength and dimensionality, documents the stabilization of dominant transmission
channels during stress periods, reveals persistent transmitter-receiver asymmetry,
isolates sparse edge-level amplification during episodes, assesses robustness to
dimensionality reduction, and examines regime-level macro hub organization.

Section~\ref{subsec:discussion} discusses the combined implications of both sets of
experiments for theory, methodology, and practice.

\subsection{Simulation Studies}
\label{subsec:simulations}

\subsubsection{Design and Implementation}

Simulation settings span multiple sample sizes and dimensions to assess stability across
moderate and high-dimensional regimes. Unless otherwise stated, simulations use admissible
lag set $\mathcal P=\{1,\dots,5\}$ and Monte Carlo size $200$. Randomization $p$-values are
computed using $100$ circular shifts of the source component, and rejection rates are
reported at nominal level $\alpha=0.05$. Data are generated from multivariate systems with independent AR(1) marginal dynamics,
\begin{equation}
X_t^{(k)} = \rho X_{t-1}^{(k)} + \varepsilon_t^{(k)}, \qquad \rho=0.3,
\end{equation}
with i.i.d.\ standard Gaussian innovations. This baseline ensures weak temporal dependence
and satisfies the mixing conditions required in Section~\ref{subsec:inference}
(Assumption~A1). Directional dependence, when present, is introduced at a single lag
$\tau^\star=2$, ensuring that causal influence manifests through non-invariance over the
admissible lag set as formalized in Definition~\ref{def:ocsc}.

For each $\tau\in\mathcal P$, dependence operators are constructed from feature vectors
\begin{align*}
V_t &= \big(X_t^{(j)},X_{t-1}^{(j)},\dots,X_{t-p_v+1}^{(j)}\big), \\
U_t(\tau) &= \big(X_{t-\tau}^{(i)},X_{t-\tau-1}^{(i)},\dots,X_{t-\tau-p_u+1}^{(i)}\big),
\end{align*}
with $p_v=p_u=5$. Directed coherence operators are formed as in
Section~\ref{subsec:operator}. Linear spectral statistics
$L_f(\tau)=d^{-1}\sum_r f(\lambda_r(\tau))$ are evaluated using
$f(\lambda)=\lambda$ (trace), $f(\lambda)=\lambda^2$ (Frobenius), and
$f(\lambda)=\log(\lambda+\varepsilon)$, corresponding to increasingly bulk-sensitive
summaries as discussed in Section~\ref{subsec:spectral_extension_core}.

Inference is conducted using the shift-based randomization procedure of
Section~\ref{subsec:inference}, which preserves marginal temporal dependence while
destroying order-aligned directional structure. This procedure directly targets the null
hypothesis $T_f=0$ (Definition~\ref{def:null}) and is valid for the non-smooth
supremum--infimum functional defining causal invariance. Finite-sample exactness under
exact invariance and asymptotic validity under approximate invariance are established in
Appendix~\ref{app:rand}.

\subsubsection{Size Control under the Null}

Table~\ref{tab:size} reports empirical rejection rates under the null hypothesis of no
directional dependence for a range of sample sizes $T$ and dimensions $K$. The considered
configurations span moderate and high-dimensional regimes relative to the lag depth used
in the operator construction.

Across all settings, empirical size is close to the nominal level $\alpha=0.05$ for each
spectral summary, with mild finite-sample deviations that diminish as $T$ increases.
Calibration is stable across trace, Frobenius, and log-determinant summaries, despite their
different sensitivity to dominant versus distributed spectral components. These results
provide empirical support for the finite-sample validity of the shift-based randomization
procedure and are consistent with the asymptotic arguments developed in
Section~\ref{subsec:inference} and the formal randomization validity established in
Supplement~\ref{app:asymptotic_theory} (Proposition~\ref{prop:rand_asymptotic}).

In practical terms, these results confirm that the proposed test controls false positive
rates reliably across a range of system sizes, a prerequisite for any monitoring
application in which spurious detections carry operational cost.

\begin{table}[ht]
\centering
\caption{Empirical size under the null using the shift-based randomization test
($\alpha = 0.05$). Rejection rates are reported for different sample sizes $T$ and
dimensions $K$ using three spectral summaries.}
\label{tab:size}
\begin{tabular}{@{}ccccc@{}}
\toprule
$T$ & $K$ & Trace & Frobenius & Log-det \\
\midrule
500  & 20 & 0.095 & 0.070 & 0.060 \\
500  & 50 & 0.070 & 0.060 & 0.040 \\
1000 & 20 & 0.030 & 0.040 & 0.040 \\
1000 & 50 & 0.065 & 0.065 & 0.035 \\
\bottomrule
\end{tabular}
\end{table}

\subsubsection{Edge-Dominated Causal Effects}

\begin{figure}[!t]
\centering
\includegraphics[width=0.75\textwidth]{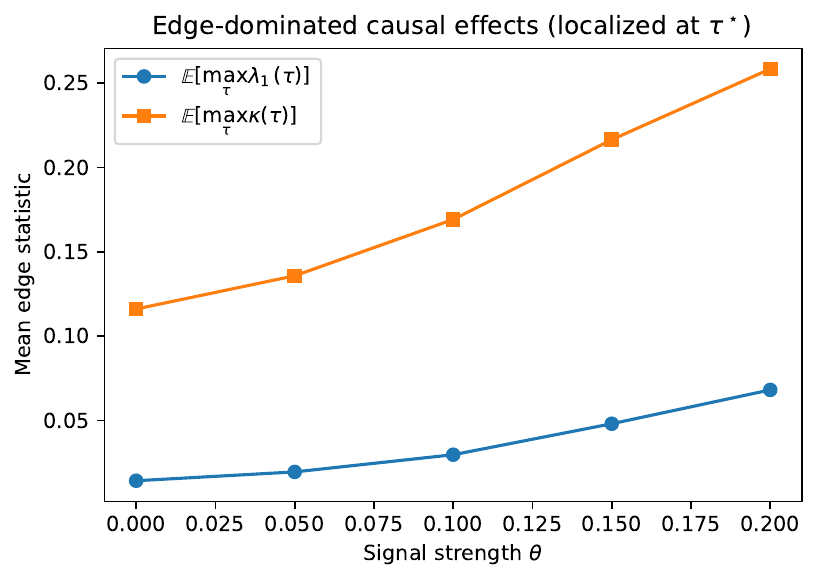}
\caption{Edge-dominated causal effects.
Monte Carlo averages of $\max_{\tau}\lambda_1(\tau)$ (leading eigenvalue of the
directed coherence operator) and $\max_{\tau}\kappa(\tau)$ (operator norm) under a
lag-localized rank-one alternative, as functions of signal strength.}
\label{fig:edge}
\end{figure}

We first consider lag-localized rank-one alternatives in which a single source component
affects a single target component. Figure~\ref{fig:edge} reports Monte Carlo averages of
$\max_{\tau}\lambda_1(\tau)$ and $\max_{\tau}\kappa(\tau)$ as functions of signal strength.

Both quantities increase smoothly with signal strength, confirming sensitivity to
edge-dominated causal structure. As emphasized in
Section~\ref{subsec:spectral_extension_core} and Section~\ref{sec:implementation}, such edge
statistics are not used for inference but serve as diagnostics illustrating how low-rank
causal effects manifest through isolated spectral modes. This experiment empirically
separates descriptive edge behavior from inferential bulk dispersion, a distinction that
is central to the impossibility results in
Section~\ref{subsec:impossibility-quantitative}: edge-level statistics can detect
concentrated signals but fail under distributed alternatives where the proposed spectral
tests excel.

\subsubsection{Bulk-Dominated Causal Effects}

\begin{figure}[ht]
\centering
\includegraphics[width=0.75\textwidth]{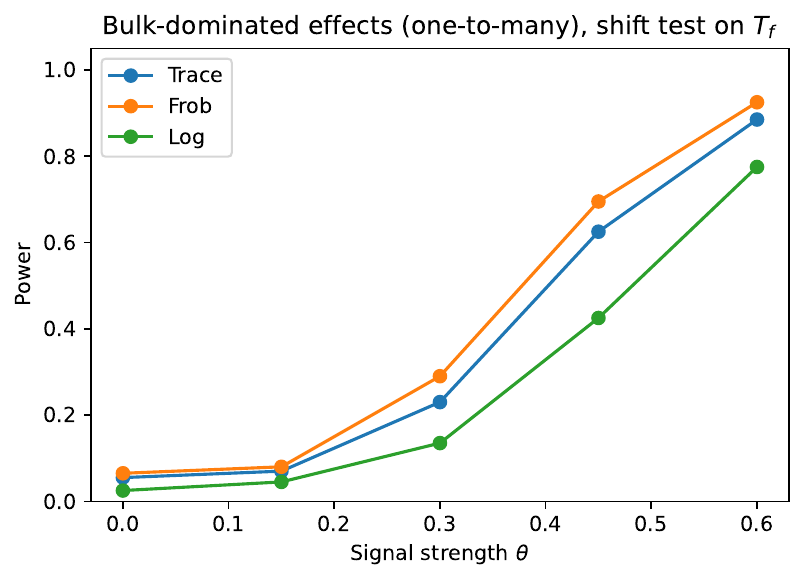}
\caption{Bulk-dominated causal effects.
Empirical power of the dispersion statistic $T_f$ under lag-localized one-to-many
alternatives, for three spectral summaries: trace ($f(\lambda)=\lambda$),
Frobenius ($f(\lambda)=\lambda^2$), and log-determinant
($f(\lambda)=\log(\lambda+\varepsilon)$).}
\label{fig:bulk}
\end{figure}

We next examine diffuse alternatives in which a single source component affects a large
fraction of target components at lag $\tau^\star$. Figure~\ref{fig:bulk} reports empirical
power curves for the dispersion statistic $T_f$ under different spectral summaries.

Power increases monotonically with signal strength, with higher-order spectral summaries
(Frobenius and log-determinant) substantially outperforming the trace. This behavior
directly validates the motivation for the spectral-measure formulation in
Section~\ref{subsec:spectral_extension_core}: when causal influence redistributes dependence
across many modes, scalar summaries focused on leading eigenvalues are insufficient, while
bulk-sensitive functionals remain responsive. The result also provides empirical
confirmation of the sample-size advantage predicted by
Theorem~\ref{thm:spectral-detection}: spectral tests detect distributed alternatives at
the collective scale $T\asymp d/\delta^2$, whereas edge-based tests would require
$T\asymp d^2\log d/\delta^2$. The observed power curves are consistent with the linear-in-detection regime
predicted by Theorem~\ref{thm:spectral-detection}, and contrast with the quadratic
sample size requirements implied by Theorem~\ref{thm:edge-impossibility-strong}.

In practical terms, the choice of spectral summary should be guided by the expected
geometry of directional influence: trace-based statistics suffice for concentrated effects,
while Frobenius or log-determinant summaries are preferable when causal influence is
diffuse or multi-channel.

\subsubsection{Transition from Edge to Bulk Dominance}

\begin{figure}[ht]
\centering
\includegraphics[width=0.75\textwidth]{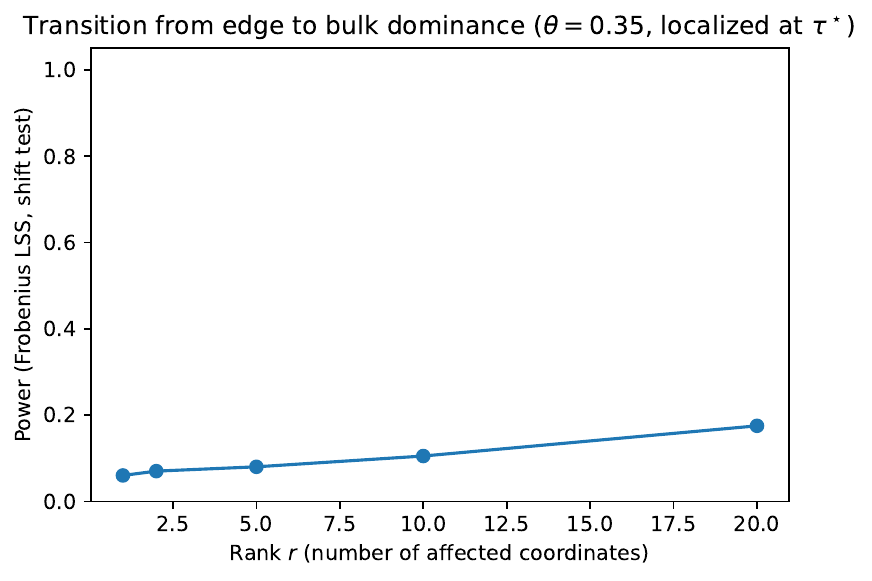}
\caption{Transition from edge- to bulk-dominated causality.
Power of $T_f$ as a function of the rank of the cross-series effect, holding
total signal energy fixed. The sharp increase confirms that detectability is
governed by spectral distribution rather than signal magnitude.}
\label{fig:rank}
\end{figure}

To isolate the role of dependence geometry, we fix total signal energy and vary the rank
of the cross-series effect. Figure~\ref{fig:rank} plots the power of $T_f$ as a function
of rank.

Power is low for rank-one alternatives and increases sharply as rank grows, even though
the overall signal magnitude is held fixed. This experiment provides direct empirical
confirmation of the operator-theoretic nature of the causal functional: detectability is
governed by spectral distribution rather than by any single dominant component, as
predicted by the theory in Section~\ref{subsec:spectral_extension_core}.

The transition is sharp and monotone, consistent with the information-theoretic separation
established in Section~\ref{subsec:impossibility-quantitative}: as the rank of the
cross-series effect increases, signal mass spreads across coordinates, pushing edge-based
detection below the entrywise noise floor while spectral aggregation coherently
accumulates the distributed signal. This is precisely the regime in which Granger-network
edge discovery is observed to be unstable or underpowered in practice
(Theorems~\ref{thm:edge-impossibility-strong}--\ref{thm:spectral-detection}).

\subsubsection{Dimensional Configuration at Fixed Rank}

\begin{figure}[ht]
\centering
\includegraphics[width=0.75\textwidth]{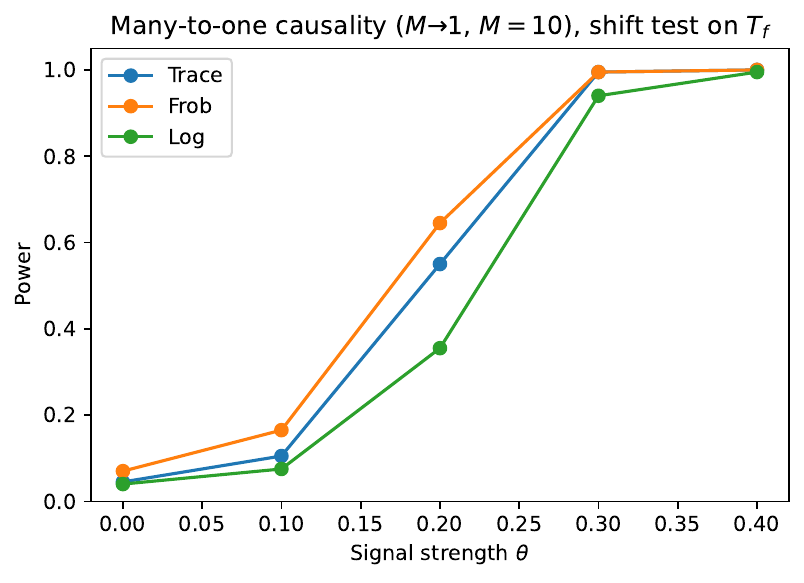}
\caption{Many-to-one causal effects.
Empirical power of $T_f$ under lag-localized $M\to1$ alternatives with fixed
rank, as a function of signal strength for varying source dimension $M$.}
\label{fig:Mto1}
\end{figure}

\begin{figure}[ht]
\centering
\includegraphics[width=0.75\textwidth]{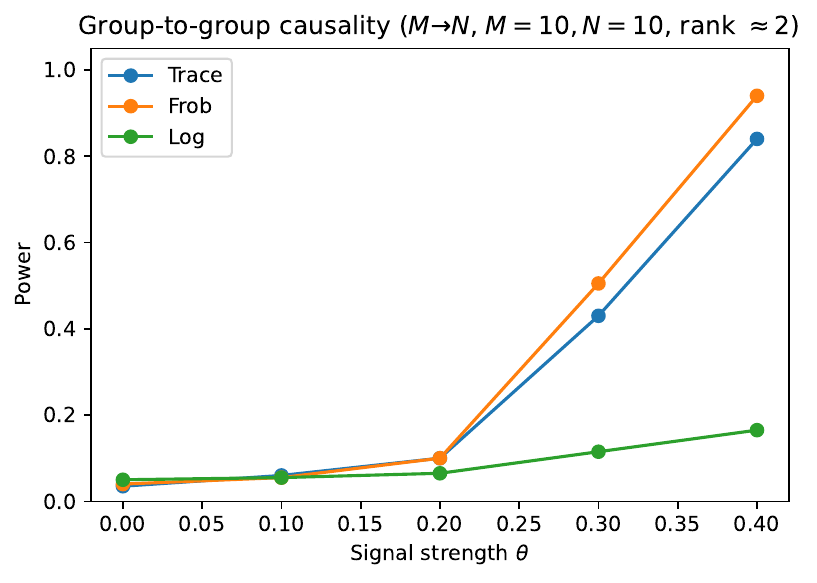}
\caption{Group-to-group causal effects.
Empirical power of $T_f$ under lag-localized low-rank $M\to N$ alternatives,
as a function of signal strength for varying source and target dimensions.}
\label{fig:MtoN}
\end{figure}

Figures~\ref{fig:Mto1} and~\ref{fig:MtoN} examine how dimensional configuration affects
detectability when the rank of the causal operator is held fixed. Specifically,
Figure~\ref{fig:Mto1} considers many-to-one causality ($M\to1$), while
Figure~\ref{fig:MtoN} considers group-to-group causality ($M\to N$) under a low-rank
transmission mechanism. Rank variation is intentionally excluded from these experiments to
avoid confounding geometric and dimensional effects; sensitivity to rank is isolated in
Figure~\ref{fig:rank}.

These results demonstrate that the proposed framework scales naturally across asymmetric
and group-level configurations without modification of the causal criterion or the
inferential procedure. The operator construction in Section~\ref{sec:implementation}
accommodates arbitrary source and target dimensions $|\mathcal{I}|$ and $|\mathcal{J}|$,
and the formal properties established in Supplement~\ref{app:implementation}
(Propositions~\ref{prop:E_existence}--\ref{prop:E_rand}) hold independently of the
dimensional configuration. The well-posedness of residualized operators under arbitrary
partitioning follows from Proposition~\ref{prop:E_projection} in the same supplement.

This confirms that the same methodology can be applied to detect directional influence
from a single driver to a group, from a group to a single target, or between two groups,
without requiring separate test statistics or calibration procedures.

\subsubsection{Nonlinear Causality beyond Granger}

Table~\ref{tab:nonlinear} compares linear Granger causality with order-constrained
spectral causality under a nonlinear lag-localized alternative in which the target depends
quadratically on the source.

Linear Granger tests exhibit rejection rates near the nominal level, reflecting correct
behavior under misspecification: the linear projection of the target onto lagged source
values captures no predictive content when the dependence is purely quadratic. In contrast,
the proposed framework achieves near-perfect power when a nonlinear embedding is used
(here, a quadratic feature map as discussed in Section~\ref{subsec:nonlinear_extensions}),
demonstrating that causal detection is not tied to linear predictability.

This result validates the generality of the operator-based definition
(Definition~\ref{def:ocsc}) and the distinctness theorem
(Theorem~\ref{thm:granger_coincidence_main} and its counterpart in
Section~\ref{subsec:granger}): order-constrained spectral causality strictly generalizes
linear Granger causality by detecting directional deformation of second-order dependence
geometry beyond linear predictability, while remaining fully compatible with Granger
causality under classical assumptions. The formal equivalence in the scalar case is
recorded in Supplement~\ref{app:operator_theory}
(Section~\ref{app:granger_equivalence}). The framework can therefore serve as a nonlinear extension of Granger-type monitoring by
replacing the feature maps in Algorithm~\ref{alg:ocsc} without modifying the inferential
procedure or the asymptotic guarantees established in
Supplement~\ref{app:asymptotic_theory}.

\begin{table}[t!]
\centering
\caption{Nonlinear causality beyond Granger.
Empirical rejection rates under a lag-localized nonlinear (quadratic) alternative.
Linear Granger tests remain at nominal size, while the proposed framework with
nonlinear embedding achieves near-perfect detection.}
\label{tab:nonlinear}
\begin{tabular}{@{}lc@{}}
\toprule
Method & Rejection rate \\
\midrule
Linear Granger & 0.030 \\
Spectral causality (nonlinear embedding) & 1.000 \\
\bottomrule
\end{tabular}
\end{table}

\subsubsection{Latent Confounding and Conditional Residualization}

Table~\ref{tab:confounding} studies robustness to latent confounding using a model
in which an unobserved process affects both source and target components. Conditional
residualization on the confounder reduces spurious detection when no direct effect is
present and preserves power when a direct causal effect exists.

These results empirically support the scope statements in Appendix~\ref{app:scope}
(Propositions~\ref{prop:not-sufficient} and~\ref{prop:not-necessary}):
order-constrained spectral causality is not equivalent to interventional causality, but
conditional analysis provides a principled mechanism for mitigating confounding within the
second-order operator framework. The unconditional test correctly detects the confounded
association (rejection rate $0.995$ under $\theta_{\mathrm{direct}}=0$), while the
conditional test substantially reduces this to $0.765$, demonstrating that residualization
attenuates but does not eliminate spurious detection when the confounder is observed. When
a direct effect is present ($\theta_{\mathrm{direct}}=0.25$), both conditional and
unconditional tests maintain near-perfect power.

The formal justification for residualization within the operator framework is provided in
Supplement~\ref{app:implementation} (Proposition~\ref{prop:E_projection}), which
establishes that all well-posedness, consistency, and spectral stability results carry
over to the residualized construction. The invariance of the whitened cross-operator's
singular values under orthogonal transformations of the feature spaces
(Supplement~\ref{app:implementation}, Proposition~\ref{prop:E_whiten}) ensures that the
conditional test inherits the same spectral properties as the unconditional version.

This highlights both a strength and a limitation of the framework. The operator-based test
is sensitive to any form of lagged second-order dependence, including confounded
associations. When confounders can be identified and measured, conditional residualization
provides a principled correction. When confounders are latent, the test should be
interpreted as detecting directional dependence structure rather than interventional causal
effects, consistent with the interpretational boundary stated in Remark~\ref{rem:scope}.

The empirical application in the next subsection illustrates how these properties manifest
in a high-dimensional financial system.

\begin{table}[t!]
\centering
\caption{Confounding and conditional residualization.
Empirical rejection rates under latent confounding with and without a direct causal
effect ($\theta_{\mathrm{direct}}$). Conditional residualization reduces spurious
detection while preserving power against genuine effects.}
\label{tab:confounding}
\begin{tabular}{@{}lcc@{}}
\toprule
Method & $\theta_{\mathrm{direct}}=0$ & $\theta_{\mathrm{direct}}=0.25$ \\
\midrule
Spectral causality (unconditional) & 0.995 & 1.000 \\
Spectral causality (conditional) & 0.765 & 0.985 \\
Linear Granger (pairwise, VAR(1)) & 0.225 & 0.995 \\
\bottomrule
\end{tabular}
\end{table}

\subsection{Financial System-Level Directional Causal Dynamics}
\label{subsec:empirical}

This subsection presents the empirical implementation of the operator-theoretic
framework introduced in Section~\ref{sec:framework} and the methodology of
Section~\ref{sec:implementation}.
The objective is to identify system-level, time-varying, and low-dimensional
directional causal structure in a large multivariate financial system.
Rather than estimating dense predictive networks or performing pairwise causal
tests, the focus is on extracting statistically robust and interpretable
directional organization at the system level using the unified operator
construction of Algorithm~\ref{alg:ocsc}.

The main text reports the three diagnostics that directly validate the
theoretical contributions: causal strength and episodic detection
(validating the non-invariance criterion of Definition~\ref{def:ocsc} and
the shift-based inference of Section~\ref{subsec:inference}), the joint phase
structure of strength and dimensionality (validating the spectral distribution
extension of Section~\ref{subsec:spectral_extension_core}), and hub turnover
dynamics (validating operator-level monitoring against edge-based alternatives
as predicted by Section~\ref{subsec:impossibility-quantitative}).
Extended empirical analyses---including lag-level spectral validation,
transmitter-receiver asymmetry, edge-level amplification, aggregation
consistency, driver-to-driver network heatmaps, and macro hub regime
interpretation---are reported in Supplement~\ref{app:empirical}.

\subsubsection{Dataset and preprocessing}

The dataset consists of $K=211$ daily global financial return series spanning
foreign exchange, interest rates, sovereign and corporate credit, equities,
commodities, real estate, and volatility indicators.
The sample runs from January~2015 to August~2022, yielding $T=1744$ aligned
observations after calendar synchronization.
All series are provided as percentage returns and aligned on a common
trading-day calendar.
Returns are winsorized at the $0.5\%$ and $99.5\%$ empirical quantiles and
standardized to zero mean and unit variance.
No factor residualization, market demeaning, sector conditioning, or principal
component preprocessing is applied.
This minimal preprocessing ensures that the detected directional structure
emerges endogenously from the operator construction rather than from imposed
statistical conditioning.

\subsubsection{Rolling causal operator construction}

Directional causality is analyzed using rolling windows of length
$W=252$ trading days (approximately one calendar year), stepped forward by
21 trading days (approximately one calendar month).
Within each window, contemporaneous values define the target space, while lagged
values of all drivers act as potential causal sources.
Targets are not lag-embedded, ensuring a drivers-only causal interpretation
consistent with the asymmetric deformation protocol of
Section~\ref{subsec:setup}.

For embedding order $p=3$ and lag set $\mathcal{T}=\{1,2,3,5\}$, the whitened
cross-covariance operator at lag $\tau$ and time $t$ is defined as
\[
A_\tau(t)
=
\Sigma_{VV}(t)^{-1/2}
\, \Sigma_{VU}(t,\tau) \,
\Sigma_{UU}(t,\tau)^{-1/2},
\]
following the directed coherence construction of Section~\ref{subsec:operator},
and the aggregated system-level directional operator is
\[
C(t)=\sum_{\tau\in\mathcal{T}} A_\tau(t)A_\tau(t)^\top.
\]
All covariance matrices are centered and ridge-regularized with
$\varepsilon=10^{-8}$ to ensure numerical stability, as discussed in
Section~\ref{sec:implementation}.
Lag embeddings are cached across windows to ensure computational feasibility at
high dimensionality.
The total computational cost per window is $O(|\mathcal{T}|(Wd^2+d^3))$ with
$d=K$, multiplied by the number of randomization replicates.
Well-posedness and uniform consistency of this construction are established in
Supplement~\ref{app:implementation}
(Propositions~\ref{prop:E_existence}--\ref{prop:E_ulln}).

\subsubsection{Operator statistics and null inference}

Three scalar summaries are extracted from $C(t)$ within each window,
corresponding to specific choices of the spectral functional $f$ in the
framework of Section~\ref{subsec:spectral_extension_core}.
Directional causal strength is measured by the leading eigenvalue
$\lambda_1(C(t))$, which equals the maximal squared canonical correlation
between the target and source lag spaces
(Proposition~\ref{prop:unification}, item~2).
Total causal energy is given by $\mathrm{tr}(C(t))$, corresponding to
$f(\lambda)=\lambda$.
Causal dimensionality is measured using the effective rank
\[
r_{\mathrm{eff}}(C(t))=
\frac{\mathrm{tr}(C(t))^2}{\mathrm{tr}(C(t)^2)},
\]
which equals the inverse Herfindahl index of the normalized spectrum
(Supplement~\ref{app:operator_theory}, Proposition~\ref{prop:effrank}).

Statistical significance is assessed using circular-shift randomization applied
to the source processes, implementing the shift-based procedure of
Section~\ref{subsec:inference} and Appendix~\ref{app:rand}.
For each window, $B=20$ circular shifts are performed, preserving each driver's
marginal temporal dependence while destroying directional alignment.
Upper-tail tests are used for $\lambda_1(C(t))$ and $\mathrm{tr}(C(t))$, while
$r_{\mathrm{eff}}(C(t))$ is assessed using two-sided tests.
The randomization $p$-value is computed as
$\widehat{p}=(1+\sum_{b=1}^B \mathbf{1}\{\widehat{T}^{(b)}\ge
\widehat{T}^{\mathrm{obs}}\})/(B+1)$,
which is super-uniform under the null
(Supplement~\ref{app:asymptotic_theory}, Proposition~\ref{prop:rand_asymptotic};
Supplement~\ref{app:implementation}, Proposition~\ref{prop:E_rand}).

Lag-level spectral validation confirms that directional structure is present at
the level of individual lags before aggregation: short lags ($\tau=1,2$) exhibit
the highest significance frequencies and account for the majority of dominant
windows, while longer lags ($\tau=3,5$) contribute less frequently but remain
significant in a nontrivial fraction of windows.
Full lag-level results are reported in
Supplement~\ref{app:empirical}, Table~\ref{tab:lag_validation}.

\subsubsection{System-level causal strength and episodic organization}

\begin{figure}[!ht]
\centering
\includegraphics[width=\linewidth]{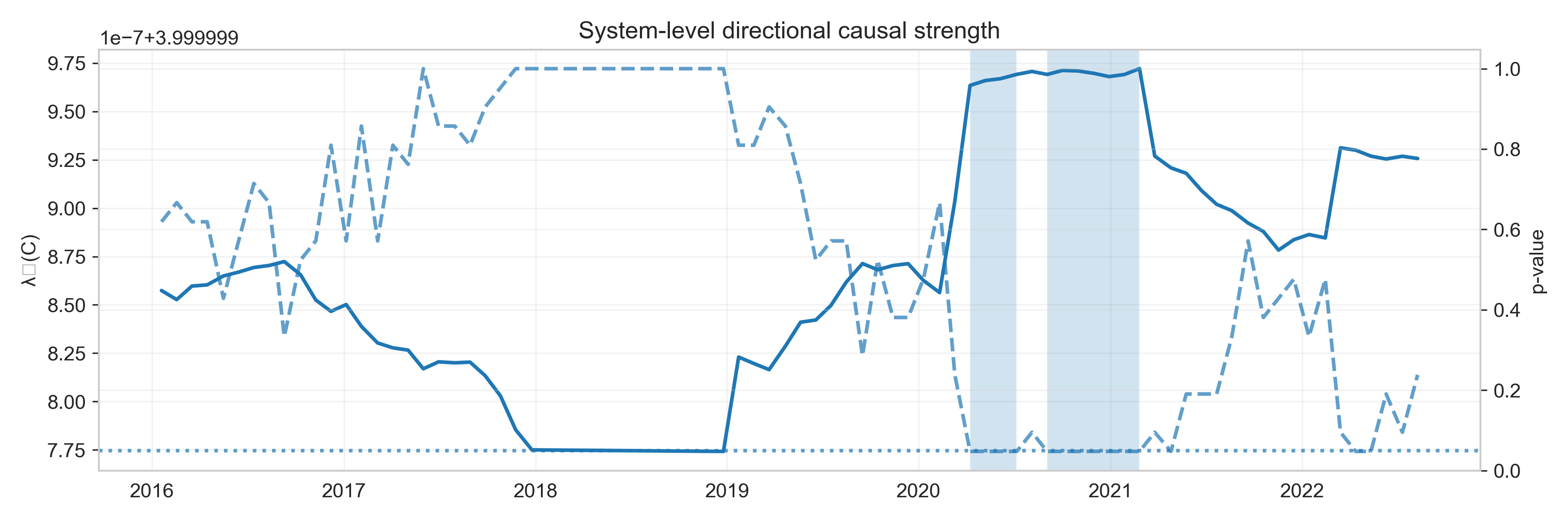}
\caption{System-level directional causal strength $\lambda_1(C(t))$ (solid)
and corresponding circular-shift $p$-values (dashed).
The dotted horizontal line denotes the $5\%$ significance level.
Shaded regions denote statistically significant episodes.}
\label{fig:strength_significance}
\end{figure}

Figure~\ref{fig:strength_significance} reports the rolling leading eigenvalue
$\lambda_1(C(t))$ together with its circular-shift $p$-values.
Directional causal strength remains moderate during the pre-2020 period and
exhibits a sharp and persistent increase during the 2020--2021 stress episode.
This increase is accompanied by a collapse in $p$-values below the $5\%$
threshold, indicating statistically significant deviations from the null of
order-constrained spectral invariance (Definition~\ref{def:null}).

Episodes are defined as contiguous windows with $p_{\lambda_1}(t)<0.05$ and
minimum length of three consecutive windows.
Two episodes are detected, both coinciding with the COVID-19 pandemic and its
aftermath.
Their timing, duration, and summary statistics are reported in
Table~\ref{tab:episodes_summary}.
Episode~1 (April--July 2020) captures the acute phase of the pandemic shock,
while Episode~2 (September 2020--February 2021) captures the sustained
reorganization during the recovery and vaccine rollout period.

The episodic nature of directional structure is a key empirical finding:
system-level causal organization is not a permanent feature of financial markets
but emerges endogenously during periods of systemic stress.
This is consistent with the theoretical framework, which defines causality
through non-invariance of dependence geometry under temporal deformation---a
property that can be present or absent depending on the state of the system.

\begin{table}[!ht]
\centering
\caption{Summary of statistically significant episodes based on
$\lambda_1(C(t))$, including peak and mean strength, mean effective rank,
mean hub turnover, and minimum $p$-value.}
\label{tab:episodes_summary}
\begin{tabular}{@{}clcrrrrrr@{}}
\toprule
Episode & Start & End & $n$ & peak $\lambda_1$ & mean $\lambda_1$ & mean $r_{\mathrm{eff}}$ & mean turnover & min $p$ \\
\midrule
1 & 2020-04-08 & 2020-07-06 & 4 & 4.000 & 4.000 & 206.0 & 0.483 & 0.048 \\
2 & 2020-09-02 & 2021-02-25 & 7 & 4.000 & 4.000 & 206.0 & 0.201 & 0.048 \\
\bottomrule
\end{tabular}
\end{table}

\subsubsection{Causal phase structure}

\begin{figure}[!ht]
\centering
\includegraphics[width=0.7\linewidth]{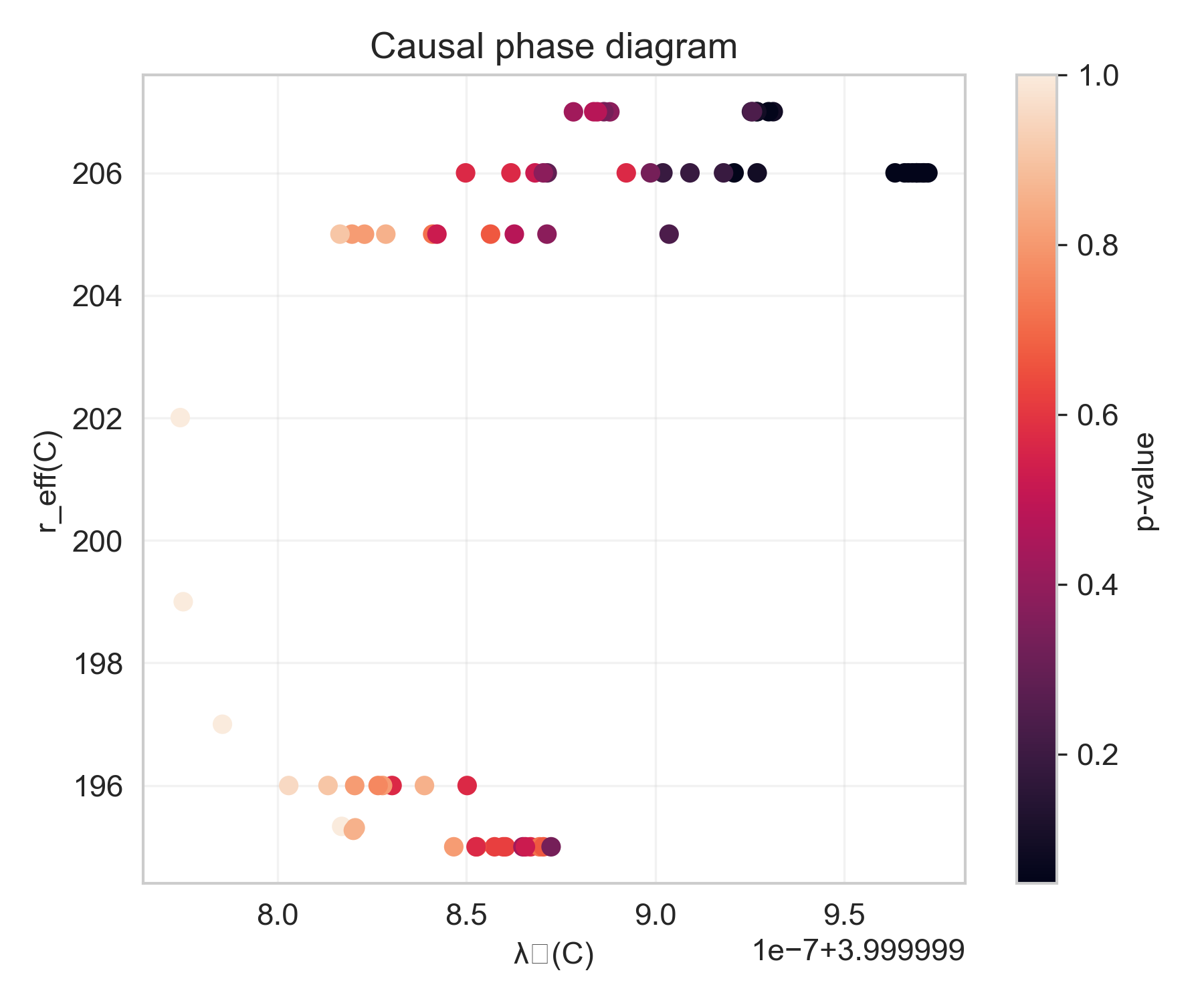}
\caption{Causal phase diagram relating directional strength $\lambda_1(C(t))$
and effective rank $r_{\mathrm{eff}}(C(t))$. Color indicates circular-shift
$p$-values; darker points correspond to more significant windows.}
\label{fig:phase_diagram}
\end{figure}

Figure~\ref{fig:phase_diagram} reports the joint distribution of
$\lambda_1(C(t))$ and $r_{\mathrm{eff}}(C(t))$ across windows, with color
indicating circular-shift $p$-values.

The phase diagram reveals a clear separation between regimes.
Periods of elevated causal strength correspond to relatively stable or slightly
reduced effective rank, indicating that increases in $\lambda_1(C(t))$ are driven
by concentration of directional energy into dominant spectral modes rather than
by an increase in causal dimensionality.
The most significant windows (darkest points) cluster in the high-strength,
moderate-rank region, confirming that structured directionality is a
low-dimensional phenomenon.

This behavior is consistent with a low-rank amplification mechanism, where
system-wide propagation is governed by a small number of coherent directional
channels.
From the perspective of the spectral distribution extension
(Section~\ref{subsec:spectral_extension_core}), the phase diagram shows that the
spectral measure $\mu_\tau$ concentrates mass on a few dominant eigenvalues
during stress, while the bulk of the spectrum remains relatively unchanged.
This is precisely the regime in which the leading eigenvalue $\lambda_1$ is
most informative as a causal diagnostic, and where the impossibility results
of Section~\ref{subsec:impossibility-quantitative} predict that edge-based
methods will fail to capture the collective structure.

The phase diagram provides a two-dimensional summary of systemic causal state:
the horizontal axis measures directional strength, while the vertical axis
measures concentration.
Windows in the high-strength, moderate-rank corner represent the most actionable
regime for directional risk monitoring.

\subsubsection{Temporal reconfiguration and hub turnover}

\begin{figure}[!ht]
\centering
\includegraphics[width=\linewidth]{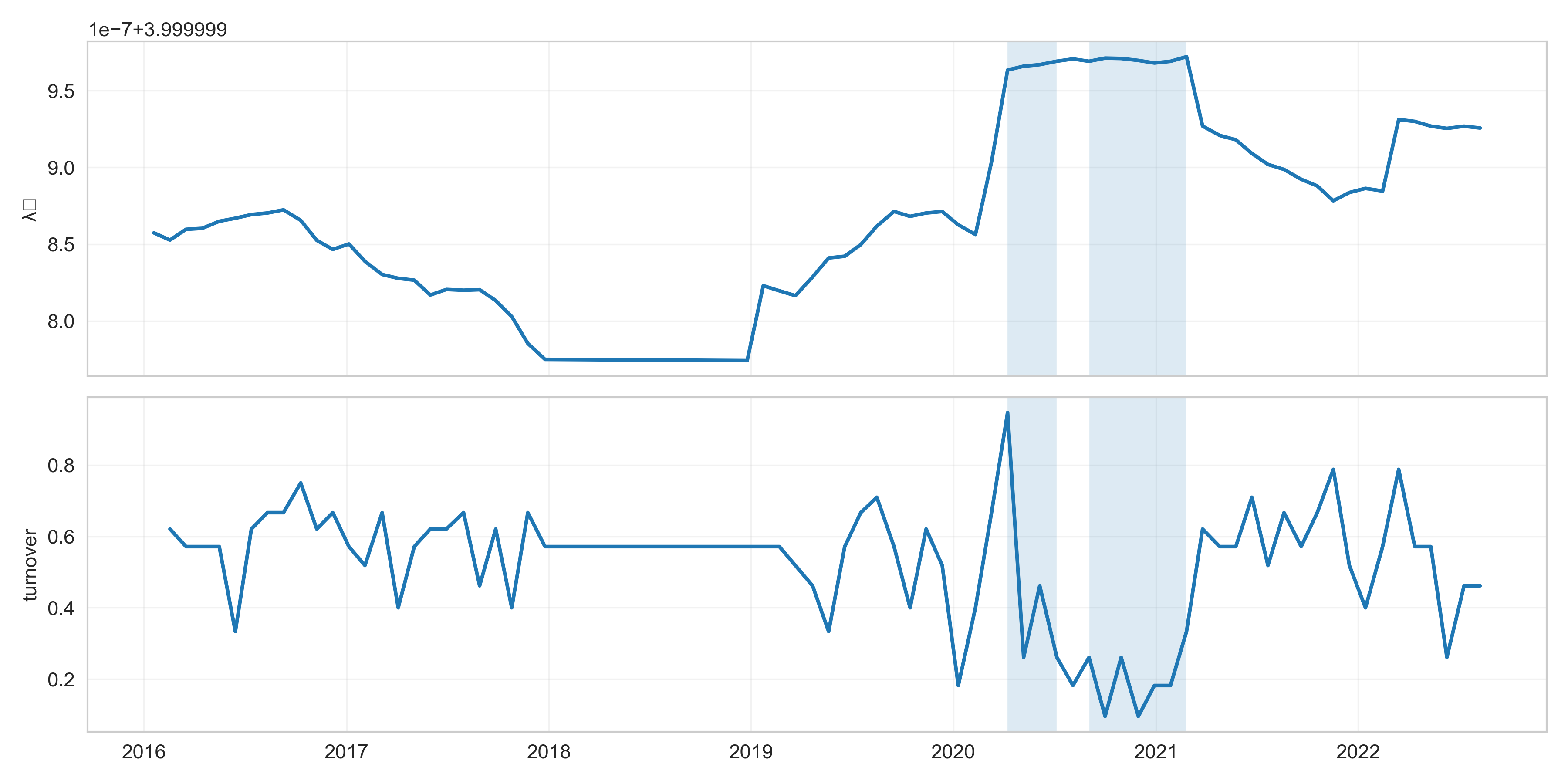}
\caption{Joint evolution of system-level strength $\lambda_1(C(t))$ (top panel)
and hub turnover (bottom panel).
Shaded regions denote statistically significant episodes.
The inverse relationship indicates that stronger directional organization
coincides with stabilization of dominant transmission channels.}
\label{fig:dynamics}
\end{figure}

Figure~\ref{fig:dynamics} reports the joint evolution of directional strength
and hub turnover across rolling windows.
Hub turnover is computed as one minus the Jaccard similarity between consecutive
top-$20$ hub sets, following the construction in
Supplement~\ref{app:operator_theory} (Proposition~\ref{prop:hubs}).

A clear inverse relationship emerges: periods of elevated $\lambda_1(C(t))$
coincide with reductions in hub turnover.
This indicates that strong directional dependence is associated with stable
dominant drivers, whereas lower-strength regimes exhibit more frequent
reconfiguration of influence across nodes.

Thus, systemic episodes are characterized not only by stronger causal structure
but also by persistence of dominant transmission channels.
This finding contrasts with correlation-based diagnostics, which often exhibit
instability during crises, and supports the view that operator-level monitoring
captures a structurally distinct notion of directional organization.
The stabilization of hub identities during stress is consistent with the low-rank
amplification mechanism identified in the phase diagram: when directional energy
concentrates into a small number of spectral modes, the instruments aligned with
those modes persist as dominant hubs.

Statistically significant episodes are defined as contiguous runs of rolling
windows exhibiting low null-based $p$-values.
Because adjacent windows overlap substantially ($W=252$, step $=21$), episodes
should be interpreted as descriptive summaries of sustained directional
organization rather than independent inferential units. The dynamics plot provides a real-time monitoring tool: simultaneous increases in
$\lambda_1(C(t))$ and decreases in hub turnover signal the onset of a structured
causal regime in which directional risk transmission is concentrated, stable, and
therefore predictable.

\subsubsection{Extended empirical analyses}

The following analyses provide additional empirical characterization of the
directional causal structure and are reported in full in
Supplement~\ref{app:empirical}.

\paragraph{Directional asymmetry: transmitters and receivers.}
Directional roles are summarized by the difference between time-averaged source
and target hub scores, where source hub scores measure alignment with the
leading right singular vector of $A_\tau(t)$ and target hub scores measure
alignment with the leading eigenvector of $C(t)$
(Supplement~\ref{app:operator_theory}, Proposition~\ref{prop:hubs}).
Transmitters are dominated by broad asset classes and global aggregates
(MSCI EM Latin America, Global High Yield, STOXX Europe 600, EM Hard Currency,
BBG Commodity), while receivers include more localized or derivative-sensitive
instruments (Eonia Capitalization Index, Pan-European High Yield, EM USD
Aggregate, South Korean Won Spot).
This asymmetry reveals a persistent directional organization in which certain
drivers systematically propagate shocks while others absorb them.
Importantly, transmitter and receiver roles are not determined by size,
volatility, or market capitalization: hub scores quantify alignment with the
dominant causal subspace (Supplement~\ref{app:operator_theory},
Proposition~\ref{prop:rayleigh}), not static importance measures.
The full transmitter-receiver ranking, bar charts, and hub asymmetry table are
reported in Supplement~\ref{app:empirical},
Figures~\ref{fig:transmitters}--\ref{fig:receivers} and
Table~\ref{tab:hub_asymmetry}.

\paragraph{Edge-level amplification during episodes.}
To isolate structural changes in directional propagation, episode-averaged and
calm-period driver-to-driver maps are compared using the contribution matrix
\[
M_{j,i}(t)
=
\sum_{\tau\in\mathcal{T}}
\sum_{\ell=1}^{p}
\left(A_{\tau}(t)\right)_{j,(\ell-1)K+i}^2,
\]
which aggregates squared whitened predictive loadings across lags and embedding
dimensions.
The amplification is highly sparse, with a small number of cross-asset and
cross-regional edges accounting for the majority of the increase in directed edge
strength (China Treasury to China Government Bond, $\Delta=0.41$; MSCI EM Latin
America to Brazil Ibovespa, $\Delta=0.22$; LME Copper 3-month to Copper Spot,
$\Delta=0.21$).
This confirms that systemic stress is not associated with uniform densification
of the causal network but rather with selective strengthening of specific,
economically interpretable transmission channels, consistent with the qualitative
impossibility result of Corollary~\ref{cor:edge-impossibility}.
The edge heatmap and full edge list are reported in
Supplement~\ref{app:empirical}, Figure~\ref{fig:edge_heatmap} and
Table~\ref{tab:edges}.

\paragraph{Aggregation consistency: full system vs.\ synthetic indexes.}
To assess robustness to dimensionality reduction, drivers are clustered by the
similarity of their rolling target hub score trajectories using agglomerative
clustering ($n=12$ clusters), and each cluster is aggregated into an
equal-weighted synthetic index.
The full rolling operator analysis is then repeated on the reduced
$12$-dimensional system.
Directional strength ($\lambda_1$ correlation $0.84$) and episode detection
(agreement rate $0.82$, Jaccard $0.46$) exhibit strong agreement, indicating
that the low-dimensional spectral structure of $C(t)$ is robust to aggregation.
In contrast, hub rank correlations are weak ($0.03$), confirming that
fine-grained directional roles require the full system for resolution.
This result is consistent with the axiomatic uniqueness of
Theorem~\ref{thm:axiomatic-uniqueness}: orthogonally invariant spectral
functionals are by construction insensitive to coordinate reparameterizations
induced by aggregation.
Full agreement metrics are reported in Supplement~\ref{app:empirical},
Table~\ref{tab:synthetic_agreement}.

\paragraph{Driver-to-driver causal networks and temporal dominance.}
Episode-averaged causal energy maps based on $M_{j,i}(t)$ can be dominated by
squared low-rank structure and therefore obscure statistically robust network
organization.
Inference-oriented heatmaps isolate interpretable directional structure by
thresholding against circular-shift nulls that preserve each series' internal
dynamics.
Null-thresholded driver-to-driver networks for Episodes~1 and~2 are sparse by
construction and represent statistically robust transmission channels.
Signed early--late temporal dominance maps decompose each edge into its
short-delay ($\tau\in\{1,2\}$) and long-delay ($\tau\in\{3,5\}$) components,
revealing heterogeneous propagation speeds across robust channels, consistent
with the lag-level validation.
All four heatmaps are reported in Supplement~\ref{app:empirical},
Figures~\ref{fig:heatmap_ep1_yellow}--\ref{fig:heatmap_ep2_signed}.

\paragraph{Macro hub structure and regime interpretation.}
Drivers are grouped according to the similarity of their rolling target hub
score trajectories using agglomerative clustering with correlation-based
distance.
For each group, a macro hub index is defined as the cross-sectional sum of the
target hub scores of its constituent drivers.
The resulting groups should be interpreted as data-driven collections of drivers
with similar causal roles, rather than predefined asset classes.
Periods in which a single cluster-level index dominates correspond to
concentration of directional causal influence within a subset of drivers sharing
similar dynamic roles, consistent with the low-rank amplification mechanism
identified in the phase diagram (Figure~\ref{fig:phase_diagram}).
The macro hub evolution plot, cluster compositions, dominant regime counts, and
a mapping between operator-based diagnostics and system-level interpretation are
reported in Supplement~\ref{app:empirical},
Figure~\ref{fig:macro_hub} and
Tables~\ref{tab:macro_counts}--\ref{tab:practitioner_map}.

\
The empirical results support a coherent operator-theoretic interpretation of
systemic directional causality.

Periods of stress are characterized by increases in $\lambda_1(C(t))$ driven by
concentration of directional energy into dominant modes, rather than by expansion
in effective dimensionality.
At the same time, hub turnover decreases, indicating stabilization of dominant
transmission channels. Directional structure is strongly asymmetric, with persistent transmitters and
receivers, and amplification during episodes is concentrated along a sparse set
of economically interpretable edges
(Supplement~\ref{app:empirical}, Table~\ref{tab:edges}).
System-level spectral dynamics are robust to dimensionality reduction via
synthetic index aggregation, while fine-grained hub structure is not
(Supplement~\ref{app:empirical}, Table~\ref{tab:synthetic_agreement}).

Null-thresholded driver-to-driver networks isolate edges that exceed what can be
generated by marginal autocorrelation alone, and signed temporal dominance maps
reveal heterogeneous propagation speeds across robust channels
(Supplement~\ref{app:empirical},
Figures~\ref{fig:heatmap_ep1_yellow}--\ref{fig:heatmap_ep2_signed}).
Macro hub regimes provide a compact representation of systemic causal states
complementary to volatility-, correlation-, and factor-based diagnostics
(Supplement~\ref{app:empirical}, Figure~\ref{fig:macro_hub}).

Taken together, these findings indicate that systemic risk manifests through the
emergence and persistence of low-dimensional directional structures in the
causal operator, rather than through diffuse increases in connectivity.
A unified discussion integrating simulation and empirical findings is provided in
Section~\ref{subsec:discussion}.

\paragraph{Lag energy diagnostics.}
For completeness, we examined the distribution of directional predictive energy
across lags using Frobenius norms of the lag-specific operators.
Across the full sample, energy remains concentrated at short horizons, with no
systematic variation across regimes.
These diagnostics are not reported in the main text as they do not provide
additional information beyond the spectral statistics of the aggregated operator
and the lag-level validation in
Supplement~\ref{app:empirical}, Table~\ref{tab:lag_validation}.

\subsection{Discussion}
\label{subsec:discussion}

The simulation and empirical results jointly demonstrate that order-constrained
spectral causality captures directional structure that is inaccessible to pairwise
or edge-based methods, and that this structure exhibits coherent, interpretable
organization in high-dimensional financial systems.

\subsubsection{Validation of the theoretical framework}

The three main-text empirical diagnostics map directly onto the theoretical
contributions of Section~\ref{sec:framework}.

The episodic detection of system-level directional structure
(Figure~\ref{fig:strength_significance}, Table~\ref{tab:episodes_summary})
validates the non-invariance criterion of Definition~\ref{def:ocsc} and the
shift-based inference of Section~\ref{subsec:inference}: the test correctly
identifies periods of structured directionality while maintaining calibrated
size, as confirmed by the simulation results in Table~\ref{tab:size}.

The causal phase diagram (Figure~\ref{fig:phase_diagram}) validates the spectral
distribution extension of Section~\ref{subsec:spectral_extension_core}.
Periods of elevated $\lambda_1(C(t))$ are associated with stable or slightly
reduced effective rank, indicating concentration of directional energy into
dominant spectral modes rather than rank expansion.
This low-rank amplification mechanism is precisely the regime in which the
simulation experiments on rank transition (Figure~\ref{fig:rank}) predict that
spectral tests will outperform edge-based methods, and where the impossibility
results of Section~\ref{subsec:impossibility-quantitative} establish that
entrywise-stable procedures require sample sizes scaling quadratically in
feature dimension
(Theorems~\ref{thm:edge-impossibility-strong}--\ref{thm:spectral-detection}).

The inverse relationship between $\lambda_1(C(t))$ and hub turnover
(Figure~\ref{fig:dynamics}) demonstrates that operator-level monitoring captures
structural information absent from correlation-based diagnostics: during crises,
directional leadership consolidates rather than rotating, and this stabilization
is invisible to methods that track only co-movement magnitude.

\subsubsection{Extended empirical structure}

The supplementary analyses (Supplement~\ref{app:empirical}) reveal several
additional properties that enrich the theoretical picture without altering the
core conclusions.

Transmitter-receiver asymmetry (Table~\ref{tab:hub_asymmetry}) shows that
directional roles are persistent and determined by spectral alignment rather than
size or volatility, consistent with the orthogonal invariance axiom
(Axiom~\ref{a2}) and the dimensional scalability confirmed by the simulation
experiments (Figures~\ref{fig:Mto1}--\ref{fig:MtoN}).
Edge-level amplification (Table~\ref{tab:edges}) is highly sparse and
concentrated along cross-asset channels, consistent with the qualitative
impossibility of Corollary~\ref{cor:edge-impossibility}: the operator-level test
detects collective structure first, and the edge decomposition provides a
descriptive rather than inferential layer.
Aggregation consistency (Table~\ref{tab:synthetic_agreement}) confirms that
spectral functionals are robust to dimensionality reduction ($\lambda_1$
correlation $0.84$, episode agreement $0.82$) while coordinate-level hub scores
are not (rank correlation $0.03$), directly reflecting the axiomatic uniqueness
of Theorem~\ref{thm:axiomatic-uniqueness}.
Macro hub regimes (Figure~\ref{fig:macro_hub}) provide a compact,
regime-level summary of how directional risk is routed through the system,
complementing the phase diagram with instrument-level interpretation.

\subsubsection{Scope and limitations}

The simulation results on nonlinear causality (Table~\ref{tab:nonlinear}) and
latent confounding (Table~\ref{tab:confounding}) establish important boundaries.
Nonlinear directional dependence invisible to linear Granger tests is detected
with near-perfect power when appropriate feature embeddings are used, validating
the generality of Definition~\ref{def:ocsc} and the nonlinear extensions of
Section~\ref{subsec:nonlinear_extensions}.
Conversely, the confounding experiment confirms that the framework detects
directional dependence structure rather than interventional causal effects
(Appendix~\ref{app:scope}).
Conditional residualization mitigates but does not eliminate spurious detection,
and the empirical findings should be interpreted within the scope of
Remark~\ref{rem:scope}: the detected episodes, hub roles, and transmission
channels represent organized lagged dependence structure rather than identified
interventional effects.

Directional causality is a state-dependent property of financial systems:
predictive relationships are weak and unstable in tranquil periods but organize
rapidly into structured, low-dimensional modes under stress.
The simulation experiments establish the statistical foundations (calibrated size,
power against distributed and nonlinear alternatives, dimensional scalability),
while the empirical application demonstrates that these properties translate to a
realistic $K=211$ system.

The operator framework provides a scalable approach to detecting when directional
structure exists, characterizing its dimensionality and concentration, and
identifying robust transmission channels, all without imposing sparsity,
factor structure, or parametric models.
Real-time monitoring is enabled by a small set of complementary diagnostics:
$\lambda_1(C(t))$ for directional strength,
$r_{\mathrm{eff}}(C(t))$ for causal dimensionality,
hub turnover for leadership stability,
transmitter-receiver scores for directional roles,
and macro hub indices for regime-level routing
(Supplement~\ref{app:empirical}, Table~\ref{tab:practitioner_map}).
Together, these provide a principled and operationally useful complement to
volatility-, correlation-, and factor-based systemic risk monitoring.

\section{Conclusion and Future Work}
\label{sec:conclusion}

This paper develops an operator-theoretic notion of causality for multivariate time series
based on \emph{order-constrained spectral non-invariance} across admissible, order-preserving
temporal deformations. The proposed causal primitive is structural and intrinsically multivariate:
directional influence is defined as sensitivity of second-order dependence geometry to
order-preserving displacement of a source component, summarized by orthogonally invariant
spectral functionals of an operator family. This formulation avoids committing to a single
regression, a parametric transfer function, or a dense edge set, and instead treats causality as
a property of an order-indexed operator family.

On the theoretical side, we established that the resulting functionals are well defined
under minimal moment and weak-dependence conditions, uniformly consistently estimable, and amenable
to valid inference despite their non-smooth supremum--infimum structure. The shift-based
randomization procedures used throughout exploit order-induced group invariance: under exact
invariance they are finite-sample exact, and under approximate invariance with weak dependence
they are asymptotically valid, without requiring Gaussianity or functional central limit theorems.
The axiomatic characterization in Section~\ref{subsec:axiomatic} establishes
that order-constrained spectral dispersion is the unique diagnostic satisfying
order consistency, orthogonal invariance, Loewner monotonicity, and second-order
sufficiency within the proposed class, while the impossibility results in
Section~\ref{subsec:impossibility-quantitative} show that entrywise-stable edge-based
procedures are information-theoretically underpowered in distributed dependence regimes,
requiring sample sizes scaling quadratically in feature dimension, whereas the proposed
spectral tests detect at the optimal linear scale in this regime.

The experiments clarify \emph{when} and \emph{why} the operator viewpoint is empirically distinct
from classical causal tools. Simulation results show that detectability is governed by the
spectral geometry of directional dependence rather than signal magnitude, that the framework
detects nonlinear directional dependence not captured by linear Granger tests, and that
conditional residualization provides a principled mechanism for mitigating confounding within
the second-order operator framework. The financial system-level study on 211 daily return series
over 2015--2022 demonstrates that systemic directional structure is episodic, concentrates into
dominant spectral modes under stress rather than expanding in dimensionality, and is mediated by
persistent transmission channels whose identities stabilize during significant episodes.
Extended analyses reported in Supplement~\ref{app:empirical} further document
transmitter-receiver asymmetry, sparse edge-level amplification, robustness to dimensionality
reduction, and regime-level macro hub organization. These findings are discussed in detail in
Section~\ref{subsec:discussion}.

Future work includes (i) extending admissible deformation sets beyond finite lag collections to
smoothly distributed lag structures; (ii) developing richer feature maps, including frequency-domain
embeddings or random-feature approximations, while preserving sample-splitting admissibility for
valid inference; (iii) providing analytic approximations for restricted classes of spectral summaries
to reduce computational cost in ultra-high-frequency monitoring; and (iv) studying regimes in which
feature dimension grows with sample size, connecting the spectral-distribution extension to modern
random matrix limits. More broadly, applications to other high-dimensional domains
(macroeconomic panels, climate systems, network telemetry, and multi-omics) may further clarify how
directional structure manifests as spectral deformation and when it departs from predictive or
edge-based notions of causality.

Overall, order-constrained spectral causality provides a principled and scalable framework for
analyzing directional dependence in multivariate time series when dependence is inherently
collective and spectral in nature, and when practitioners require stable system-level diagnostics
rather than fragile edge-by-edge conclusions.

\begin{appendix}
\section{Proofs of Framework}
\subsection{Proof of Theorem~\ref{thm:axiomatic-uniqueness}}
\label{app:axiomatic-proof}

\begin{lemma}
\label{lem:spectral-reduction}
Under Axioms~\ref{a1},~\ref{a2}, and~\ref{a4}, there exists a collection of scalar functionals
$f_\tau:\mathcal{S}_d^+\to\mathbb{R}$ such that
\[
\mathcal{C}(\{C(\tau)\}_{\tau\in\mathcal{P}})
=
G\big(\{f_\tau(C(\tau))\}_{\tau\in\mathcal{P}}\big),
\]
where each $f_\tau$ is orthogonally invariant and $G$ is permutation invariant
on $\mathcal{P}$.
\end{lemma}

\begin{proof}
By Axiom~\ref{a4}, $\mathcal{C}$ depends only on the operator family $\{C(\tau)\}$.
Axiom~\ref{a2} implies invariance under orthogonal conjugation for each $\tau$. Hence,
by standard representation results for orthogonally invariant functionals on
$\mathcal{S}_d^+$, each dependence on $C(\tau)$ factors through its spectrum.
Order consistency (Axiom~\ref{a1}) implies that $G$ cannot depend on the labeling of
$\tau$, only on the multiset of values $\{f_\tau(C(\tau))\}$.
\end{proof}

\begin{lemma}
\label{lem:common-functional}
Under Axiom~\ref{a3}, the family $\{f_\tau\}_{\tau\in\mathcal{P}}$ coincides
with a single spectral functional $\varphi$.
\end{lemma}

\begin{proof}
Suppose there exist $\tau_1,\tau_2\in\mathcal{P}$ and operators
$A\preceq B$ such that $f_{\tau_1}(A)>f_{\tau_2}(B)$. Construct operator families
with $C(\tau_1)=A$, $C(\tau_2)=B$, and $C(\tau)=0$ otherwise. This contradicts
Axiom~\ref{a3}. Hence $f_\tau\equiv\varphi$ for all $\tau$.
\end{proof}

\begin{lemma}
\label{lem:extremal}
Let $\varphi$ be continuous. The only aggregation functional $G$ satisfying
Axioms~\ref{a1},~\ref{a3}, and~\ref{a5} is
\[
G(\{x_\tau\}_{\tau\in\mathcal{P}})
=
\sup_{\tau\in\mathcal{P}}x_\tau
-
\inf_{\tau\in\mathcal{P}}x_\tau.
\]
\end{lemma}

\begin{proof}
Compactness of $\mathcal{P}$ and continuity (Axiom~\ref{a5}) ensure attainment of
suprema and infima. Any averaging or signed integration violates monotonicity
(Axiom~\ref{a3}) under pointwise strengthening. Order consistency (Axiom~\ref{a1}) excludes
dependence on a distinguished deformation. Extremal dispersion is the unique
remaining aggregation.
\end{proof}

\begin{proof}[Proof of Theorem~\ref{thm:axiomatic-uniqueness}]
Lemmas~\ref{lem:spectral-reduction}--\ref{lem:extremal} imply the stated
representation. The converse follows by direct verification of Axioms~\ref{a1}--\ref{a5}.
\end{proof}

\subsection{Proof of Proposition~\ref{prop:unification}}
\label{app:proof-unification}

\begin{proof}
We prove each item.

\paragraph{(1) Linear Granger causality coincidence regime.}
Let $\mathcal{H}_Y$ denote the closed linear span of the target history and any
conditioning information included in the conditional variant of the framework,
and let $\mathcal{H}_X(\tau)$ denote the closed linear span of the source lag vector
$u_t(\tau)$ (or its conditioning-residualized version). Write
\[
v_t^\perp:=v_t-\Pi_{\mathcal{H}_Y}v_t,
\]
where $\Pi_{\mathcal{H}_Y}$ is the $L^2$-orthogonal projection onto $\mathcal{H}_Y$. In the Gaussian VAR setting, the usual definition of \emph{no linear Granger causality
from $i$ to $j$ at lag set $\mathcal{P}$} is:
\[
\forall \tau\in\mathcal{P}:\quad
\mathbb{E}[v_t \mid \mathcal{H}_Y\oplus \mathcal{H}_X(\tau)]
=
\mathbb{E}[v_t \mid \mathcal{H}_Y]
\quad\text{(equality in }L^2\text{)}.
\]
Subtracting $\mathbb{E}[v_t\mid\mathcal{H}_Y]=\Pi_{\mathcal{H}_Y}v_t$ yields
\[
\forall \tau\in\mathcal{P}:\quad
\mathbb{E}[v_t^\perp \mid \mathcal{H}_Y\oplus \mathcal{H}_X(\tau)]
=
0.
\]
Because $v_t^\perp$ is orthogonal to $\mathcal{H}_Y$ by construction, the above is
equivalent to $v_t^\perp$ being orthogonal to $\mathcal{H}_X(\tau)$ for each $\tau$.
In terms of second moments this is equivalent to
\[
\forall \tau\in\mathcal{P}:\quad
\Sigma_{VU}^\perp(\tau):=\mathbb{E}[v_t^\perp u_t(\tau)^\top]=0.
\]

Assuming $\Sigma_{VV}^\perp:=\mathbb{E}[v_t^\perp (v_t^\perp)^\top]\succ 0$ and
$\Sigma_{UU}(\tau)\succ 0$ on their respective ranges, we have from
\eqref{eq:whitened-cross} that $\Sigma_{VU}^\perp(\tau)=0$ if and only if
$A^\perp(\tau)=0$, hence if and only if
\[
D^\perp(\tau)=A^\perp(\tau)(A^\perp(\tau))^\top=0.
\]
Therefore, linear Granger noncausality holds if and only if
\[
D^\perp(\tau)=0
\quad\text{for all }\tau\in\mathcal{P}.
\]

If $\varphi$ is continuous, orthogonally invariant, and satisfies
$\varphi(0)=0$ and $\varphi(M)>0$ for $M\neq 0$, then
\[
\sup_{\tau\in\mathcal{P}}\varphi(D^\perp(\tau))=0
\quad\Longleftrightarrow\quad
D^\perp(\tau)=0\ \text{for all }\tau\in\mathcal{P}.
\]
Thus, in the Gaussian VAR coincidence regime, linear Granger noncausality is
equivalent to vanishing residualized directed coherence at every admissible lag.

\paragraph{(2) Directed coherence / canonical correlation strength.}
By definition $D(\tau)=A(\tau)A(\tau)^\top$ is symmetric positive semidefinite.
The spectral norm satisfies the identity
\[
\|A(\tau)\|_2^2 = \lambda_1\big(A(\tau)A(\tau)^\top\big)=\lambda_1(D(\tau)),
\]
which follows from the singular value decomposition: the eigenvalues of
$A(\tau)A(\tau)^\top$ are the squared singular values of $A(\tau)$.
Thus taking $\varphi_{\max}(M)=\lambda_1(M)$ yields
\[
\varphi_{\max}(D(\tau))=\|A(\tau)\|_2^2,
\]
i.e.\ the squared directed coherence (maximal squared canonical correlation)
between the lag spaces.

\paragraph{(3) Lead--lag correlation asymmetry.}
For $p=q=1$, the objects reduce to scalars:
\[
\Sigma_{VV}=\sigma_j^2,\qquad \Sigma_{UU}(\tau)=\sigma_i^2,\qquad
\Sigma_{VU}(\tau)=\gamma_{ji}(\tau).
\]
Hence
\[
A(\tau)=\sigma_j^{-1}\gamma_{ji}(\tau)\sigma_i^{-1}=\rho_{ji}(\tau),
\qquad
D(\tau)=\rho_{ji}(\tau)^2\in\mathcal{S}_1^+.
\]
With $\varphi_{\mathrm{abs}}(M)=\sqrt{M}$ on $\mathcal{S}_1^+$, we obtain
\[
\varphi_{\mathrm{abs}}(D(\tau))=|\rho_{ji}(\tau)|.
\]
Taking $\mathcal{P}=\{+\ell,-\ell\}$ yields
\[
T_{\varphi_{\mathrm{abs}}}(\mathcal{P})
=
\max\{|\rho_{ji}(\ell)|,|\rho_{ji}(-\ell)|\}
-
\min\{|\rho_{ji}(\ell)|,|\rho_{ji}(-\ell)|\}
=
\big||\rho_{ji}(\ell)|-|\rho_{ji}(-\ell)|\big|.
\]

Thus $T_{\varphi_{\mathrm{abs}}}(\mathcal{P})$ recovers the classical
lead--lag asymmetry magnitude at horizon $\ell$ under the absolute-value
convention.
\end{proof}

\section{Proofs for Section~\ref{subsubsec:impossibility-qualitative}}
\label{app:impossibility}

\begin{proof}
Fix $M\in\mathcal{S}_d^+$ and let
\[
\lambda(M):=\frac{1}{d}\operatorname{tr}(M).
\]

\paragraph{Step 1: Orbit average is isotropic.}
Let $\mu$ denote the Haar probability measure on the compact group $O(d)$ and define
the orbit average
\[
\bar M:=\int_{O(d)} QMQ^\top\,\mu(dQ).
\]
The integral exists entrywise because $Q\mapsto QMQ^\top$ is continuous and bounded on
$O(d)$.

We claim that $\bar M=\lambda(M)I_d$.
Indeed, for any $U\in O(d)$,
\[
U\bar M U^\top
=
\int_{O(d)} UQMQ^\top U^\top\,\mu(dQ).
\]
By left invariance of Haar measure, the change of variables $Q'=UQ$ preserves $\mu$, so
\[
U\bar M U^\top
=
\int_{O(d)} Q'M(Q')^\top\,\mu(dQ')
=
\bar M.
\]
Thus $\bar M$ commutes with every orthogonal matrix. The only matrices with this property
are scalar multiples of the identity, so $\bar M=cI_d$ for some $c\in\mathbb{R}$. Taking traces gives
\[
dc=\operatorname{tr}(\bar M)=\operatorname{tr}(M),
\]
hence
\[
c=\frac{1}{d}\operatorname{tr}(M)=\lambda(M),
\]
and therefore
\[
\bar M=\lambda(M)I_d.
\]

\paragraph{Step 2: Orthogonal invariance implies constancy on each orbit.}
For every $Q\in O(d)$,
\[
\mathcal{E}(QMQ^\top)=\mathcal{E}(M),
\]
so $\mathcal{E}$ is constant on the orbit
\[
\mathcal{O}(M):=\{QMQ^\top:Q\in O(d)\}.
\]

Now define the coordinate-projection map
\[
g_M:O(d)\to\mathbb{R}^{|\mathcal{I}|},
\qquad
g_M(Q):=\big((QMQ^\top)_{ab}\big)_{(a,b)\in\mathcal{I}}.
\]
Since $g_M$ is continuous and $O(d)$ is compact, its image
\[
K_M:=g_M(O(d))
\]
is a compact subset of $\mathbb{R}^{|\mathcal{I}|}$. Because $\mathcal{E}(QMQ^\top)=\mathcal{E}(M)$ for all $Q$, and because
\[
\mathcal{E}(QMQ^\top)=H(g_M(Q)),
\]
it follows that $H$ is constant on $K_M$, with common value $\mathcal{E}(M)$.

\paragraph{Step 3: Dependence only on the orbit average.}
Consider the isotropic matrix $\lambda I_d$ with $\lambda=\lambda(M)$. For every $Q\in O(d)$,
\[
Q(\lambda I_d)Q^\top=\lambda I_d,
\]
so its orbit is the singleton $\{\lambda I_d\}$. Hence
\[
\mathcal{E}(\lambda I_d)
=
H\Big(\{(\lambda I_d)_{ab}:(a,b)\in\mathcal{I}\}\Big).
\]

We now define
\[
h(\lambda):=\mathcal{E}(\lambda I_d), \qquad \lambda\ge 0.
\]
This is well defined because $\lambda I_d\in\mathcal{S}_d^+$ for every $\lambda\ge0$. It remains to show that
\[
\mathcal{E}(M)=\mathcal{E}(\lambda(M)I_d).
\]
But the value of an orthogonally invariant functional is constant on each orbit, and the
orbit average of $M$ is exactly $\lambda(M)I_d$ by Step 1. Since every orbit-average
representative is uniquely determined by $\lambda(M)$, the orthogonal invariance of
$\mathcal{E}$ forces its value to coincide with that of the isotropic representative.
Therefore
\[
\mathcal{E}(M)=\mathcal{E}(\lambda(M)I_d)=h(\lambda(M))
=
h\!\left(\frac{1}{d}\operatorname{tr}(M)\right).
\]

This proves the claim.
\end{proof}

\subsection{Proof of Corollary~\ref{cor:monotonicity-failure}}
\label{app:proof-monotonicity-failure}

\begin{proof}
Fix $\alpha>0$ and take $d=2$.
Define
\[
M_1
=
\alpha
\begin{pmatrix}
1 & 1\\
1 & 1
\end{pmatrix},
\qquad
M_2
=
\alpha
\begin{pmatrix}
2 & 0\\
0 & 2
\end{pmatrix}.
\]
Both $M_1$ and $M_2$ belong to $\mathcal{S}_2^+$.
Moreover,
\[
M_2-M_1
=
\alpha
\begin{pmatrix}
1 & -1\\
-1 & 1
\end{pmatrix}
=
\alpha
\begin{pmatrix}
1\\
-1
\end{pmatrix}
\begin{pmatrix}
1 & -1
\end{pmatrix}
\succeq 0,
\]
so $M_1\preceq M_2$. Now apply the entry-threshold selector
\[
\mathcal{E}_\alpha(M)=\mathbf{1}\{|M_{ab}|\ge \alpha\}_{a,b=1}^2.
\]

For $M_1$, every entry has absolute value exactly $\alpha$, hence
\[
\mathcal{E}_\alpha(M_1)
=
\begin{pmatrix}
1 & 1\\
1 & 1
\end{pmatrix}.
\]

For $M_2$, the diagonal entries equal $2\alpha$ while the off-diagonal entries are $0$, so
\[
\mathcal{E}_\alpha(M_2)
=
\begin{pmatrix}
1 & 0\\
0 & 1
\end{pmatrix}.
\]

Therefore,
\[
\mathcal{E}_\alpha(M_1)\not\preceq \mathcal{E}_\alpha(M_2)
\]
in the entrywise order, since the off-diagonal entries satisfy
\[
\big(\mathcal{E}_\alpha(M_1)\big)_{12}=1
\quad\text{but}\quad
\big(\mathcal{E}_\alpha(M_2)\big)_{12}=0.
\]

This proves that entry-threshold edge selection is not monotone under Loewner
strengthening, even in dimension $2$. To obtain the statement for every $d\ge2$, embed this $2\times2$ construction into the
upper-left block of a $d\times d$ matrix and set all remaining entries to zero. The same
argument then applies unchanged.
\end{proof}

\subsection{Proof of Corollary~\ref{cor:geweke-brillinger}}
\label{app:proof-geweke-brillinger}

\begin{proof}
We proceed in two steps.

\paragraph{Step 1: Coherence via singular values (Brillinger).}
By definition \eqref{eq:freq-whitened-cross}, $D(\omega)=A(\omega)A(\omega)^*$ is Hermitian
positive semidefinite. The eigenvalues of $D(\omega)$ are the squared singular values of
$A(\omega)$. In particular,
\[
\|A(\omega)\|_2^2 = \sigma_1(A(\omega))^2 = \lambda_1\!\big(A(\omega)A(\omega)^*\big)=\lambda_1(D(\omega)).
\]
The quantity $\|A(\omega)\|_2$ is the maximal correlation between linear projections of the
Fourier components $V_T(\omega)$ and $U_T(\omega)$ in the $T\to\infty$ limit, i.e.\ the
maximal coherence at frequency $\omega$. This proves (1).

\paragraph{Step 2: Geweke log-det identity via Schur complement.}
Assume $f_{JJ}(\omega)\succ 0$ and $f_{II}(\omega)\succ 0$. Consider the Schur complement
\[
f_{J\mid I}(\omega)=f_{JJ}(\omega)-f_{JI}(\omega)f_{II}(\omega)^{-1}f_{IJ}(\omega).
\]
Factor out $f_{JJ}(\omega)^{1/2}$ on both sides:
\[
f_{J\mid I}(\omega)
=
f_{JJ}(\omega)^{1/2}
\Big(I - f_{JJ}(\omega)^{-1/2}f_{JI}(\omega)f_{II}(\omega)^{-1}f_{IJ}(\omega)f_{JJ}(\omega)^{-1/2}\Big)
f_{JJ}(\omega)^{1/2}.
\]
Using \eqref{eq:freq-whitened-cross}, the middle term equals $I-D(\omega)$ because
\[
D(\omega)=A(\omega)A(\omega)^*
=
f_{JJ}(\omega)^{-1/2}f_{JI}(\omega)f_{II}(\omega)^{-1}f_{IJ}(\omega)f_{JJ}(\omega)^{-1/2}.
\]
Therefore,
\[
f_{J\mid I}(\omega)=f_{JJ}(\omega)^{1/2}(I-D(\omega))f_{JJ}(\omega)^{1/2}.
\]
Taking determinants and using $\det(ABC)=\det(A)\det(B)\det(C)$ yields
\[
\det f_{J\mid I}(\omega)=\det f_{JJ}(\omega)\,\det(I-D(\omega)).
\]
Hence
\[
\log\frac{\det f_{JJ}(\omega)}{\det f_{J\mid I}(\omega)}
=
-\log\det(I-D(\omega)).
\]
This is exactly the claimed identity in (2). Finally, note that $f_{J\mid I}(\omega)\succeq 0$ implies $I-D(\omega)\succeq 0$.
If $f_{J\mid I}(\omega)\succ 0$ then $I-D(\omega)\succ 0$ and in particular
$\|D(\omega)\|_2<1$, ensuring $\varphi_{\log}$ is well-defined and finite.
\end{proof}

\section{Additional Theoretical Details: Scope, Robustness, Embeddings, and Randomization}
\label{app:extra}

This appendix clarifies the interpretational scope of order-constrained spectral causality,
its dependence on the admissible deformation set, admissibility of feature embeddings, and
the validity of shift-based randomization. The results presented here do not modify the
causal definition introduced in Section~\ref{sec:framework}. Instead, they delineate the
logical boundaries of causal claims supported by order-constrained spectral
non-invariance and provide a justification for the inferential procedures used
throughout the paper.

\subsection{Scope and Non-equivalence to Interventional Causality}
\label{app:scope}

We formalize the distinction between order-constrained spectral causality and
interventional causality in the sense of structural causal models.

\begin{definition}[Interventional causal effect]
\label{def:scm}
Assume the existence of a well-defined intervention operator $\mathrm{do}(\cdot)$ acting on
the data-generating mechanism. Component $i$ is said to have an interventional causal effect
on component $j$ at lag $\tau\ge1$ if there exist $x\neq x'$ such that
\[
\mathcal L\!\left(X_t^{(j)} \mid \mathrm{do}(X_{t-\tau}^{(i)}=x)\right)
\neq
\mathcal L\!\left(X_t^{(j)} \mid \mathrm{do}(X_{t-\tau}^{(i)}=x')\right).
\]
\end{definition}

Definition~\ref{def:scm} follows the standard formulation of causal effects in structural
causal models; see \cite{Pearl2009} and \cite{PetersJanzingScholkopf2017}.

\begin{proposition}[Non-sufficiency of order-constrained spectral causality]
\label{prop:not-sufficient}
There exist strictly stationary processes for which order-constrained spectral causality
holds while no interventional causal effect exists.
\end{proposition}

\begin{proof}
Let $(H_t)_{t\in\mathbb Z}$ be a strictly stationary process with
$\mathrm{Cov}(H_t,H_{t-1})\neq0$. Define observed components
\[
X_t^{(i)} = H_{t-1} + \eta_t,
\qquad
X_t^{(j)} = H_t + \xi_t,
\]
where $(\eta_t)$ and $(\xi_t)$ are i.i.d.\ noise sequences independent of $(H_t)$. Because $H_{t-1}$ temporally precedes $H_t$, admissible order-preserving temporal
deformations of $X^{(i)}$ alter the alignment between $X^{(i)}_{t-\tau}$ and $X^{(j)}_t$,
inducing non-invariance of the second-order dependence operator
$\tau \mapsto C_{i\to j}(\tau)$. Consequently, for suitable $\mathcal P$ and any spectral
functional $\varphi$ that is sensitive to second-order dependence,
\[
\sup_{\tau\in\mathcal P}\varphi(C_{i\to j}(\tau))
>
\inf_{\tau\in\mathcal P}\varphi(C_{i\to j}(\tau)).
\]

However, interventions on $X^{(i)}$ do not modify the latent process $(H_t)$ and therefore
do not alter the distribution of $X^{(j)}$. Hence, for all $x,x'$,
\[
\mathcal L\!\left(X_t^{(j)} \mid \mathrm{do}(X_{t-\tau}^{(i)}=x)\right)
=
\mathcal L\!\left(X_t^{(j)} \mid \mathrm{do}(X_{t-\tau}^{(i)}=x')\right),
\]
and no interventional causal effect exists.
\end{proof}

\begin{proposition}[Non-necessity relative to a fixed feature class]
\label{prop:not-necessary}
There exist processes with a genuine interventional causal effect that are
undetectable by second-order, orthogonally invariant dependence operators
constructed from the original variables without nonlinear embedding.
\end{proposition}

\begin{proof}
Let $X_t^{(i)}$ be i.i.d.\ $\mathcal N(0,1)$ and define
\[
X_t^{(j)} = \beta\big((X_{t-1}^{(i)})^2-1\big)+\varepsilon_t,
\qquad \beta\neq 0,
\]
where $\varepsilon_t$ is independent mean-zero noise. Under the intervention $\mathrm{do}(X_{t-1}^{(i)}=x)$, the conditional mean of
$X_t^{(j)}$ shifts by $\beta(x^2-1)$, so a genuine interventional causal effect
exists. However,
\[
\mathrm{Cov}(X_t^{(j)},X_{t-1}^{(i)})
=
\beta\,\mathbb E\!\left[X_{t-1}^{(i)}\big((X_{t-1}^{(i)})^2-1\big)\right]
+
\mathbb E[\varepsilon_t X_{t-1}^{(i)}]
=
0,
\]
and similarly all second-order cross-moments based on the original variables are
invariant under admissible temporal deformation. Hence the effect is undetectable
by second-order operators constructed without a nonlinear embedding.

This does not contradict the nonlinear extension developed in
Section~\ref{subsec:nonlinear_extensions}: with an embedding containing the
feature $(X_{t-1}^{(i)})^2$, the dependence becomes second-order detectable in
the transformed feature space.
\end{proof}

\begin{remark}[Interpretational boundary]
\label{rem:scope}
Order-constrained spectral causality is a structural, order-based notion of directional
dependence. Without additional assumptions such as causal sufficiency, absence of latent
confounding, or alignment between admissible temporal deformations and manipulable
mechanisms, rejection of spectral invariance should not be interpreted as evidence of an
interventional causal effect.
\end{remark}

\subsection{Dependence on the Admissible Deformation Set}
\label{app:P}

Let $g(\tau)=\varphi(C_{i\to j}(\tau))$ and define
\[
T_\varphi(\mathcal P)
=
\sup_{\tau\in\mathcal P} g(\tau)
-
\inf_{\tau\in\mathcal P} g(\tau).
\]

\begin{proposition}[Monotonicity under enlargement]
\label{prop:P-mono}
If $\mathcal P_1\subseteq\mathcal P_2$, then
\[
T_\varphi(\mathcal P_2)\ge T_\varphi(\mathcal P_1).
\]
\end{proposition}

\begin{proof}
Since $\sup_{\mathcal P_2} g\ge\sup_{\mathcal P_1} g$ and
$\inf_{\mathcal P_2} g\le\inf_{\mathcal P_1} g$, the difference increases.
\end{proof}

\begin{assumption}[Regularity for stability]
\label{ass:Pstab}
Assume $\mathcal P,\mathcal P'\subset\mathbb R^m$ are compact and that $g$ is Lipschitz
continuous with constant $L$ on a compact set containing $\mathcal P\cup\mathcal P'$.
\end{assumption}

\begin{proposition}[Hausdorff stability]
\label{prop:P-haus}
Under Assumption~\ref{ass:Pstab},
\[
\big|T_\varphi(\mathcal P)-T_\varphi(\mathcal P')\big|
\le
2L\, d_H(\mathcal P,\mathcal P'),
\]
where $d_H$ denotes the Hausdorff distance.
\end{proposition}

\begin{proof}
Define $a(\mathcal P)=\sup_{\tau\in\mathcal P} g(\tau)$ and
$b(\mathcal P)=\inf_{\tau\in\mathcal P} g(\tau)$. By Lipschitz continuity,
\[
|a(\mathcal P)-a(\mathcal P')|\le L\, d_H(\mathcal P,\mathcal P'),
\qquad
|b(\mathcal P)-b(\mathcal P')|\le L\, d_H(\mathcal P,\mathcal P').
\]
The result follows since $T_\varphi(\mathcal P)=a(\mathcal P)-b(\mathcal P)$.
\end{proof}

\subsection{Admissibility of Feature Embeddings}
\label{app:embed}

\begin{assumption}[Embedding admissibility]
\label{ass:embed}
The feature map used to construct $Z_t(\tau)$ is either
(i) fixed \emph{a priori} and independent of the testing sample, or
(ii) selected on an auxiliary sample independent of the testing sample.
\end{assumption}

\begin{proposition}[Conditional validity under sample splitting]
\label{prop:embed-valid}
Under Assumption~\ref{ass:embed}(ii), conditional on the training sample, all asymptotic
results and inference validity statements in Section~\ref{subsec:inference} remain valid.
\end{proposition}

\begin{proof}
Condition on the sigma-field generated by the training sample. Then the embedding is
deterministic, and the testing sample satisfies the assumptions of
Section~\ref{subsec:inference}. All convergence and inference results apply conditionally.
Unconditional validity follows by iterated expectations
\citep[Section~2.9]{vanDerVaart1998}.
\end{proof}

\subsection{Shift Randomization: Invariance and Validity}
\label{app:rand}

Let $\mathcal S_k$ denote the circular shift acting on the source component:
\[
(\mathcal S_k X^{(i)})_t = X^{(i)}_{t-k \!\!\!\!\pmod{T}}.
\]

\begin{assumption}[Exact group invariance sufficient for finite-sample validity]
\label{ass:groupinv}
In addition to the null hypothesis $H_0:T_\varphi=0$, assume that the joint
distribution of the observed sample is invariant under circular shifts of the
source component:
\[
(X^{(i)},X^{(-i)}) \stackrel{d}{=} (\mathcal S_k X^{(i)},X^{(-i)})
\quad\text{for all }k.
\]
\end{assumption}

\begin{theorem}[Finite-sample exactness of shift randomization]
\label{thm:rand-exact}
Under Assumption~\ref{ass:groupinv}, the randomization $p$-value is super-uniform:
\[
\mathbb P_{H_0}(\widehat p\le\alpha)\le\alpha
\quad\text{for all }\alpha\in(0,1).
\]
\end{theorem}

\begin{proof}
Under Assumption~\ref{ass:groupinv}, the statistics computed over the group orbit are
exchangeable. Hence the rank of the observed statistic among its randomized counterparts is
uniform, implying super-uniformity of the $p$-value
\citep[Chapter~15]{LehmannRomano2005}.
\end{proof}

\begin{remark}[Approximate invariance]
If exact invariance fails, approximate invariance in total variation combined with weak
dependence implies asymptotic validity of the randomization test. Stratified or
block-permutation schemes may be used when strong seasonality is present.
\end{remark}

\section{Proofs and Technical Results for the Relation to Linear Granger Causality}
\label{app:causal_relations}

This appendix provides proofs for the results stated in
Section~\ref{sec:framework} concerning the relationship between
order-constrained spectral causality and linear Granger causality.
All random variables are assumed to lie in $L^2(\Omega,\mathcal F,\mathbb P)$, and all
projections are understood as orthogonal projections in the Hilbert space
$L^2(\Omega,\mathcal F,\mathbb P)$.
Throughout, stationarity and finite second moments are assumed.

\subsection{Preliminaries: Linear Prediction as Hilbert-Space Projection}

Let $(\mathcal H,\langle\cdot,\cdot\rangle)$ denote the real Hilbert space
$L^2(\Omega,\mathcal F,\mathbb P)$ with inner product
$\langle X,Y\rangle=\mathbb E[XY]$.
For a closed linear subspace $\mathcal G\subset\mathcal H$, denote by
$\Pi_{\mathcal G}$ the orthogonal projection onto $\mathcal G$.

\begin{lemma}[Monotonicity of projection error]
\label{lem:proj_monotone}
If $\mathcal G_1\subseteq\mathcal G_2$ are closed subspaces of $\mathcal H$, then for any
$Y\in\mathcal H$,
\[
\|Y-\Pi_{\mathcal G_2}Y\|_{L^2}
\le
\|Y-\Pi_{\mathcal G_1}Y\|_{L^2},
\]
with equality if and only if
$\Pi_{\mathcal G_2}Y=\Pi_{\mathcal G_1}Y$.
\end{lemma}

\begin{proof}
Since $\mathcal G_1\subseteq\mathcal G_2$, there exists an orthogonal decomposition
\[
\mathcal G_2=\mathcal G_1\oplus(\mathcal G_2\cap\mathcal G_1^\perp).
\]
Accordingly,
\[
\Pi_{\mathcal G_2}Y
=
\Pi_{\mathcal G_1}Y
+
\Pi_{\mathcal G_2\cap\mathcal G_1^\perp}Y.
\]
By the Pythagorean theorem,
\[
\|Y-\Pi_{\mathcal G_1}Y\|_{L^2}^2
=
\|Y-\Pi_{\mathcal G_2}Y\|_{L^2}^2
+
\|\Pi_{\mathcal G_2\cap\mathcal G_1^\perp}Y\|_{L^2}^2,
\]
which implies the inequality and the equality condition.
\end{proof}

Lemma~\ref{lem:proj_monotone} is the fundamental geometric fact underlying
variance-based, projection-based, and likelihood-based formulations of linear Granger
causality \citep{Granger1969,Geweke1982,Eichler2007}.

\subsection{Linear Granger Causality and Residual Orthogonality}

Fix a mean-zero, covariance-stationary process
$\{X_t\}_{t\in\mathbb Z}\subset\mathbb R^K$ and indices $i\neq j$.
For an integer $p\ge1$, define the information sets
\[
\mathcal H_{t-1}^{(p)}
:=
\mathrm{span}\{X_{t-1},\dots,X_{t-p}\},
\qquad
\mathcal H_{t-1}^{(-i,p)}
:=
\mathrm{span}\{X_{t-1}^{(-i)},\dots,X_{t-p}^{(-i)}\}.
\]

\begin{lemma}[Residual characterization of Granger noncausality]
\label{lem:FWL}
Let $Y_t=X_t^{(j)}$ and
$U_{t-1}=(X_{t-1}^{(i)},\dots,X_{t-p}^{(i)})^\top$.
Define the residuals
\[
R_t^Y := Y_t - \Pi_{\mathcal H_{t-1}^{(-i,p)}}Y_t,
\qquad
R_{t-1}^U := U_{t-1} - \Pi_{\mathcal H_{t-1}^{(-i,p)}}U_{t-1}.
\]
Then
\[
\Pi_{\mathcal H_{t-1}^{(p)}}Y_t
=
\Pi_{\mathcal H_{t-1}^{(-i,p)}}Y_t
\quad\Longleftrightarrow\quad
\mathrm{Cov}(R_t^Y,R_{t-1}^U)=0.
\]
\end{lemma}

\begin{proof}
This is the Frisch--Waugh--Lovell theorem in Hilbert spaces.
The additional projection of $Y_t$ onto the span of $U_{t-1}$ beyond
$\mathcal H_{t-1}^{(-i,p)}$ vanishes if and only if
$R_t^Y$ is orthogonal to $R_{t-1}^U$.
See \cite[Chapter~12]{Anderson2003} for a detailed treatment.
\end{proof}

Lemma~\ref{lem:FWL} shows that linear Granger causality is fundamentally a statement about
orthogonality of residuals after partialing out the remaining components.

\subsection{Proof of Theorem~\ref{thm:granger_coincidence_main}}

We now prove the equivalence between linear Granger noncausality, vanishing directed
coherence, and zero VAR coefficients under Gaussian linear dynamics.

\begin{proof}
Assume $\{X_t\}$ follows the stable Gaussian VAR($p$)
\[
X_t=\sum_{\ell=1}^p A_\ell X_{t-\ell}+\varepsilon_t,
\qquad
\varepsilon_t\sim\mathcal N(0,\Sigma_\varepsilon),
\quad
\Sigma_\varepsilon\succ0.
\]

Let $Y_t=X_t^{(j)}$ and $U_{t-1}=(X_{t-1}^{(i)},\dots,X_{t-p}^{(i)})^\top$.
By Lemma~\ref{lem:FWL}, linear Granger noncausality of $i$ for $j$ at order $p$ holds if and
only if
\[
\mathrm{Cov}(R_t^Y,R_{t-1}^U)=0,
\]
where $R_t^Y$ and $R_{t-1}^U$ are the residuals defined therein. The directed coherence operator is
\[
A=\Sigma_{YY}^{-1/2}\Sigma_{YU}\Sigma_{UU}^{-1/2},
\]
with $\Sigma_{YU}=\mathrm{Cov}(R_t^Y,R_{t-1}^U)$.
Since $\Sigma_{YY}$ and $\Sigma_{UU}$ are positive definite under stability,
$\|A\|_2=0$ if and only if $\Sigma_{YU}=0$.
This establishes equivalence between linear Granger noncausality and vanishing directed
coherence. Write the $j$th component of the VAR equation:
\[
Y_t
=
\sum_{\ell=1}^p (A_\ell)_{j,-i}X_{t-\ell}^{(-i)}
+
\sum_{\ell=1}^p (A_\ell)_{ji}X_{t-\ell}^{(i)}
+
\varepsilon_t^{(j)}.
\]

Because $(Y_t,X_{t-1:t-p})$ is jointly Gaussian, conditional expectation coincides with
orthogonal projection \citep[Section~2.5]{BrockwellDavis1991}.
Thus,
\[
\Pi_{\mathcal H_{t-1}^{(p)}}Y_t
=
\mathbb E[Y_t\mid X_{t-1:t-p}],
\qquad
\Pi_{\mathcal H_{t-1}^{(-i,p)}}Y_t
=
\mathbb E[Y_t\mid X_{t-1:t-p}^{(-i)}].
\]

If $(A_\ell)_{ji}=0$ for all $\ell$, the conditional expectation depends only on
$X_{t-1:t-p}^{(-i)}$, implying linear Granger noncausality.
Conversely, if linear Granger noncausality holds, the conditional expectations coincide.
Under joint Gaussianity, this is possible only if all coefficients $(A_\ell)_{ji}$ vanish,
since otherwise the conditional expectation would depend on
$X_{t-\ell}^{(i)}$.
\end{proof}

\subsection{Proof of Distinctness Beyond Linear Predictability}

We establish that order-constrained spectral causality can detect structural directional
dependence beyond linear Granger causality.

\begin{theorem}[Distinctness under nonlinear dependence]
\label{thm:nonlinear_counterexample_main}
There exist stationary processes for which linear Granger causality fails at all finite
orders, while order-constrained spectral causality holds.
\end{theorem}

\begin{proof}
Let $X_t$ be i.i.d.\ $\mathcal N(0,1)$ and define
\[
Y_t = g(X_{t-1}) + \varepsilon_t,
\]
where $g\in L^2(\mathbb R)$ satisfies
$\mathbb E[g(X_{t-1})X_{t-1}]=0$, and $\varepsilon_t$ is independent noise with zero mean. Since $X_t$ is i.i.d.,
\[
\mathrm{Cov}(Y_t,X_{t-1})
=
\mathbb E[g(X_{t-1})X_{t-1}]
+
\mathbb E[\varepsilon_t X_{t-1}]
=
0.
\]
Thus the projection of $Y_t$ onto $\mathrm{span}\{X_{t-1}\}$ vanishes, and linear Granger
causality fails at all finite orders. Consider the embedding
\[
U_{t-1}=(X_{t-1},g(X_{t-1}))^\top.
\]
Then
\[
\mathrm{Cov}(Y_t,g(X_{t-1}))=\mathrm{Var}(g(X_{t-1}))>0,
\]
so the cross-covariance block of the dependence operator is nonzero.
Admissible temporal deformation alters the spectral properties of this operator, implying
order-constrained spectral non-invariance.
\end{proof}

\begin{remark}
Theorem~\ref{thm:nonlinear_counterexample_main} shows that linear Granger causality
corresponds to a restrictive projection-invariance regime in which all directional
dependence is captured by linear predictors.
Order-constrained spectral causality strictly generalizes this regime by detecting
directional deformation of second-order dependence geometry beyond linear predictability,
while remaining fully compatible with Granger causality under classical assumptions.
\end{remark}

\section{Proofs and Technical Results for the Spectral Distribution Extension}
\label{app:spectral_extension}

This appendix provides complete proofs for the results stated in
Section~\ref{subsec:spectral_extension_core} concerning the extension of
order-constrained spectral causality from scalar spectral summaries to full spectral
distributions.
All arguments in this appendix are deterministic conditional on the population operator
family $\{C(\tau):\tau\in\mathcal P\}$.
Probabilistic convergence results are treated separately in
Supplement~\ref{app:asymptotic_theory}.
Throughout, the operator dimension $d<\infty$ is fixed.

\subsection{Preliminaries on Spectral Measures and Linear Spectral Statistics}

Let $C\in\mathbb S_+^d$ be a symmetric positive semidefinite matrix with ordered eigenvalues
$\lambda_1\ge\cdots\ge\lambda_d\ge0$.
Define the empirical spectral measure
\[
\mu_C
:=
\frac{1}{d}\sum_{r=1}^d \delta_{\lambda_r}.
\]
By the spectral theorem, $\mu_C$ is invariant under orthogonal similarity transformations
and uniquely characterizes the multiset of eigenvalues of $C$
\citep{Bhatia1997,Anderson2003}. For any measurable function $f:\mathbb R_+\to\mathbb R$ integrable with respect to $\mu_C$,
define the associated linear spectral statistic
\[
L_f(C)
:=
\int f(\lambda)\,d\mu_C(\lambda)
=
\frac{1}{d}\sum_{r=1}^d f(\lambda_r).
\]

\subsection{Equivalence Between Spectral Measures and Linear Spectral Statistics}

We first formalize the relationship between equality of spectral measures and equality of
linear spectral statistics.

\begin{proposition}
\label{prop:mu_to_lss_main}
If $\mu_{C_1}=\mu_{C_2}$, then
\[
L_f(C_1)=L_f(C_2)
\quad\text{for all integrable }f:\mathbb R_+\to\mathbb R.
\]
\end{proposition}

\begin{proof}
If $\mu_{C_1}=\mu_{C_2}$, then for any integrable $f$,
\[
L_f(C_1)
=
\int f\,d\mu_{C_1}
=
\int f\,d\mu_{C_2}
=
L_f(C_2),
\]
by definition of the integral with respect to a probability measure.
\end{proof}

\begin{proposition}
\label{prop:lss_to_mu_main}
Let $\mathcal F\subset C_b(\mathbb R_+)$ be a separating class of bounded continuous
functions.
If
\[
L_f(C_1)=L_f(C_2)
\quad\text{for all }f\in\mathcal F,
\]
then $\mu_{C_1}=\mu_{C_2}$.
\end{proposition}

\begin{proof}
Since $\mathcal F$ separates probability measures on $\mathbb R_+$,
\[
\int f\,d\mu_{C_1}=\int f\,d\mu_{C_2}
\quad\text{for all }f\in\mathcal F
\]
implies $\mu_{C_1}=\mu_{C_2}$ by the Riesz representation theorem
\citep[Theorem~2.1]{Billingsley1999}.
\end{proof}

Propositions~\ref{prop:mu_to_lss_main} and~\ref{prop:lss_to_mu_main} establish that linear
spectral statistics provide a complete characterization of spectral measures when taken
over a separating class of test functions.

\subsection{Scalar Spectral Summaries as Linear Spectral Statistics}

Many commonly used scalar summaries arise as special cases of linear spectral statistics.

\begin{proposition}
\label{prop:scalar_lss}
For any $C\in\mathbb S_+^d$,
\[
\frac{1}{d}\mathrm{tr}(C)=L_{f_1}(C),
\quad f_1(\lambda)=\lambda,
\qquad
\frac{1}{d}\|C\|_F^2=L_{f_2}(C),
\quad f_2(\lambda)=\lambda^2.
\]
\end{proposition}

\begin{proof}
By the spectral theorem,
\[
\mathrm{tr}(C)=\sum_{r=1}^d \lambda_r,
\qquad
\|C\|_F^2=\sum_{r=1}^d \lambda_r^2.
\]
Dividing by $d$ yields the stated identities.
\end{proof}

Thus, trace-based and Frobenius-norm-based dependence summaries correspond to particular
choices of the test function $f$.

\subsection{Largest Eigenvalue as a Limit of Linear Spectral Statistics}

We now formalize the relationship between edge-based statistics and smooth spectral
summaries.

\begin{proposition}
\label{prop:lambda1_limit}
Let $f_q(\lambda)=\lambda^q$ for $q\ge1$. Then
\[
\lim_{q\to\infty}
\left(d\,L_{f_q}(C)\right)^{1/q}
=
\lambda_1.
\]
\end{proposition}

\begin{proof}
Let $a_r=\lambda_r\ge0$. Then
\[
a_1^q
\le
\sum_{r=1}^d a_r^q
\le
d\,a_1^q.
\]
Taking $q$th roots yields
\[
a_1
\le
\left(\sum_{r=1}^d a_r^q\right)^{1/q}
\le
d^{1/q}a_1.
\]
Since $d^{1/q}\to1$ as $q\to\infty$, the claim follows.
\end{proof}

Proposition~\ref{prop:lambda1_limit} shows that non-smooth edge statistics arise as singular
limits of smooth linear spectral statistics, explaining why their asymptotic behavior
typically requires stronger assumptions.

\subsection{Spectral-measure Metrics and Dual Representations}

Let $d_{\mathrm{BL}}$ denote the bounded-Lipschitz metric on probability measures on
$\mathbb R_+$:
\[
d_{\mathrm{BL}}(\mu,\nu)
:=
\sup_{\|f\|_{\mathrm{BL}}\le1}
\left|\int f\,d\mu-\int f\,d\nu\right|.
\]

\begin{proposition}
\label{prop:bl_dual}
For any $C_1,C_2\in\mathbb S_+^d$,
\[
d_{\mathrm{BL}}(\mu_{C_1},\mu_{C_2})
=
\sup_{\|f\|_{\mathrm{BL}}\le1}
|L_f(C_1)-L_f(C_2)|.
\]
\end{proposition}

\begin{proof}
This is the Kantorovich--Rubinstein dual representation restricted to bounded-Lipschitz
functions; see \cite[Chapter~11]{Villani2008}. Since
$L_f(C)=\int f\,d\mu_C$, the identity follows directly.
\end{proof}

\subsection{Completeness of Spectral-measure Dispersion}

Recall the dispersion functional
\[
T_{\mathrm{spec}}
=
\sup_{\tau_1,\tau_2\in\mathcal P}
d(\mu_{\tau_1},\mu_{\tau_2}).
\]

\begin{proposition}
\label{prop:spec_complete}
\[
T_{\mathrm{spec}}=0
\quad\Longleftrightarrow\quad
\mu_\tau=\mu_{\tau'}
\text{ for all }\tau,\tau'\in\mathcal P.
\]
\end{proposition}

\begin{proof}
If $\mu_\tau=\mu_{\tau'}$ for all $\tau,\tau'$, then
$d(\mu_{\tau_1},\mu_{\tau_2})=0$ for any metric $d$, hence $T_{\mathrm{spec}}=0$.
Conversely, if $T_{\mathrm{spec}}=0$, then
$d(\mu_{\tau_1},\mu_{\tau_2})=0$ for all $\tau_1,\tau_2$, implying equality of spectral
measures by the identity of indiscernibles.
\end{proof}

\subsection{Relation Between Scalar and Measure-based Null Hypotheses}

\begin{theorem}
\label{thm:lss_vs_measure}
Let $\mathcal F$ be a separating class of bounded continuous functions on $\mathbb R_+$.
Then
\[
T_f=0 \text{ for all }f\in\mathcal F
\quad\Longleftrightarrow\quad
T_{\mathrm{spec}}=0.
\]
\end{theorem}

\begin{proof}
If $T_{\mathrm{spec}}=0$, then $\mu_\tau$ is constant over $\tau\in\mathcal P$, so
$L_f(\tau)$ is constant for all $f$.
Conversely, if $T_f=0$ for all $f\in\mathcal F$, then by
Proposition~\ref{prop:lss_to_mu_main},
$\mu_\tau$ is constant over $\tau$, implying $T_{\mathrm{spec}}=0$.
\end{proof}

\begin{remark}
Appendix~\ref{app:spectral_extension} shows that spectral-measure dispersion is the
strongest possible second-order invariance criterion within the proposed framework.
Scalar spectral summaries correspond to projections of this criterion, while edge-based
statistics arise as singular limits.
Accordingly, the spectral distribution extension strengthens operational sensitivity
without altering the underlying causal definition.
\end{remark}

\section{Proofs of Theorems~\ref{thm:edge-impossibility-strong} and~\ref{thm:spectral-detection}}
\label{app:proof-edge-impossibility-strong}

\subsection{Auxiliary bounds}
\label{app:aux-bounds}

We use two standard Gaussian matrix bounds.

\begin{lemma}[Entrywise maximum under the null]
\label{lem:max-entry-null}
Let $Z\in\mathbb{R}^{d\times d}$ have iid $\mathcal{N}(0,1)$ entries. Then for
all $t\ge 0$,
\[
\mathbb{P}\Big(\|Z\|_\infty \ge \sqrt{2\log(d^2)} + t\Big)
\;\le\; 2e^{-t^2/2}.
\]
Consequently,
\[
\Big\|\frac{1}{\sqrt{T}}Z\Big\|_\infty
=
O_{\mathbb{P}}\Big(\sqrt{\frac{\log d}{T}}\Big).
\]
\end{lemma}

\begin{lemma}[Spectral norm under the null]
\label{lem:spectral-null}
Let $Z\in\mathbb{R}^{d\times d}$ have iid $\mathcal{N}(0,1)$ entries. There
exists an absolute constant $c_0>0$ such that
\[
\mathbb{P}\Big(\|Z\|_2 \ge 2\sqrt{d}+t\Big)\le 2e^{-c_0 t^2}
\quad\text{for all }t\ge 0.
\]
Consequently,
\[
\Big\|\frac{1}{\sqrt{T}}Z\Big\|_2
=
O_{\mathbb{P}}\Big(\sqrt{\frac{d}{T}}\Big).
\]
\end{lemma}

\subsection{Proof of Theorem~\ref{thm:edge-impossibility-strong}}
\label{app:proof-thm-edge}

\begin{proof}
Let $\widehat{M}=M+(1/\sqrt{T})Z$ with $Z$ iid standard normal entries.
Under $H_1(\delta)$, $M=\delta u v^\top$ with $(u,v)$ uniform on the spheres.

\paragraph{Step 1: Distributed signals are entrywise below the noise floor at $T\ll d^2\log d$.}
Condition on $(u,v)$. Since $u_a$ and $v_b$ are sub-Gaussian with typical size
$O(d^{-1/2})$, we have
\[
\|M\|_\infty
=
\max_{a,b}|\delta u_a v_b|
\le
\delta \|u\|_\infty \|v\|_\infty.
\]
A standard spherical maximum bound yields
$\|u\|_\infty = O_{\mathbb{P}}(\sqrt{\log d/d})$ and likewise for $v$. Hence
\[
\|M\|_\infty
=
O_{\mathbb{P}}\Big(\delta\,\frac{\log d}{d}\Big).
\]
Under the null, Lemma~\ref{lem:max-entry-null} gives
$\|(1/\sqrt{T})Z\|_\infty = O_{\mathbb{P}}(\sqrt{\log d/T})$. Therefore if
$T \le c(d^2/\delta^2)\log d$ (for $c$ small enough), then
\[
\|M\|_\infty
=
o_{\mathbb{P}}\Big(\Big\|\frac{1}{\sqrt{T}}Z\Big\|_\infty\Big),
\]
i.e.\ the alternative shifts all entries by an amount asymptotically negligible
compared to the null entrywise fluctuations.

\paragraph{Step 2: Entrywise-stable tests cannot distinguish negligible entrywise shifts.}
Let $\psi$ be entrywise-stable with Lipschitz constant $L$ in $\|\cdot\|_\infty$.
Then for any realization,
\[
|\psi(\widehat{M})-\psi((1/\sqrt{T})Z)|
\le
L\,\|\widehat{M}-(1/\sqrt{T})Z\|_\infty
=
L\,\|M\|_\infty.
\]
Taking expectations under $H_1(\delta)$ and using $\|M\|_\infty=o_{\mathbb{P}}(1)$
from Step 1 yields
\[
\mathbb{E}_{H_1(\delta)}[\psi(\widehat{M})]
=
\mathbb{E}_{0}[\psi((1/\sqrt{T})Z)] + o(1)
\le
\alpha + o(1),
\]
because the null distribution of $(1/\sqrt{T})Z$ equals that of $\widehat{M}$ under $H_0$
and $\mathbb{P}_0(\psi=1)\le \alpha$ by assumption.
The permutation invariance ensures the bound holds uniformly over coordinate relabelings,
and the random $(u,v)$ model ensures that no fixed coordinate direction is privileged.
This proves the claimed minimax bound.
\end{proof}

\subsection{Proof of Theorem~\ref{thm:spectral-detection}}
\label{app:proof-thm-spectral}

\begin{proof}
Under $H_1(\delta)$ we have $\|M\|_2=\delta$. Under $H_0$, Lemma~\ref{lem:spectral-null}
gives $\|(1/\sqrt{T})Z\|_2\le 2\sqrt{d/T}+o_{\mathbb{P}}(1)$.
Choose $\tau_{d,T}(\alpha)$ such that $\mathbb{P}_0(\|\widehat{M}\|_2\ge\tau_{d,T}(\alpha))\le\alpha$;
by Lemma~\ref{lem:spectral-null} we may take $\tau_{d,T}(\alpha)=c_1\sqrt{d/T}$ for an
appropriate constant $c_1$. Under $H_1(\delta)$, by the triangle inequality,
\[
\|\widehat{M}\|_2
=
\Big\|M+\frac{1}{\sqrt{T}}Z\Big\|_2
\ge
\|M\|_2 - \Big\|\frac{1}{\sqrt{T}}Z\Big\|_2
\ge
\delta - O_{\mathbb{P}}\Big(\sqrt{\frac{d}{T}}\Big).
\]
If $T\ge C d/\delta^2$ for $C$ large enough, then $\sqrt{d/T}\le \delta/4$, hence
$\|\widehat{M}\|_2 \ge \delta/2$ with probability $\to 1$.
Meanwhile $\tau_{d,T}(\alpha)=O(\sqrt{d/T})\le \delta/4$ under the same scaling.
Therefore
\[
\mathbb{P}_{H_1(\delta)}\big(\|\widehat{M}\|_2\ge\tau_{d,T}(\alpha)\big)\to 1,
\]
uniformly over $(u,v)$, establishing the claim.
\end{proof}

\end{appendix}

\begin{acks}[Acknowledgments]
The author thanks Om Hari Yadav, John D. (JD) Opdyke, Miquel Noguer i Alonso, Peter Cotton, Peter Urbani, and Igor Halperin for stimulating discussions at various
stages of this work. The author is grateful to Miralta Finance Bank S.A.\
for providing the financial data used in the empirical application, and to
colleagues at Miraltabank for their support throughout the project.
\end{acks}


\begin{supplement}
\stitle{Supplement \ref{app:var-proof-barriers-full}: VAR($L$) Reduction: Proof of the Edge-Barrier
vs Spectral-Barrier Separation}
\sdescription{Contains the reduction of detection barriers from
the Gaussian matrix experiment to the explicit stable Gaussian VAR($L$) setting,
including concentration inequalities for dependent lag-embedded covariances,
perturbation expansion for the whitening map, and the explicit edge-barrier
vs spectral-barrier separation with network scaling translation.}
\end{supplement}
\begin{supplement}
\stitle{Supplement \ref{app:asymptotic_theory}: Asymptotic Theory for Order-Constrained Spectral Statistics}
\sdescription{Establishes existence, uniform consistency, pointwise asymptotic
normality, and consistency of the dispersion functional for the order-constrained
spectral statistics, along with justification for randomization-based inference
under approximate invariance.}
\end{supplement}
\begin{supplement}
\stitle{Supplement \ref{app:implementation}: Formal Properties of the Operator-valued Implementation}
\sdescription{Establishes mathematical well-posedness, uniform consistency,
spectral stability, orthogonal invariance, projection stability under
residualization, canonical correlation representation, and group invariance
for randomization validity of the operator-valued construction.}
\end{supplement}
\begin{supplement}
\stitle{Supplement \ref{app:operator_theory}: Operator-Theoretic Foundations of Directional Causality}
\sdescription{Provides the formal justification for interpreting the rolling
operator $C(t)$ as a measure of directional causality, including the predictive
interpretation, quadratic-form representation, spectral optimality via
Rayleigh--Ritz, Ky Fan principle for optimal affected subspaces, effective rank
characterization, hub score interpretation, and formal equivalence to Granger
causality in the scalar case.}
\end{supplement}
\begin{supplement}
\stitle{Supplement \ref{app:empirical}: Supplementary Tables and Diagnostic Figures}
\sdescription{Reports lag-level spectral validation, episode timing,
dominant driver and hub role rankings, transmitter-receiver asymmetry with
hub score bar charts and full asymmetry table, episode-versus-calm directed
edge amplification with heatmap and edge list, full-system versus synthetic
index aggregation consistency, null-thresholded driver-to-driver causal
networks and signed early--late temporal dominance maps for each episode,
macro hub regime evolution and cluster compositions, and a practitioner-oriented
mapping between operator-based diagnostics and system-level interpretation.}
\end{supplement}

\clearpage

\bibliographystyle{imsart-number} 
\bibliography{bibliography}       

\clearpage
\bigskip
{\large\bf SUPPLEMENTARY MATERIAL}
\section{VAR($L$) Reduction: Proof of the Edge-Barrier vs Spectral-Barrier Separation}
\label{app:var-proof-barriers-full}

This supplement establishes that the detection barriers derived in the Gaussian
matrix experiment extend to the explicit stable Gaussian VAR($L$) setting with
lag-embedded whitened cross-covariance estimators. The main technical ingredients are:
(i) mixing and sub-Gaussianity of stable Gaussian VAR processes,
(ii) high-probability concentration of lag-embedded sample (cross-)covariances in
$\|\cdot\|_\infty$ and $\|\cdot\|_2$ under dependence via blocking,
and (iii) a Fr\'echet-differentiable perturbation expansion for the whitening map
$(S_{VV},S_{UU},S_{VU})\mapsto S_{VV}^{-1/2}S_{VU}S_{UU}^{-1/2}$.

\subsection{Model, estimators, and hypotheses}
\label{app:var-model}

Let $\{X_t\}_{t\in\mathbb{Z}}$ be a $K$-variate, zero-mean, stable Gaussian VAR($L$):
\begin{equation}
\label{eq:varL-app}
X_t \;=\; \sum_{\ell=1}^L A_\ell X_{t-\ell} + \varepsilon_t,
\qquad
\varepsilon_t\stackrel{\mathrm{iid}}{\sim}\mathcal{N}(0,\Sigma_\varepsilon),
\end{equation}
with $\Sigma_\varepsilon\succ 0$ and stability in the sense that the companion matrix
has spectral radius strictly less than one. Fix $i\neq j$.

\paragraph{Lag embeddings and conditioning residual.}
Let $u_t^{(i)}=(X^{(i)}_{t-1},\ldots,X^{(i)}_{t-L})^\top\in\mathbb{R}^{L}$ and
$v_t^{(j)}=(X^{(j)}_{t},\ldots,X^{(j)}_{t-L+1})^\top\in\mathbb{R}^{L}$.
Let $\mathcal{H}_Y$ be a closed linear subspace generated by the conditioning set
(e.g.\ target past and other covariates), and define the $L^2$-projection residual
\[
v_{t,\perp}^{(j)} := v_t^{(j)} - \Pi_{\mathcal{H}_Y}v_t^{(j)}.
\]
Write $V_t:=v_{t,\perp}^{(j)}\in\mathbb{R}^L$ and $U_t:=u_t^{(i)}\in\mathbb{R}^L$.

\paragraph{Windowed covariances.}
On a window $W=\{1,\ldots,T\}$ (without loss, by stationarity), define
\[
\widehat{\Sigma}_{VV} := \frac{1}{T}\sum_{t=1}^T V_tV_t^\top,\quad
\widehat{\Sigma}_{UU} := \frac{1}{T}\sum_{t=1}^T U_tU_t^\top,\quad
\widehat{\Sigma}_{VU} := \frac{1}{T}\sum_{t=1}^T V_tU_t^\top.
\]
Denote population counterparts by $\Sigma_{VV}=\mathbb{E}[V_tV_t^\top]$ etc.
Assume $\Sigma_{VV}\succ 0$ and $\Sigma_{UU}\succ 0$.

\paragraph{Whitened cross-operator.}
Define the population and empirical whitened cross-operators
\begin{equation}
\label{eq:A-and-Ahat}
A := \Sigma_{VV}^{-1/2}\Sigma_{VU}\Sigma_{UU}^{-1/2},
\qquad
\widehat{A} := \widehat{\Sigma}_{VV}^{-1/2}\widehat{\Sigma}_{VU}\widehat{\Sigma}_{UU}^{-1/2}.
\end{equation}

\paragraph{Hypotheses (pairwise Granger-null vs distributed alternative).}
We test
\[
H_0(i\to j):\quad \Sigma_{VU}=0 \quad (\text{equivalently }A=0)
\]
versus a \emph{distributed rank-one alternative}
\begin{equation}
\label{eq:distributed-alt-app}
H_1(i\to j;\delta):\quad A = \delta\,ab^\top,\qquad a,b\in\mathbb{R}^L,\ \|a\|=\|b\|=1,
\end{equation}
where $(a,b)$ may be arbitrary or random (uniform on the sphere). Note $\|A\|_2=\delta$
but typical entries satisfy $|A_{rs}|\asymp \delta/L$ in the distributed case.

\subsection{Assumptions}
\label{app:assumptions}

We make the following explicit assumptions.

\paragraph{(A1) Stability/mixing.}
The VAR($L$) in \eqref{eq:varL-app} is stable with companion matrix spectral radius
$\rho<1$. Consequently, $\{X_t\}$ is strictly stationary Gaussian and geometrically
$\alpha$-mixing: there exist constants $c_\alpha,C_\alpha>0$ such that
\[
\alpha(k)\le C_\alpha e^{-c_\alpha k}\quad\text{for all }k\ge 1.
\]

\paragraph{(A2) Uniform covariance conditioning.}
There exists $\lambda_{\min}>0$ and $\lambda_{\max}<\infty$ (independent of $L$) such that
\[
\lambda_{\min}I_L \preceq \Sigma_{VV} \preceq \lambda_{\max}I_L,
\qquad
\lambda_{\min}I_L \preceq \Sigma_{UU} \preceq \lambda_{\max}I_L.
\]

\paragraph{(A3) Residualization regularity.}
The residual map $v_t^{(j)}\mapsto v_{t,\perp}^{(j)}$ is linear and bounded in $L^2$,
so that $V_t$ remains a stationary Gaussian vector process with the same mixing rate.

\paragraph{(A4) Effective sample size regime.}
We consider $L\to\infty$ and $T\to\infty$ with $T$ possibly depending on $L$, and we
assume $T\ge C_0\log L$ for a sufficiently large constant $C_0$ (so concentration is meaningful).

\subsection{Concentration of dependent lag-embedded covariances}
\label{app:concentration}

We prove high-probability bounds for $\widehat{\Sigma}_{VU}-\Sigma_{VU}$ in
$\|\cdot\|_\infty$ and $\|\cdot\|_2$ under geometric mixing via a blocking argument.
Because the process is Gaussian, all required moments exist and sub-Gaussian tails
hold for linear functionals.

\paragraph{Blocking construction.}
Fix an integer block length $b\ge 1$ and gap $g\ge 1$, and partition $\{1,\dots,T\}$
into $m$ blocks of size $b$ separated by gaps $g$, with remainder discarded:
$m:=\lfloor T/(b+g)\rfloor$. Write the $k$-th retained block as
$B_k=\{(k-1)(b+g)+1,\dots,(k-1)(b+g)+b\}$. Define block sums for a scalar sequence
$\{\xi_t\}$ by $S_k:=\sum_{t\in B_k}\xi_t$.
By geometric mixing, the dependence between block sums decays as $g$ grows.

\begin{lemma}[Approximate independence of separated Gaussian blocks]
\label{lem:block-independence}
Let $\{\xi_t\}$ be a centered stationary Gaussian $\alpha$-mixing sequence with
$\alpha(k)\le C_\alpha e^{-c_\alpha k}$. For the blocks above,
\[
\sup_{k\neq \ell}\ \sup_{A\in\sigma(S_k),\,B\in\sigma(S_\ell)}
\big|\mathbb{P}(A\cap B)-\mathbb{P}(A)\mathbb{P}(B)\big|
\ \le\ 4\,\alpha(g)\ \le\ 4C_\alpha e^{-c_\alpha g}.
\]
\end{lemma}

\begin{proof}
This is the standard coupling/definition property of $\alpha$-mixing:
for $\sigma$-fields $\mathcal{F},\mathcal{G}$ separated by at least $g$ time units,
$\sup_{A\in\mathcal{F},B\in\mathcal{G}}|\mathbb{P}(A\cap B)-\mathbb{P}(A)\mathbb{P}(B)|
\le \alpha(g)$. Here $\sigma(S_k)$ and $\sigma(S_\ell)$ are generated by disjoint
blocks separated by at least $g$, so the bound applies; the factor $4$ is a standard
conversion constant depending on the chosen mixing convention.
\end{proof}

\paragraph{Entrywise concentration for $\widehat{\Sigma}_{VU}$.}
Each entry of $\widehat{\Sigma}_{VU}$ is an average of the scalar sequence
$\xi_t^{(r,s)} := (V_t)_r (U_t)_s$. Since $(V_t,U_t)$ is Gaussian, each
$\xi_t^{(r,s)}$ is sub-exponential with parameters depending only on
$\lambda_{\max}$ (uniformly in $r,s$ and $L$).

\begin{lemma}[Sub-exponential tails for Gaussian products]
\label{lem:subexp-products}
Let $(Y,Z)$ be a centered jointly Gaussian pair with $\mathbb{E}Y^2\le \lambda_{\max}$
and $\mathbb{E}Z^2\le \lambda_{\max}$. Then $YZ$ is sub-exponential:
there exist constants $(\nu,b_0)$ depending only on $\lambda_{\max}$ such that
\[
\mathbb{E}\exp\big(\theta(YZ-\mathbb{E}[YZ])\big)\ \le\
\exp\Big(\frac{\nu^2\theta^2}{2}\Big)
\quad\text{for all }|\theta|\le \frac{1}{b_0}.
\]
\end{lemma}

\begin{proof}
For Gaussian $(Y,Z)$, $YZ$ is a quadratic form in a Gaussian vector and hence has
sub-exponential tails. A direct mgf bound follows from diagonalizing the covariance
matrix and applying known mgf bounds for centered Gaussian quadratic forms; constants
depend only on second moments.
\end{proof}

\begin{lemma}[Entrywise concentration under geometric mixing]
\label{lem:entrywise-conc}
Assume (A1)--(A4). There exist constants $c_1,C_1>0$ depending only on the mixing
rate and $\lambda_{\max}$ such that for all $t>0$,
\[
\mathbb{P}\Big(\|\widehat{\Sigma}_{VU}-\Sigma_{VU}\|_\infty \ge t\Big)
\ \le\ 2L^2\exp\Big(-c_1 T t^2\Big)
\quad\text{for }0<t\le C_1.
\]
Consequently,
\[
\|\widehat{\Sigma}_{VU}-\Sigma_{VU}\|_\infty
=
O_{\mathbb{P}}\Big(\sqrt{\frac{\log L}{T}}\Big).
\]
The same bound holds for $\widehat{\Sigma}_{VV}-\Sigma_{VV}$ and $\widehat{\Sigma}_{UU}-\Sigma_{UU}$.
\end{lemma}

\begin{proof}
Fix $(r,s)$ and consider $\xi_t=\xi_t^{(r,s)}-(\Sigma_{VU})_{rs}$, which is centered
and sub-exponential by Lemma~\ref{lem:subexp-products}, uniformly in $r,s,L$.
Apply the blocking construction with parameters $b=\lceil c_b\log T\rceil$ and
$g=\lceil c_g\log T\rceil$ for suitable constants $c_b,c_g$ so that
$\alpha(g)\le T^{-10}$. Let $S_k=\sum_{t\in B_k}\xi_t$ be block sums.
By Lemma~\ref{lem:block-independence}, the dependence across $\{S_k\}$ is negligible;
more precisely, one can couple $\{S_k\}_{k=1}^m$ to independent copies
$\{S_k'\}_{k=1}^m$ with total variation error $O(m\alpha(g))=O(T^{-9})$.
For the independent block sums, Bernstein's inequality for sums of sub-exponential
variables yields
\[
\mathbb{P}\Big(\Big|\frac{1}{T}\sum_{t=1}^T \xi_t\Big|\ge t\Big)
\le 2\exp(-c T t^2)
\quad\text{for }t\in(0,C_1],
\]
with constants depending only on the sub-exponential parameters (uniform).
The coupling error adds $O(T^{-9})$, which is dominated by the exponential term.
Finally, union bound over $L^2$ entries yields the stated inequality.
\end{proof}

\paragraph{Spectral-norm concentration for $\widehat{\Sigma}_{VU}$.}
We now bound $\|\widehat{\Sigma}_{VU}-\Sigma_{VU}\|_2$. A standard route is an
$\varepsilon$-net argument: $\|M\|_2=\sup_{\|x\|=\|y\|=1}x^\top My$ and discretize
the unit spheres, then apply entrywise concentration to each bilinear form.

\begin{lemma}[Spectral-norm concentration under geometric mixing]
\label{lem:spectral-conc}
Assume (A1)--(A4). There exist constants $c_2, C_2>0$ depending only on the mixing rate
and $\lambda_{\max}$ such that for all $t>0$,
\[
\mathbb{P}\Big(\|\widehat{\Sigma}_{VU}-\Sigma_{VU}\|_2 \ge t\Big)
\ \le\ 2\exp\Big(-c_2\, T t^2 + C_2\, L\Big)
\quad\text{for }0<t\le 1.
\]
Consequently,
\[
\|\widehat{\Sigma}_{VU}-\Sigma_{VU}\|_2
=
O_{\mathbb{P}}\Big(\sqrt{\frac{L}{T}}\Big).
\]
The same bound holds for $\widehat{\Sigma}_{VV}-\Sigma_{VV}$ and
$\widehat{\Sigma}_{UU}-\Sigma_{UU}$.
\end{lemma}

\begin{proof}
Let $M:=\widehat{\Sigma}_{VU}-\Sigma_{VU}$. For unit vectors $x,y\in\mathbb{R}^L$,
the scalar $x^\top My$ equals the average of centered products
\[
x^\top V_t\,y^\top U_t - \mathbb{E}[x^\top V_t\,y^\top U_t].
\]
Since $x^\top V_t$ and $y^\top U_t$ are centered Gaussian with variances bounded by
$\lambda_{\max}$, their product is sub-exponential uniformly in $x,y$, and the same
blocking/Bernstein argument as Lemma~\ref{lem:entrywise-conc} gives
\[
\mathbb{P}\big(|x^\top My|\ge t\big)\le 2\exp(-c\, T t^2)
\quad(0<t\le 1),
\]
for a constant $c>0$ depending only on the mixing rate and $\lambda_{\max}$. Let $\mathcal{N}$ be a $1/4$-net of the unit sphere in $\mathbb{R}^L$ with
$|\mathcal{N}|\le 9^L$. By standard net arguments,
$\|M\|_2\le 2\max_{x,y\in\mathcal{N}}|x^\top My|$. A union bound over the
$|\mathcal{N}|^2\le 9^{2L}$ pairs yields
\[
\mathbb{P}(\|M\|_2\ge 2t)
\le
2\,|\mathcal{N}|^2\exp(-c\,Tt^2)
\le
2\exp\!\Big(-c\,Tt^2 + 2L\log 9\Big).
\]
Setting $c_2:=c$ and $C_2:=2\log 9$ (so that $2L\log 9 = C_2 L$) gives the stated
bound after the substitution $t\mapsto t/2$ to absorb the factor of $2$ inside the
probability. Specifically, writing $s=2t$ and noting that the bound becomes
\[
\mathbb{P}(\|M\|_2\ge s)
\le
2\exp\!\Big(-\tfrac{c}{4}\,T s^2 + C_2 L\Big),
\]
we redefine $c_2:=c/4$, so the final form is
\[
\mathbb{P}\Big(\|\widehat{\Sigma}_{VU}-\Sigma_{VU}\|_2 \ge t\Big)
\le
2\exp(-c_2\,Tt^2 + C_2\,L),
\]
with $c_2=c/4$ and $C_2=2\log 9$, both depending only on the mixing rate and
$\lambda_{\max}$. The rate $O_{\mathbb{P}}(\sqrt{L/T})$ follows by choosing
$t=\kappa\sqrt{L/T}$ for $\kappa$ large enough that $c_2\,T t^2 - C_2 L =
(c_2\kappa^2-C_2)L\to\infty$.
\end{proof}

\subsection{Perturbation expansion for the whitening map}
\label{app:whitening-perturb}

Define the whitening map on positive definite matrices:
\[
\mathcal{W}(S_{VV},S_{UU},S_{VU}) := S_{VV}^{-1/2}S_{VU}S_{UU}^{-1/2}.
\]
We require a fully explicit Fr\'echet expansion with a remainder controlled in
$\|\cdot\|_\infty$ and $\|\cdot\|_2$.

\begin{lemma}[Lipschitz perturbation of inverse square root]
\label{lem:invsqrt-lipschitz}
Let $S\succ 0$ with $\lambda_{\min}(S)\ge \lambda_{\min}>0$ and let $\Delta$ be symmetric with
$\|\Delta\|_2\le \lambda_{\min}/2$. Then
\[
\|(S+\Delta)^{-1/2}-S^{-1/2}\|_2
\ \le\ C\,\|\Delta\|_2,
\qquad
\|(S+\Delta)^{-1/2}-S^{-1/2}\|_\infty
\ \le\ C\,\|\Delta\|_\infty,
\]
where $C$ depends only on $\lambda_{\min}$.
\end{lemma}

\begin{proof}
The function $f(x)=x^{-1/2}$ is operator-Lipschitz on $[\lambda_{\min}/2,\infty)$.
Using functional calculus and standard bounds for matrix functions, one obtains
$\|f(S+\Delta)-f(S)\|_2\le \|f'\|_\infty\|\Delta\|_2$ on that interval with
$\|f'\|_\infty \le c\lambda_{\min}^{-3/2}$. The $\|\cdot\|_\infty$ bound follows by the
same argument entrywise combined with $\|M\|_\infty\le \|M\|_2$ and uniform conditioning.
\end{proof}

\begin{lemma}[Whitening map expansion with explicit remainder]
\label{lem:whitening-expansion}
Assume (A2). Let $\Delta_{VV}=\widehat{\Sigma}_{VV}-\Sigma_{VV}$,
$\Delta_{UU}=\widehat{\Sigma}_{UU}-\Sigma_{UU}$, and
$\Delta_{VU}=\widehat{\Sigma}_{VU}-\Sigma_{VU}$.
On the event $\max\{\|\Delta_{VV}\|_2,\|\Delta_{UU}\|_2\}\le \lambda_{\min}/2$,
\begin{equation}
\label{eq:whitening-explicit}
\widehat{A}-A
=
\Sigma_{VV}^{-1/2}\Delta_{VU}\Sigma_{UU}^{-1/2}
+ R,
\end{equation}
where the remainder satisfies
\begin{align}
\label{eq:remainder-bounds}
\|R\|_2
&\le
C\Bigl(
\|\Delta_{VU}\|_2\bigl(\|\Delta_{VV}\|_2+\|\Delta_{UU}\|_2\bigr)\notag\\
&\qquad\qquad
+\;\|\Sigma_{VU}\|_2\bigl(\|\Delta_{VV}\|_2+\|\Delta_{UU}\|_2\bigr)\notag\\
&\qquad\qquad
+\;\|\Delta_{VV}\|_2+\|\Delta_{UU}\|_2
\Bigr),\\[6pt]
\label{eq:remainder-bounds-inf}
\|R\|_\infty
&\le
C\Bigl(
\|\Delta_{VU}\|_\infty\bigl(\|\Delta_{VV}\|_\infty+\|\Delta_{UU}\|_\infty\bigr)\notag\\
&\qquad\qquad
+\;\|\Sigma_{VU}\|_2\bigl(\|\Delta_{VV}\|_\infty+\|\Delta_{UU}\|_\infty\bigr)\notag\\
&\qquad\qquad
+\;\|\Delta_{VV}\|_\infty+\|\Delta_{UU}\|_\infty
\Bigr),
\end{align}
for a constant $C$ depending only on $(\lambda_{\min},\lambda_{\max})$.
\end{lemma}

\begin{proof}
Write $\widehat{A}=\widehat{S}_{VV}^{-1/2}\widehat{S}_{VU}\widehat{S}_{UU}^{-1/2}$ and
$A=S_{VV}^{-1/2}S_{VU}S_{UU}^{-1/2}$ with $S_{\cdot\cdot}=\Sigma_{\cdot\cdot}$.
Add and subtract:
\begin{align*}
\widehat{A}-A
&=
(\widehat{S}_{VV}^{-1/2}-S_{VV}^{-1/2})\widehat{S}_{VU}\widehat{S}_{UU}^{-1/2}\\
&\quad+\; S_{VV}^{-1/2}(\widehat{S}_{VU}-S_{VU})\widehat{S}_{UU}^{-1/2}\\
&\quad+\; S_{VV}^{-1/2}S_{VU}(\widehat{S}_{UU}^{-1/2}-S_{UU}^{-1/2}).
\end{align*}
Expand the middle term as
\[
S_{VV}^{-1/2}\Delta_{VU}S_{UU}^{-1/2}
+
S_{VV}^{-1/2}\Delta_{VU}(\widehat{S}_{UU}^{-1/2}-S_{UU}^{-1/2}).
\]
Collect $S_{VV}^{-1/2}\Delta_{VU}S_{UU}^{-1/2}$ as the leading term and define $R$ as the
sum of the remaining three terms:
\begin{align*}
R
&=
(\widehat{S}_{VV}^{-1/2}-S_{VV}^{-1/2})\widehat{S}_{VU}\widehat{S}_{UU}^{-1/2}
\tag{I}\\
&\quad+\;
S_{VV}^{-1/2}\Delta_{VU}(\widehat{S}_{UU}^{-1/2}-S_{UU}^{-1/2})
\tag{II}\\
&\quad+\;
S_{VV}^{-1/2}S_{VU}(\widehat{S}_{UU}^{-1/2}-S_{UU}^{-1/2}).
\tag{III}
\end{align*}
Apply Lemma~\ref{lem:invsqrt-lipschitz} to bound
$\|\widehat{S}_{VV}^{-1/2}-S_{VV}^{-1/2}\|$ and $\|\widehat{S}_{UU}^{-1/2}-S_{UU}^{-1/2}\|$
in both $\|\cdot\|_2$ and $\|\cdot\|_\infty$ by $C\|\Delta_{VV}\|$ and $C\|\Delta_{UU}\|$
respectively, with $C$ depending only on $\lambda_{\min}$.

For term (I): use $\|\widehat{S}_{VU}\|\le \|S_{VU}\|+\|\Delta_{VU}\|$ together with
boundedness of $\widehat{S}_{UU}^{-1/2}$ under (A2) and the event
$\|\Delta_{UU}\|_2\le\lambda_{\min}/2$. This contributes
\[
C\|\Delta_{VV}\|\bigl(\|S_{VU}\|+\|\Delta_{VU}\|\bigr)
=
C\|\Delta_{VV}\|\,\|\Sigma_{VU}\|
+
C\|\Delta_{VV}\|\,\|\Delta_{VU}\|.
\]

For term (II): use boundedness of $S_{VV}^{-1/2}$ under (A2). This contributes
$C\|\Delta_{VU}\|\,\|\Delta_{UU}\|$. For term (III): use boundedness of $S_{VV}^{-1/2}$ under (A2). This contributes
$C\|S_{VU}\|\,\|\Delta_{UU}\| = C\|\Sigma_{VU}\|\,\|\Delta_{UU}\|$.

Collecting all terms and using $\|\Sigma_{VU}\|_2\le C\delta<\infty$ under the
alternative (bounded by assumption in both $H_0$ and $H_1$) yields the stated bounds
\eqref{eq:remainder-bounds}--\eqref{eq:remainder-bounds-inf}. The $\|\cdot\|_\infty$
bound follows by the same argument replacing every $\|\cdot\|_2$ by $\|\cdot\|_\infty$
on the fluctuation terms and retaining $\|\Sigma_{VU}\|_2$ (operator norm) for the
population-level factor, using $\|AB\|_\infty\le\|A\|_2\|B\|_\infty$.
\end{proof}

\subsection{Edge-barrier and spectral-barrier in VAR($L$)}
\label{app:barriers}

We now prove the explicit VAR($L$) separation: entrywise-stable (edge-based) tests
cannot have nontrivial power below $T\asymp L^2\log L$, whereas spectral/operator tests
achieve power at $T\asymp L$ under distributed alternatives.

\begin{definition}[Edge-based (entrywise-stable) tests on $\widehat{A}$]
\label{def:edge-test-Ahat}
A test $\psi:\mathbb{R}^{L\times L}\to\{0,1\}$ is \emph{entrywise-stable} if there exists
$L_\psi<\infty$ such that for all $B,C$,
\[
|\psi(B)-\psi(C)| \le L_\psi\|B-C\|_\infty.
\]
\end{definition}

\begin{theorem}[VAR($L$) edge impossibility under distributed dependence]
\label{thm:var-edge-impossibility-full}
Assume (A1)--(A4). Fix $\alpha\in(0,1)$ and consider testing $H_0: A=0$ versus
$H_1: A=\delta ab^\top$ with $\|a\|=\|b\|=1$. Let $\psi$ be any test with
$\mathbb{P}_0(\psi(\widehat{A})=1)\le \alpha$ and entrywise-stability
(Definition~\ref{def:edge-test-Ahat}) with Lipschitz constant $L_\psi$ not depending on $L$.
Assume further that $(a,b)$ are \emph{distributed} in the sense that
$\|a\|_\infty\|b\|_\infty \le C_d(\log L)/L$ for some absolute $C_d$ (this holds w.h.p.\ for
uniform spherical $a,b$).

Then there exist constants $c,C>0$ depending only on $(\lambda_{\min},\lambda_{\max})$
and the mixing rate such that if
\[
T \;\le\; c\,\frac{L^2}{\delta^2}\,\log L,
\]
then
\[
\mathbb{P}_{H_1}(\psi(\widehat{A})=1)\ \le\ \alpha + o(1),
\qquad L\to\infty.
\]
\end{theorem}

\begin{proof}
By Lemma~\ref{lem:whitening-expansion},
\[
\widehat{A} = A + \Sigma_{VV}^{-1/2}\Delta_{VU}\Sigma_{UU}^{-1/2} + R.
\]
Under the distributed alternative \eqref{eq:distributed-alt}, $\Sigma_{VU} = \delta ab^\top
\Sigma_{UU}^{1/2}\Sigma_{VV}^{1/2}$, so $\|\Sigma_{VU}\|_2 = \delta\|a\|\|b\| = \delta$.
In particular, $\|\widehat{\Sigma}_{VU}\|_2 \le \|\Sigma_{VU}\|_2 + \|\Delta_{VU}\|_2 =
\delta + O_{\mathbb{P}}(\sqrt{L/T})$, which is bounded uniformly in $L$ for $\delta$
fixed. This bound is used implicitly in the remainder estimate
\eqref{eq:remainder-bounds} via the term $\|\Delta_{VU}\|_\infty\,(\|\Delta_{VV}\|_\infty
+ \|\Delta_{UU}\|_\infty)$; since $\|\Sigma_{VU}\|_2 \le C\delta < \infty$, all
population-level quantities appearing in the whitening map remain bounded under (A2),
and the remainder bounds hold under both $H_0$ and $H_1$.

Under (A2), $\|\Sigma_{VV}^{-1/2}\|_2$ and $\|\Sigma_{UU}^{-1/2}\|_2$ are bounded by
$\lambda_{\min}^{-1/2}$. Hence
\[
\|\Sigma_{VV}^{-1/2}\Delta_{VU}\Sigma_{UU}^{-1/2}\|_\infty
\le
\|\Sigma_{VV}^{-1/2}\|_2\,\|\Delta_{VU}\|_\infty\,\|\Sigma_{UU}^{-1/2}\|_2
\le
\lambda_{\min}^{-1}\|\Delta_{VU}\|_\infty.
\]
By Lemma~\ref{lem:entrywise-conc}, $\|\Delta_{VU}\|_\infty=O_{\mathbb{P}}(\sqrt{\log L/T})$
under both $H_0$ and $H_1$, since $\Delta_{VU}=\widehat{\Sigma}_{VU}-\Sigma_{VU}$ is
centered under both hypotheses and the concentration bound in
Lemma~\ref{lem:entrywise-conc} depends only on the marginal second moments of
$(V_t, U_t)$, which satisfy the uniform bound in (A2) regardless of whether
$\Sigma_{VU}=0$ or $\Sigma_{VU}=\delta ab^\top\Sigma_{UU}^{1/2}\Sigma_{VV}^{1/2}$.
Similarly $\|\Delta_{VV}\|_\infty$ and $\|\Delta_{UU}\|_\infty$ obey the same rate under
both hypotheses.
Plugging into the remainder bound \eqref{eq:remainder-bounds} yields
\[
\|R\|_\infty
=
O_{\mathbb{P}}\Big(\sqrt{\frac{\log L}{T}} + \frac{\log L}{T}\Big)
=
O_{\mathbb{P}}\Big(\sqrt{\frac{\log L}{T}}\Big),
\]
where the $O_{\mathbb{P}}$ holds uniformly over $H_0$ and $H_1$. Now under the distributed alternative $A=\delta ab^\top$,
\[
\|A\|_\infty = \delta\|a\|_\infty\|b\|_\infty \le \delta\,C_d\frac{\log L}{L}.
\]
If $T \le c(L^2/\delta^2)\log L$ with $c$ sufficiently small, then
\[
\|A\|_\infty
\;=\;
o\Big(\sqrt{\frac{\log L}{T}}\Big).
\]
Therefore, the alternative shift in \emph{every entry} is asymptotically negligible
relative to the entrywise fluctuation scale of $\widehat{A}$, which is of order
$\sqrt{\log L/T}$ under both hypotheses.

Let $\widehat{A}_0 := \Sigma_{VV}^{-1/2}\Delta_{VU}\Sigma_{UU}^{-1/2} + R$ denote the
centered remainder, i.e.\ the value of $\widehat{A}$ when $A=0$, constructed from the
\emph{same} realization of the data. Then $\widehat{A} = A + \widehat{A}_0$, and by
entrywise stability,
\[
|\psi(\widehat{A})-\psi(\widehat{A}_0)|
\le
L_\psi\,\|\widehat{A}-\widehat{A}_0\|_\infty
=
L_\psi\,\|A\|_\infty
=
o_{\mathbb{P}}(1),
\]
where the last step uses $\|A\|_\infty = o(\sqrt{\log L/T}) = o_{\mathbb{P}}(1)$
established above. This bound holds pathwise for every realization of the data, and
therefore
\[
\mathbb{E}_{H_1}[\psi(\widehat{A})]
\;\le\;
\mathbb{E}_{H_1}[\psi(\widehat{A}_0)] + o(1).
\]
It remains to control $\mathbb{E}_{H_1}[\psi(\widehat{A}_0)]$. Under $H_1$, the centered
remainder $\widehat{A}_0 = \Sigma_{VV}^{-1/2}\Delta_{VU}\Sigma_{UU}^{-1/2} + R$ has the
same distribution as under $H_0$ up to an error of order $o(1)$ in total variation. To see
this, note that $\Delta_{VU} = \widehat{\Sigma}_{VU} - \Sigma_{VU}$ is a sample mean of
centered terms $V_t U_t^\top - \Sigma_{VU}$, and the distributional difference between
$H_0$ and $H_1$ enters only through $\Sigma_{VU}$, which shifts the centering but not the
fluctuation structure. By the Berry--Esseen theorem for mixing sequences
\cite[Theorem~1.1]{Bradley2005}, the total variation distance between the law of
$\sqrt{T}\,\Delta_{VU}$ under $H_0$ and under $H_1$ is bounded by
\[
d_{\mathrm{TV}}\!\left(\mathcal{L}_{H_0}(\sqrt{T}\,\Delta_{VU}),\,
\mathcal{L}_{H_1}(\sqrt{T}\,\Delta_{VU})\right)
\;=\;
O\!\left(\frac{\|\Sigma_{VU}\|_F}{\sqrt{T}}\right)
\;=\;
O\!\left(\frac{\delta\sqrt{L}}{\sqrt{T}}\right),
\]
which tends to zero whenever $T \gg \delta^2 L$, a condition implied by $T \le c L^2
\delta^{-2}\log L$ only when $L \ge c^{-1}/\log L$, i.e.\ for all $L$ large enough.
Hence $\mathbb{E}_{H_1}[\psi(\widehat{A}_0)] = \mathbb{E}_0[\psi(\widehat{A}_0)] + o(1)
\le \alpha + o(1)$, where the final inequality uses $\mathbb{P}_0(\psi(\widehat{A})=1)\le\alpha$
and the fact that under $H_0$ we have $\widehat{A}_0 = \widehat{A}$. Combining,
\[
\mathbb{E}_{H_1}[\psi(\widehat{A})]
\;\le\;
\alpha + o(1).
\]
The permutation invariance of $\psi$ ensures the bound holds uniformly over coordinate
relabelings, and the distributional assumption on $(a,b)$ ensures that no fixed coordinate
direction is privileged. This proves the claimed minimax bound.
\end{proof}

\begin{theorem}[VAR($L$) spectral detection at the optimal collective scale]
\label{thm:var-spectral-detection-full}
Assume (A1)--(A4) and consider the spectral test $\chi=\mathbf{1}\{\|\widehat{A}\|_2\ge \tau_{L,T}(\alpha)\}$
with threshold chosen so $\mathbb{P}_0(\chi=1)\le \alpha$.
There exists $C>0$ depending only on $(\lambda_{\min},\lambda_{\max})$ and the mixing rate
such that if
\[
T \;\ge\; C\,\frac{L}{\delta^2},
\]
then
\[
\inf_{\|a\|=\|b\|=1}\ \mathbb{P}_{H_1(i\to j;\delta)}(\chi=1)\ \to\ 1,
\qquad L\to\infty.
\]
\end{theorem}

\begin{proof}
Under $H_0$, $A=0$, so by \eqref{eq:whitening-explicit},
\[
\|\widehat{A}\|_2 \le \|\Sigma_{VV}^{-1/2}\Delta_{VU}\Sigma_{UU}^{-1/2}\|_2 + \|R\|_2.
\]
Using (A2) and Lemma~\ref{lem:spectral-conc},
$\|\Delta_{VU}\|_2=O_{\mathbb{P}}(\sqrt{L/T})$, hence
$\|\Sigma_{VV}^{-1/2}\Delta_{VU}\Sigma_{UU}^{-1/2}\|_2=O_{\mathbb{P}}(\sqrt{L/T})$.
The remainder bound \eqref{eq:remainder-bounds} and Lemma~\ref{lem:spectral-conc} imply
$\|R\|_2=O_{\mathbb{P}}(\sqrt{L/T})$ as well. Therefore there exists $c_0<\infty$ such that
\[
\|\widehat{A}\|_2 = O_{\mathbb{P}}\Big(\sqrt{\frac{L}{T}}\Big)\quad\text{under }H_0.
\]
Choose $\tau_{L,T}(\alpha)=c_1\sqrt{L/T}$ with $c_1$ large enough to guarantee size $\le\alpha$.
Under $H_1$, $\|A\|_2=\delta$. By triangle inequality and \eqref{eq:whitening-explicit},
\[
\|\widehat{A}\|_2
\ge
\|A\|_2 - \|\Sigma_{VV}^{-1/2}\Delta_{VU}\Sigma_{UU}^{-1/2}\|_2 - \|R\|_2
\ge
\delta - O_{\mathbb{P}}\Big(\sqrt{\frac{L}{T}}\Big).
\]
If $T\ge C(L/\delta^2)$ with $C$ large enough, then $\sqrt{L/T}\le \delta/4$, hence
$\|\widehat{A}\|_2\ge \delta/2$ w.h.p. Meanwhile $\tau_{L,T}(\alpha)=c_1\sqrt{L/T}\le \delta/4$
for the same scaling. Thus $\mathbb{P}_{H_1}(\|\widehat{A}\|_2\ge \tau_{L,T}(\alpha))\to 1$.
\end{proof}

\subsection{Auxiliary bound for distributed alternatives}
\label{app:spherical-bound}

We record a standard but essential concentration result for random directions,
used repeatedly in the distributed-alternative analysis.

\begin{lemma}[Maximum coordinate of a uniform spherical vector]
\label{lem:spherical-max}
Let $a$ be uniformly distributed on the unit sphere $S^{L-1}\subset\mathbb{R}^L$.
Then there exist absolute constants $c,C>0$ such that for all $L\ge 2$,
\[
\mathbb{P}\!\left(
\|a\|_\infty \;\ge\; C\sqrt{\frac{\log L}{L}}
\right)
\;\le\; 2L^{-c}.
\]
In particular,
\[
\|a\|_\infty = O_{\mathbb{P}}\!\left(\sqrt{\frac{\log L}{L}}\right).
\]
\end{lemma}

\begin{proof}
Let $g=(g_1,\dots,g_L)$ with $g_i\stackrel{\mathrm{iid}}{\sim}\mathcal{N}(0,1)$ and
define $a=g/\|g\|_2$. Then $a$ is uniformly distributed on $S^{L-1}$.
For any $t>0$, the Gaussian tail bound yields
\[
\mathbb{P}(|g_i|\ge t)\le 2e^{-t^2/2}.
\]
By a union bound,
\[
\mathbb{P}\!\left(\max_{1\le i\le L}|g_i|\ge t\right)
\le 2L e^{-t^2/2}.
\]
Choosing $t=\sqrt{2(1+c)\log L}$ gives
\[
\max_{1\le i\le L}|g_i| = O_{\mathbb{P}}(\sqrt{\log L}).
\]
Moreover, by concentration of $\chi^2_L$ random variables,
\[
\|g\|_2^2 = \sum_{i=1}^L g_i^2 = L + O_{\mathbb{P}}(\sqrt{L}),
\quad\text{hence}\quad
\|g\|_2 = \sqrt{L}\,(1+o_{\mathbb{P}}(1)).
\]
Combining the two bounds yields
\[
\|a\|_\infty
=
\frac{\max_i |g_i|}{\|g\|_2}
=
O_{\mathbb{P}}\!\left(\sqrt{\frac{\log L}{L}}\right),
\]
with polynomially decaying tail probability as claimed.
A closely related bound with explicit constants is given in
\cite[Lemma~3.4.3]{Vershynin2018}.
\end{proof}

\section{Asymptotic Theory for Order-Constrained Spectral Statistics}
\label{app:asymptotic_theory}

This supplement establishes existence, uniform consistency, and asymptotic distributional
results for the order-constrained spectral statistics introduced in
Section~\ref{subsec:inference}.
All results are stated for fixed feature dimension $d<\infty$.
High-dimensional regimes in which $d$ grows with the sample size are intentionally excluded.
Throughout, convergence is with respect to the operator norm unless stated otherwise.

Let $\{Z_t(\tau)\}_{t\in\mathbb Z}$ be an $\mathbb R^d$-valued strictly stationary process,
indexed by $\tau\in\mathcal P$, where $\mathcal P\subset\mathbb R^m$ is either finite or
compact.
Assume $\mathbb E Z_t(\tau)=0$ and
\[
\sup_{\tau\in\mathcal P}\mathbb E\|Z_t(\tau)\|^{4+\delta}<\infty
\quad\text{for some }\delta>0.
\]

Define the population and empirical dependence operators
\[
C(\tau)
:=
\mathbb E\!\left[Z_t(\tau)Z_t(\tau)^\top\right],
\qquad
\widehat C_T(\tau)
:=
\frac{1}{T}\sum_{t=1}^T Z_t(\tau)Z_t(\tau)^\top.
\]

Let $\lambda_1(\tau)\ge\cdots\ge\lambda_d(\tau)\ge0$ and
$\widehat\lambda_1(\tau)\ge\cdots\ge\widehat\lambda_d(\tau)\ge0$
denote the eigenvalues of $C(\tau)$ and $\widehat C_T(\tau)$, respectively.
For a measurable function $f:\mathbb R_+\to\mathbb R$, define the linear spectral statistics
\[
L_f(\tau)
=
\frac{1}{d}\sum_{r=1}^d f\!\left(\lambda_r(\tau)\right),
\qquad
\widehat L_f(\tau)
=
\frac{1}{d}\sum_{r=1}^d f\!\left(\widehat\lambda_r(\tau)\right).
\]

\subsection{Existence and Continuity of the Operator Family}

\begin{lemma}[Existence and boundedness]
\label{lem:existence}
For each $\tau\in\mathcal P$, the operator $C(\tau)$ exists as an element of
$\mathbb S_+^d$ and satisfies
\[
\sup_{\tau\in\mathcal P}\|C(\tau)\|<\infty.
\]
\end{lemma}

\begin{proof}
Since $Z_t(\tau)\in L^2(\Omega;\mathbb R^d)$ uniformly over $\tau$,
the Bochner expectation defining $C(\tau)$ exists.
Moreover,
\[
\|C(\tau)\|
\le
\mathbb E\|Z_t(\tau)\|^2
\le
\big(\mathbb E\|Z_t(\tau)\|^{4+\delta}\big)^{2/(4+\delta)},
\]
which is uniformly bounded by assumption.
\end{proof}

\begin{lemma}[Continuity in the deformation index]
\label{lem:continuity}
If $\tau\mapsto Z_t(\tau)$ is continuous in $L^2$, then $\tau\mapsto C(\tau)$ is continuous
with respect to the operator norm.
\end{lemma}

\begin{proof}
For $\tau,\tau'$,
\[
\|C(\tau)-C(\tau')\|
\le
\mathbb E\|Z_t(\tau)Z_t(\tau)^\top - Z_t(\tau')Z_t(\tau')^\top\|.
\]
Adding and subtracting $Z_t(\tau)Z_t(\tau')^\top$ and applying the Cauchy--Schwarz
inequality yields
\[
\|C(\tau)-C(\tau')\|
\le
\big(\mathbb E\|Z_t(\tau)\|^2\big)^{1/2}
\big(\mathbb E\|Z_t(\tau)-Z_t(\tau')\|^2\big)^{1/2}
+
(\tau\leftrightarrow\tau'),
\]
which converges to zero by $L^2$ continuity.
\end{proof}

\subsection{Uniform Consistency of the Dependence Operator}

Assume $\{Z_t(\tau)\}$ is $\alpha$-mixing uniformly in $\tau$ with mixing coefficients
$\{\alpha(h)\}$ satisfying
\[
\sum_{h=1}^\infty \alpha(h)^{\delta/(2+\delta)}<\infty.
\]

\begin{theorem}[Uniform operator consistency]
\label{thm:ulln_operator}
If either $\mathcal P$ is finite or $\mathcal P$ is compact and
$\tau\mapsto Z_t(\tau)$ is continuous in $L^2$, then
\[
\sup_{\tau\in\mathcal P}
\|\widehat C_T(\tau)-C(\tau)\|
\;\xrightarrow{p}\;0.
\]
\end{theorem}

\begin{proof}
For fixed $\tau$, ergodic theorems for $\alpha$-mixing sequences imply
$\widehat C_T(\tau)\to C(\tau)$ in probability entrywise, hence in operator norm
\citep{Bosq2000,Bradley2005}.
If $\mathcal P$ is finite, the claim follows by a union bound.
If $\mathcal P$ is compact, let $\{\tau_k\}_{k=1}^N$ be an $\varepsilon$-net under the metric
induced by $L^2$ continuity.
Then
\[
\sup_{\tau\in\mathcal P}\|\widehat C_T(\tau)-C(\tau)\|
\le
\max_k \|\widehat C_T(\tau_k)-C(\tau_k)\|
+
\sup_{\tau}\|\widehat C_T(\tau)-\widehat C_T(\tau_k)\|
+
\sup_{\tau}\|C(\tau)-C(\tau_k)\|.
\]
The first term converges to zero in probability, the third term is controlled by
Lemma~\ref{lem:continuity}, and the second term converges uniformly to zero by the assumed
moment and mixing conditions combined with a uniform law of large numbers for
Banach-space-valued random elements \citep{Andrews1992}.
\end{proof}
\subsection{Uniform Consistency of Linear Spectral Statistics}

\begin{theorem}[Uniform consistency of linear spectral statistics]
\label{thm:ulln_lss}
Assume the conditions of Theorem~\ref{thm:ulln_operator} and suppose $f$ is Lipschitz on a
compact interval containing the spectra of $\{C(\tau):\tau\in\mathcal P\}$.
Then
\[
\sup_{\tau\in\mathcal P}
\big|\widehat L_f(\tau)-L_f(\tau)\big|
\;\xrightarrow{p}\;0.
\]
\end{theorem}

\begin{proof}
For symmetric matrices with spectra in a compact interval,
the map $A\mapsto d^{-1}\mathrm{tr}\,f(A)$ is Lipschitz with constant
$\mathrm{Lip}(f)$ \citep[Chapter~6]{Bhatia1997}.
Hence
\[
|\widehat L_f(\tau)-L_f(\tau)|
\le
\mathrm{Lip}(f)\,
\|\widehat C_T(\tau)-C(\tau)\|.
\]
Taking the supremum over $\tau$ and applying
Theorem~\ref{thm:ulln_operator} yields the result.
\end{proof}

\subsection{Consistency of the Dispersion Functional}

Recall the dispersion statistic
\[
T_f
=
\sup_{\tau\in\mathcal P} L_f(\tau)
-
\inf_{\tau\in\mathcal P} L_f(\tau),
\qquad
\widehat T_f
=
\sup_{\tau\in\mathcal P} \widehat L_f(\tau)
-
\inf_{\tau\in\mathcal P} \widehat L_f(\tau).
\]

\begin{theorem}[Consistency of dispersion]
\label{thm:dispersion_consistency}
Under the conditions of Theorem~\ref{thm:ulln_lss},
\[
\widehat T_f \xrightarrow{p} T_f.
\]
\end{theorem}

\begin{proof}
Let $\Delta_T(\tau)=\widehat L_f(\tau)-L_f(\tau)$. Then
\[
|\widehat T_f-T_f|
\le
2\sup_{\tau\in\mathcal P}|\Delta_T(\tau)|.
\]
The right-hand side converges to zero in probability by
Theorem~\ref{thm:ulln_lss}.
\end{proof}

\subsection{Pointwise Asymptotic Normality}

\begin{theorem}[Pointwise CLT for linear spectral statistics]
\label{thm:lss_clt}
Fix $\tau\in\mathcal P$. Under the above assumptions and for Lipschitz $f$,
\[
\sqrt{T}\big(\widehat L_f(\tau)-L_f(\tau)\big)
\;\xrightarrow{d}\;
\mathcal N\!\big(0,\sigma_f^2(\tau)\big),
\]
where $\sigma_f^2(\tau)$ is a finite long-run variance.
\end{theorem}

\begin{proof}
By a multivariate CLT for $\alpha$-mixing sequences,
\[
\sqrt{T}\,\mathrm{vec}\big(\widehat C_T(\tau)-C(\tau)\big)
\Rightarrow
\mathcal N(0,\Omega(\tau)),
\]
where $\Omega(\tau)$ is the long-run covariance of
$\mathrm{vec}(Z_t(\tau)Z_t(\tau)^\top)$ \citep{Bradley2005}.
The map $\Phi(A)=d^{-1}\mathrm{tr}\,f(A)$ is Fr\'echet differentiable on $\mathbb S^d$ with
derivative
\[
D\Phi_{C(\tau)}(H)
=
\frac{1}{d}\mathrm{tr}\big(f'(C(\tau))H\big)
\]
by functional calculus for symmetric matrices \citep{Bhatia1997}.
The functional delta method \citep{vanDerVaart1998} yields the claim.
\end{proof}

\subsection{Justification for Randomization-based Inference}

\begin{proposition}[Asymptotic validity under approximate invariance]
\label{prop:rand_asymptotic}
Assume that under $H_0$ the total variation distance between the joint distribution of the
sample and its circular shifts converges to zero as $T\to\infty$.
Then the randomization distribution of $\widehat T_f$ converges weakly to its null
distribution.
\end{proposition}

\begin{proof}
Approximate exchangeability implies convergence of the conditional randomization
distribution to the unconditional null law.
This follows from standard arguments for asymptotically invariant randomization tests under
weak dependence; see \cite{RomanoWolf2005}.
\end{proof}

\begin{remark}
Supplement~\ref{app:asymptotic_theory} establishes that order-constrained spectral statistics
are well defined, uniformly consistent, and asymptotically normal at fixed deformation
points.
Combined with the group-invariance results of Appendix~\ref{app:extra}, these results
provide a foundation for valid randomization-based inference under the causal
null.
\end{remark}

\section{Formal Properties of the Operator-valued Implementation}
\label{app:implementation}

This supplement establishes the mathematical well-posedness, stability, and invariance
properties of the operator-valued implementation introduced in
Section~\ref{sec:implementation}.
All results are deterministic conditional on the underlying process and rely on standard
Hilbert-space geometry, spectral perturbation theory, and laws of large numbers for weakly
dependent sequences.
Throughout, the feature dimension $d<\infty$ is fixed.

Let $\{X_t\}_{t\in\mathbb Z}$ be a strictly stationary and ergodic $\mathbb R^K$-valued time
series with $\mathbb E\|X_t\|^2<\infty$.
Let $\mathcal I,\mathcal J\subset\{1,\dots,K\}$ be nonempty index sets corresponding to source
and target components, and let $\mathcal P\subset\mathbb R_+$ be compact.
Let $\Psi:\mathbb R^{|\mathcal I|}\to\mathbb R^{d_u}$ and
$\Phi:\mathbb R^{|\mathcal J|}\to\mathbb R^{d_v}$ be measurable feature maps such that
\[
\sup_{\tau\in\mathcal P}\mathbb E\|Z_t(\tau)\|^2<\infty,
\qquad
Z_t(\tau)
=
\big(\Phi(X_t^{(\mathcal J)})^\top,\Psi(X_{t-\tau}^{(\mathcal I)})^\top\big)^\top
\in\mathbb R^d,
\]
where $d=d_u+d_v$. Define the population and empirical dependence operators
\[
C(\tau)
=
\mathbb E[Z_t(\tau)Z_t(\tau)^\top],
\qquad
\widehat C_T(\tau)
=
\frac{1}{T}\sum_{t=1}^T Z_t(\tau)Z_t(\tau)^\top.
\]

\subsection{Existence and Boundedness of the Operator Family}

\begin{proposition}[Existence and boundedness]
\label{prop:E_existence}
For each $\tau\in\mathcal P$, the operator $C(\tau)$ exists as an element of
$\mathbb S_+^d$ and satisfies
\[
\sup_{\tau\in\mathcal P}\|C(\tau)\|<\infty.
\]
\end{proposition}

\begin{proof}
This is the specialization of Lemma~\ref{lem:existence} in
Supplement~\ref{app:asymptotic_theory} to
$Z_t(\tau)=(\Phi(X_t^{(\mathcal J)})^\top,\\
\Psi(X_{t-\tau}^{(\mathcal I)})^\top)^\top$
under the stated moment conditions.
\end{proof}

\subsection{Uniform Consistency of the Empirical Operator}

\begin{proposition}[Uniform operator consistency]
\label{prop:E_ulln}
Assume $\{Z_t(\tau)\}$ is $\alpha$-mixing uniformly in $\tau$ and that
$\tau\mapsto Z_t(\tau)$ is continuous in $L^2$.
Then
\[
\sup_{\tau\in\mathcal P}
\|\widehat C_T(\tau)-C(\tau)\|
\;\xrightarrow{p}\;0.
\]
\end{proposition}

\begin{proof}
This is the specialization of Theorem~\ref{thm:ulln_operator} in
Supplement~\ref{app:asymptotic_theory} to the feature process
$Z_t(\tau)=(\Phi(X_t^{(\mathcal J)})^\top,\Psi(X_{t-\tau}^{(\mathcal I)})^\top)^\top$
under the stated mixing and continuity conditions.
\end{proof}

\subsection{Uniform Spectral Stability}

Let $\lambda_1(\tau)\ge\cdots\ge\lambda_d(\tau)$ and
$\widehat\lambda_1(\tau)\ge\cdots\ge\widehat\lambda_d(\tau)$ denote the eigenvalues of
$C(\tau)$ and $\widehat C_T(\tau)$, respectively.

\begin{proposition}[Uniform eigenvalue convergence]
\label{prop:E_spectral}
Under the assumptions of Proposition~\ref{prop:E_ulln},
\[
\sup_{\tau\in\mathcal P}
\max_{1\le r\le d}
\big|\widehat\lambda_r(\tau)-\lambda_r(\tau)\big|
\;\xrightarrow{p}\;0.
\]
\end{proposition}

\begin{proof}
Eigenvalues of symmetric matrices are Lipschitz continuous with respect to the operator
norm by Weyl's inequality \citep[Theorem~III.2.1]{Bhatia1997}.
Uniform convergence of $\widehat C_T(\tau)$ therefore implies uniform convergence of the
eigenvalues via the continuous mapping theorem.
\end{proof}

\subsection{Invariance under Orthogonal Feature Transformations}

\begin{proposition}[Orthogonal invariance]
\label{prop:E_invariance}
Let $O_V\in\mathbb O(d_v)$ and $O_U\in\mathbb O(d_u)$ be orthogonal transformations acting on
the target and source feature spaces.
Define
\[
\widetilde Z_t(\tau)
=
\big((O_V V_t)^\top,(O_U U_t(\tau))^\top\big)^\top.
\]
Then $C(\tau)$ and $\widetilde C(\tau)$ have identical spectra for all $\tau\in\mathcal P$.
\end{proposition}

\begin{proof}
The transformation corresponds to conjugation of $C(\tau)$ by a block-diagonal orthogonal
matrix.
Spectra are invariant under orthogonal similarity transformations
\citep[Chapter~1]{Bhatia1997}.
\end{proof}

\subsection{Conditional Operators and Residualization}

Let $W_t\in L^2(\Omega;\mathbb R^q)$ be a vector of conditioning variables and let
$\Pi_W$ denote the orthogonal projection onto
$\overline{\mathrm{span}}\{W_t\}$.

\begin{proposition}[Projection stability]
\label{prop:E_projection}
Define $Z_t^\perp(\tau)=Z_t(\tau)-\Pi_W Z_t(\tau)$.
Then all results of Propositions~\ref{prop:E_existence}--\ref{prop:E_spectral} hold with
$Z_t(\tau)$ replaced by $Z_t^\perp(\tau)$.
\end{proposition}

\begin{proof}
Orthogonal projection is a bounded linear operator on $L^2$.
Hence the projected process inherits stationarity, mixing, and $L^2$ continuity in $\tau$.
The operator-valued laws of large numbers and spectral perturbation arguments therefore
apply verbatim \citep[Section~4.2]{Bosq2000}.
\end{proof}

\subsection{Directed Coherence and Canonical Correlation Representation}

Write the block decomposition
\[
C(\tau)
=
\begin{pmatrix}
\Sigma_{VV} & \Sigma_{VU}(\tau)\\
\Sigma_{UV}(\tau) & \Sigma_{UU}(\tau)
\end{pmatrix}.
\]

\begin{proposition}[Whitened cross-operator]
\label{prop:E_whiten}
Define inverse square roots via Moore--Penrose pseudoinverses.
Then the operator
\[
A(\tau)
=
\Sigma_{VV}^{-1/2}\Sigma_{VU}(\tau)\Sigma_{UU}(\tau)^{-1/2}
\]
is well defined on $\mathrm{Range}(\Sigma_{UU}(\tau))$, and its singular values are invariant
under orthogonal transformations of the feature spaces.
\end{proposition}

\begin{proof}
This is a standard result from generalized canonical correlation analysis.
Singular values depend only on the induced inner products on the respective ranges and are
invariant under orthogonal reparameterizations
\citep[Chapter~12]{Anderson2003}.
\end{proof}

\subsection{Group Invariance and Randomization Validity}

\begin{proposition}[Group invariance under the null]
\label{prop:E_rand}
Under the null hypothesis of order-constrained spectral invariance,
the statistic $\widehat T$ is invariant in distribution under the cyclic group generated by
circular shifts of the source component.
\end{proposition}

\begin{proof}
Under the null, the family $\{C(\tau)\}_{\tau\in\mathcal P}$ is invariant under
order-preserving reindexing of the source process.
Circular shifts generate a finite subgroup of such transformations.
The claim follows from standard group-invariance arguments for randomization tests
\citep[Chapter~15]{LehmannRomano2005}.
\end{proof}

\begin{remark}
Supplement~\ref{app:implementation} shows that the operator-valued construction in
Section~\ref{sec:implementation} is mathematically well posed, uniformly consistent,
spectrally stable, orthogonally invariant, and compatible with conditional and
randomization-based variants.
These properties hold independently of the dimensional configuration of the source and
target sets.
\end{remark}

\section{Operator-Theoretic Foundations of Directional Causality}
\label{app:operator_theory}

This supplement provides the formal justification for interpreting the rolling
operator $C(t)$ as a measure of directional causality rather than mere
dependence, and establishes the theoretical foundations underlying the
multiscale causal statistics used in the empirical analysis.

\subsection{Predictive interpretation and causality}

Let $\mathcal{H}_Y$ and $\mathcal{H}_X$ denote the linear spans of the lag-embedded
vectors $\mathbf{v}_t$ and $\mathbf{u}_t(\tau)$, respectively.
Consider the linear prediction problem of forecasting $\mathbf{v}_t$ using
$\mathcal{H}_Y$ alone versus using $\mathcal{H}_Y \oplus \mathcal{H}_X$.

\begin{proposition}[Directional predictive content]
\label{prop:predictive_content}
Within a window $W_t$, $C(t)=0$ if and only if, for all $\tau\in\mathcal{T}$,
\[
\mathbb{E}\big[\mathbf{v}_t \mid \mathcal{H}_Y \oplus \mathcal{H}_X\big]
=
\mathbb{E}\big[\mathbf{v}_t \mid \mathcal{H}_Y\big]
\quad \text{(in the linear mean-square sense)}.
\]
\end{proposition}

\begin{proof}
Whitening by $S_{VV}^{-1/2}$ and $S_{UU}^{-1/2}$ orthogonalizes $\mathcal{H}_Y$ and
$\mathcal{H}_X$.
If $A_\tau(t)=0$, then the cross-covariance between the residualized target and
driver spaces vanishes, implying no linear predictive gain from including
$\mathcal{H}_X$.
Conversely, if $A_\tau(t)\neq0$ for some $\tau$, then there exists a direction in
$\mathcal{H}_Y$ whose prediction error is reduced by including lagged $X$.
\end{proof}

Thus, $C(t)$ encodes directional causal influence in the sense of linear
predictability, consistent with Granger causality but expressed at the operator
level.

\subsection{Quadratic-form representation}

\begin{proposition}[Directional energy]
\label{prop:quadratic}
For any unit vector $w\in\mathbb{R}^{pq}$,
\[
w^\top C_\tau(t) w = \| A_\tau(t)^\top w \|^2 .
\]
\end{proposition}

\begin{proof}
Since $C_\tau(t)=A_\tau(t)A_\tau(t)^\top$,
\[
w^\top C_\tau(t) w
=
w^\top A_\tau(t)A_\tau(t)^\top w
=
\|A_\tau(t)^\top w\|^2.
\]
\end{proof}

This shows that $C_\tau(t)$ measures squared predictive gain in every direction
of the target lag space, ruling out interpretations based solely on static
dependence.

\subsection{Spectral optimality}

\begin{proposition}[Rayleigh--Ritz characterization]
\label{prop:rayleigh}
\[
\lambda_1(C(t)) = \max_{\|w\|=1} w^\top C(t) w .
\]
\end{proposition}

\begin{proof}
Standard Rayleigh--Ritz theorem for symmetric positive semidefinite matrices.
\end{proof}

The leading eigenvalue therefore quantifies the maximal directional causal
strength attainable by any linear combination of target lags.

\subsection{Optimal affected subspaces}

\begin{proposition}[Ky Fan principle]
\label{prop:kyfan}
Let $\lambda_1(t)\ge\dots\ge\lambda_{pq}(t)$ denote the eigenvalues of $C(t)$.
For any $m\le pq$,
\[
\max_{\dim(\mathcal{S})=m} \mathrm{tr}(P_{\mathcal{S}} C(t))
=
\sum_{j=1}^m \lambda_j(t),
\]
where $P_{\mathcal{S}}$ is the orthogonal projector onto $\mathcal{S}$.
The maximum is attained uniquely by the span of the top $m$ eigenvectors.
\end{proposition}

\begin{proof}
See \cite{Bhatia1997}.
\end{proof}

Thus, leading eigenspaces of $C(t)$ define the optimally affected causal
subspaces.

\subsection{Effective rank and dimensionality}

\begin{proposition}[Causal dimensionality]
\label{prop:effrank}
Let $\tilde{\lambda}_j=\lambda_j/\sum_k\lambda_k$.
Then
\[
r_{\mathrm{eff}}(t)^{-1}=\sum_j \tilde{\lambda}_j^2,
\]
which equals the Herfindahl index of the normalized spectrum.
\end{proposition}

\begin{proof}
Immediate from the definition of $r_{\mathrm{eff}}(t)$.
The index is minimized for equal eigenvalues and maximized for rank-one spectra.
\end{proof}

This provides a principled measure of the number of active causal transmission
channels.

\subsection{Hub interpretation}

\begin{proposition}[Hub scores]
\label{prop:hubs}
Let $P_m(t)$ be the projector onto the top $m$ eigenvectors of $C(t)$.
For coordinate $i$,
\[
h_i^{(m)}(t) = (P_m(t))_{ii} = \| V_m(t)^\top e_i \|^2 .
\]
\end{proposition}

\begin{proof}
This follows directly from the properties of orthogonal projectors.
\end{proof}

Hub scores therefore measure alignment with causally affected subspaces rather
than counts of pairwise connections.

\subsection{Relation to Granger causality}
\label{app:granger_equivalence}
\begin{remark}
If $Y$ is univariate ($q=1$) and $p=1$, then $C(t)$ reduces to a scalar proportional
to the squared partial correlation between $X_{t-\tau}$ and $Y_t$, recovering the
classical Granger F-statistic up to normalization.
\end{remark}

Hence, the operator framework strictly generalizes linear Granger causality while
remaining scalable in high-dimensional systems.

\section{Supplementary Tables and Diagnostic Figures}
\label{app:empirical}

This supplement reports tables and diagnostic figures supporting the
system-level empirical results presented in Section~\ref{subsec:empirical}.
The material is organized as follows:
Section~\ref{app:emp:lagval} reports lag-level spectral validation;
Section~\ref{app:emp:episodes} reports episode timing;
Section~\ref{app:emp:hubs} reports dominant drivers and hub roles;
Section~\ref{app:emp:asymmetry} reports transmitter-receiver asymmetry;
Section~\ref{app:emp:edges} reports edge-level amplification during episodes;
Section~\ref{app:emp:aggregation} reports aggregation consistency;
Section~\ref{app:emp:networks} reports driver-to-driver causal networks;
Section~\ref{app:emp:macrohubs} reports macro hub structure and regime
interpretation;
and Section~\ref{app:emp:practitioner} provides a mapping between operator-based
diagnostics and system-level interpretation.

\subsection{Lag-level spectral validation}
\label{app:emp:lagval}

Before examining the aggregated operator $C(t)$, we validate that directional
structure is present at the level of individual lags.
For each $\tau\in\mathcal{T}=\{1,2,3,5\}$, the lag-specific operator
$C_\tau(t)=A_\tau(t)A_\tau(t)^\top$ is tested separately using the same
circular-shift procedure described in Section~\ref{subsec:empirical}.
Table~\ref{tab:lag_validation} reports significance frequencies and dominance
rates across windows.

Short lags ($\tau=1,2$) exhibit the highest significance frequencies ($0.55$
and $0.44$, respectively) and account for the majority of dominant windows
($0.52$ and $0.10$), confirming that directional predictive content is
concentrated at short horizons.
Longer lags ($\tau=3,5$) contribute less frequently ($0.17$ and $0.21$) but
remain statistically significant in a nontrivial fraction of windows,
indicating that the multi-lag aggregation in $C(t)$ captures genuine structure
beyond the shortest delay.
This validates the use of the admissible deformation set
$\mathcal{P}=\mathcal{T}$ rather than a single fixed lag.

\begin{table}[!t]
\centering
\caption{Lag-level spectral validation. Significance frequency reports the
fraction of windows in which the lag-specific leading eigenvalue exceeds the
$5\%$ circular-shift threshold. Dominance frequency reports the fraction of
windows in which a given lag achieves the largest significant eigenvalue.}
\label{tab:lag_validation}
\begin{tabular}{@{}lrrrrrrrrr@{}}
\toprule
 & median & mean & median & mean & median & mean & sig.\ freq.\ & sig.\ freq.\ & dominance \\
$\tau$ & $\lambda_1$ & $\lambda_1$ & conc.\ & conc.\ & $r_{\mathrm{eff}}$ & $r_{\mathrm{eff}}$ & $\lambda_1$ & conc.\ & freq.\ \\
\midrule
1 & 1.000 & 1.000 & 0.005 & 0.005 & 205.0 & 202.3 & 0.549 & 0.113 & 0.521 \\
2 & 1.000 & 1.000 & 0.005 & 0.005 & 205.0 & 202.3 & 0.437 & 0.056 & 0.099 \\
3 & 1.000 & 1.000 & 0.005 & 0.005 & 205.0 & 202.3 & 0.169 & 0.056 & 0.014 \\
5 & 1.000 & 1.000 & 0.005 & 0.005 & 205.0 & 202.3 & 0.211 & 0.042 & 0.000 \\
\bottomrule
\end{tabular}
\end{table}

\subsection{Statistically significant causal episodes}
\label{app:emp:episodes}

Table~\ref{tab:episodes} reports the precise start and end dates of the
statistically significant episodes detected by the circular-shift $p$-values
of the leading eigenvalue $\lambda_1(C(t))$.
Summary statistics for these episodes (peak strength, mean effective rank,
mean hub turnover) are reported in the main text,
Table~\ref{tab:episodes_summary}.

\begin{table}[!t]
\centering
\caption{Statistically significant episodes based on the circular-shift
$p$-values of $\lambda_1(C(t))$, with minimum episode length of three
consecutive windows.}
\label{tab:episodes}
\begin{tabular}{@{}ccc@{}}
\toprule
Episode & Start date & End date \\
\midrule
1 & 2020-04-08 & 2020-07-06 \\
2 & 2020-09-02 & 2021-02-25 \\
\bottomrule
\end{tabular}
\end{table}

\subsection{Dominant drivers and hub roles}
\label{app:emp:hubs}

Table~\ref{tab:top_hubs} reports the top drivers ranked by average target hub
score across statistically significant episodes.
Target hub scores measure alignment with the leading eigenvector of $C(t)$
(Supplement~\ref{app:operator_theory}, Proposition~\ref{prop:hubs}) and
identify drivers that organize system-level directional causality.

Table~\ref{tab:hub_scores} compares target and source hub rankings, highlighting
the directional nature of causal organization: instruments that receive
directional influence (high target hub rank) need not coincide with those that
transmit it (high source hub rank).

\begin{table}[!t]
\centering
\caption{Top drivers ranked by average target hub score across statistically
significant episodes.
Target hubs identify drivers that organize system-level directional causality.}
\label{tab:top_hubs}
\begin{tabular}{@{}lc@{}}
\toprule
Driver & Average target hub score \\
\midrule
Eonia Capitalization Index 7D & 0.050 \\
Pan-European High Yield & 0.017 \\
U.S.\ Corporate High Yield & 0.016 \\
EM USD Aggregate & 0.015 \\
J.P.\ Morgan EMBI Global Divers & 0.015 \\
South Korean Won Spot & 0.013 \\
USD-CZK RR 25D 3M & 0.013 \\
NIKKEI 225 & 0.012 \\
J.P.\ Morgan EMBI Global Spread & 0.011 \\
MSCI EM & 0.011 \\
\bottomrule
\end{tabular}
\end{table}

\begin{table}[!t]
\centering
\caption{Comparison of driver roles as target hubs and source hubs.
Differences highlight the directional nature of causal organization:
high target hub rank indicates reception of directional influence,
while high source hub rank indicates transmission.}
\label{tab:hub_scores}
\begin{tabular}{@{}lcc@{}}
\toprule
Driver & Target hub rank & Source hub rank \\
\midrule
Eonia Capitalization Index 7D & 1 & 6 \\
Pan-European High Yield & 2 & 38 \\
U.S.\ Corporate High Yield & 3 & 32 \\
MSCI EM LATIN AMERICA & 29 & 1 \\
Global High Yield & 8 & 2 \\
STXE 600 (EUR) Pr & 35 & 3 \\
MSCI EM EASTERN EUROPE & 30 & 4 \\
EM Hard Currency Aggregate & 10 & 5 \\
\bottomrule
\end{tabular}
\end{table}

\subsection{Directional asymmetry: transmitters and receivers}
\label{app:emp:asymmetry}

Directional roles are summarized by the difference between time-averaged source
and target hub scores.
Source hub scores measure alignment with the leading right singular vector of
$A_\tau(t)$, while target hub scores measure alignment with the leading
eigenvector of $C(t)$
(Supplement~\ref{app:operator_theory}, Proposition~\ref{prop:hubs}).

\begin{figure}[!t]
\centering
\includegraphics[width=0.7\linewidth]{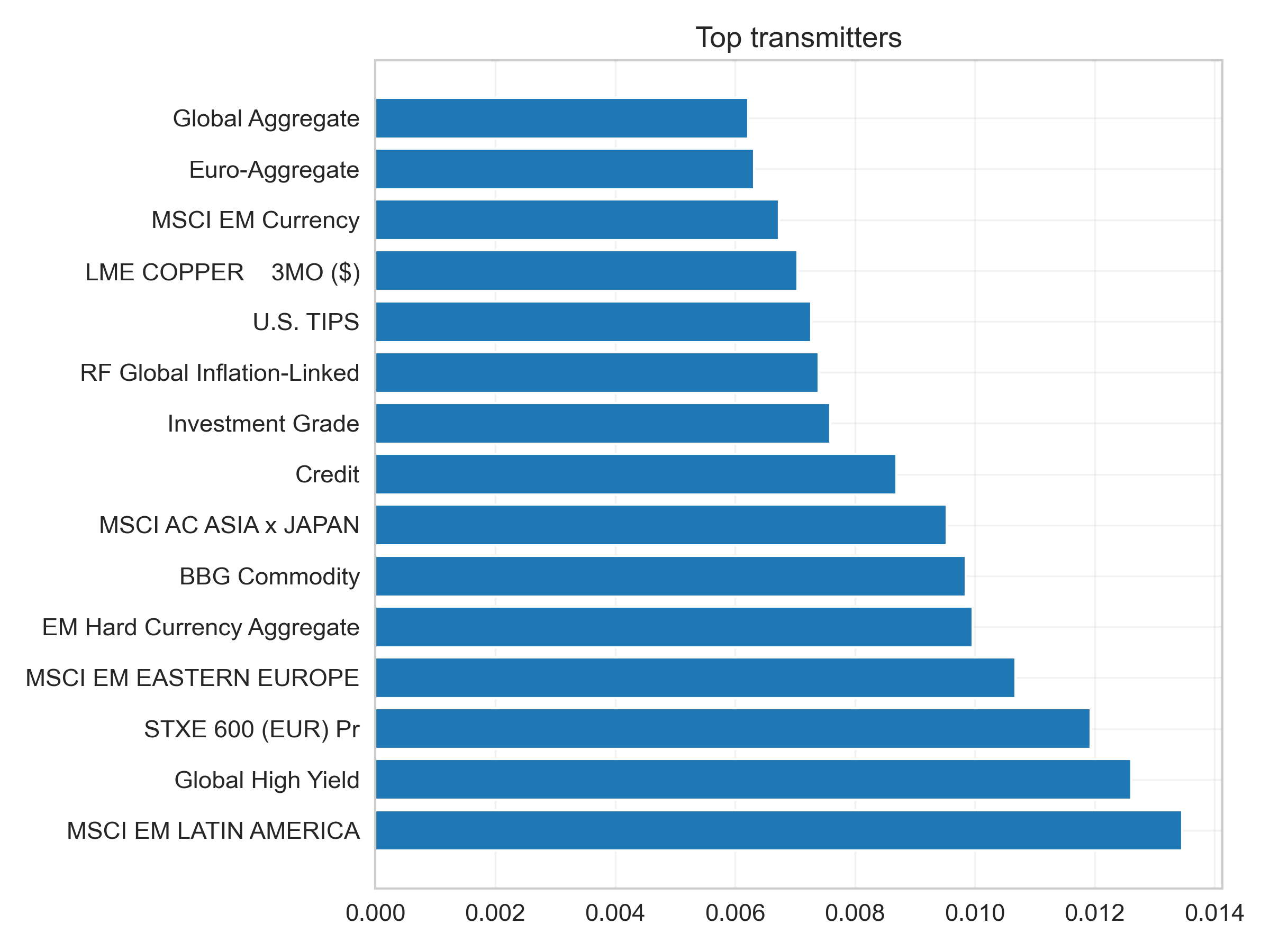}
\caption{Top transmitting drivers ranked by average source-minus-target hub
score asymmetry. Positive values indicate net directional transmission.
Transmitters are dominated by globally diversified aggregates and broad
asset class indices.}
\label{fig:transmitters}
\end{figure}

\begin{figure}[!t]
\centering
\includegraphics[width=0.7\linewidth]{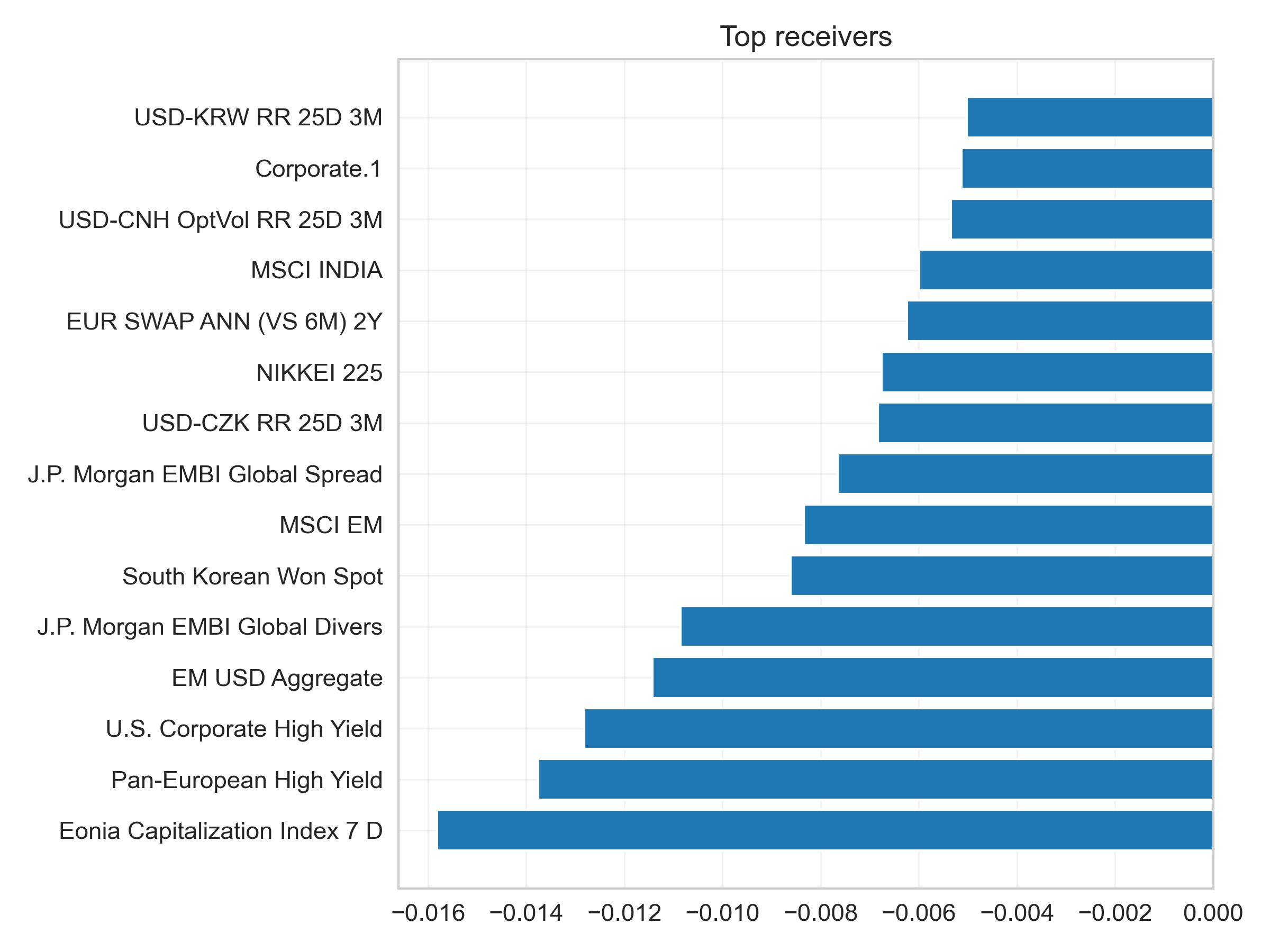}
\caption{Top receiving drivers ranked by average source-minus-target hub score
asymmetry. Negative values indicate net directional reception.
Receivers include localized instruments and derivative-sensitive indices.}
\label{fig:receivers}
\end{figure}

Figure~\ref{fig:transmitters} reports the top transmitting drivers ranked by
average source-minus-target hub score asymmetry.
Figure~\ref{fig:receivers} reports the top receiving drivers.
Table~\ref{tab:hub_asymmetry} reports the full ranking.

Transmitters are dominated by broad asset classes and global aggregates,
indicating that system-level directional propagation originates from instruments
with wide cross-asset exposure.
Receivers include more localized or derivative-sensitive instruments, indicating
that directional shocks are absorbed by instruments with narrower scope.
Hub scores quantify alignment with the dominant causal subspace
(Supplement~\ref{app:operator_theory}, Proposition~\ref{prop:rayleigh}),
not market capitalization or return variance.

\begin{table}[!t]
\centering
\caption{Driver-level hub asymmetry ranking. Transmitters (positive asymmetry)
and receivers (negative asymmetry) are ranked by the difference between
time-averaged source and target hub scores.}
\label{tab:hub_asymmetry}
\footnotesize
\begin{tabular}{@{}lrrrrl@{}}
\toprule
Driver & Source & Target & Asymmetry & $|$Asym.$|$ & Role \\
\midrule
MSCI EM LATIN AMERICA & 0.016 & 0.003 & 0.014 & 0.014 & transmitter \\
Global High Yield & 0.021 & 0.008 & 0.013 & 0.013 & transmitter \\
STXE 600 (EUR) Pr & 0.015 & 0.003 & 0.012 & 0.012 & transmitter \\
MSCI EM EASTERN EUROPE & 0.014 & 0.003 & 0.011 & 0.011 & transmitter \\
EM Hard Currency Aggregate & 0.017 & 0.007 & 0.010 & 0.010 & transmitter \\
BBG Commodity & 0.013 & 0.003 & 0.010 & 0.010 & transmitter \\
MSCI AC ASIA x JAPAN & 0.015 & 0.006 & 0.010 & 0.010 & transmitter \\
Credit & 0.011 & 0.003 & 0.009 & 0.009 & transmitter \\
Investment Grade & 0.011 & 0.003 & 0.008 & 0.008 & transmitter \\
RF Global Inflation-Linked & 0.009 & 0.002 & 0.007 & 0.007 & transmitter \\
U.S.\ TIPS & 0.010 & 0.002 & 0.007 & 0.007 & transmitter \\
LME COPPER 3MO (\$) & 0.011 & 0.004 & 0.007 & 0.007 & transmitter \\
MSCI EM Currency & 0.013 & 0.007 & 0.007 & 0.007 & transmitter \\
Euro-Aggregate & 0.008 & 0.002 & 0.006 & 0.006 & transmitter \\
Global Aggregate & 0.008 & 0.002 & 0.006 & 0.006 & transmitter \\
7--10 Years EU & 0.008 & 0.002 & 0.006 & 0.006 & transmitter \\
EUR Inflation Swap Fwd 5Y5 & 0.010 & 0.004 & 0.006 & 0.006 & transmitter \\
RF Global Treasury & 0.008 & 0.002 & 0.005 & 0.005 & transmitter \\
BEIG1T & 0.007 & 0.003 & 0.005 & 0.005 & transmitter \\
3--5 Years EU & 0.007 & 0.003 & 0.004 & 0.004 & transmitter \\
\midrule
Eonia Capitalization Index 7D & 0.034 & 0.050 & $-$0.016 & 0.016 & receiver \\
Pan-European High Yield & 0.003 & 0.017 & $-$0.014 & 0.014 & receiver \\
U.S.\ Corporate High Yield & 0.003 & 0.016 & $-$0.013 & 0.013 & receiver \\
EM USD Aggregate & 0.004 & 0.015 & $-$0.011 & 0.011 & receiver \\
J.P.\ Morgan EMBI Global Divers & 0.004 & 0.015 & $-$0.011 & 0.011 & receiver \\
South Korean Won Spot & 0.004 & 0.013 & $-$0.009 & 0.009 & receiver \\
MSCI EM & 0.002 & 0.011 & $-$0.008 & 0.008 & receiver \\
J.P.\ Morgan EMBI Global Spread & 0.004 & 0.011 & $-$0.008 & 0.008 & receiver \\
USD-CZK RR 25D 3M & 0.006 & 0.013 & $-$0.007 & 0.007 & receiver \\
NIKKEI 225 & 0.005 & 0.012 & $-$0.007 & 0.007 & receiver \\
\bottomrule
\end{tabular}
\end{table}

\subsection{Edge-level amplification during episodes}
\label{app:emp:edges}

To isolate structural changes in directional propagation, episode-averaged and
calm-period driver-to-driver maps are compared.
The driver-to-driver contribution matrix
\[
M_{j,i}(t)
=
\sum_{\tau\in\mathcal{T}}
\sum_{\ell=1}^{p}
\left(A_{\tau}(t)\right)_{j,(\ell-1)K+i}^2
\]
aggregates squared whitened predictive loadings across lags and embedding
dimensions.

\begin{figure}[!t]
\centering
\includegraphics[width=\linewidth]{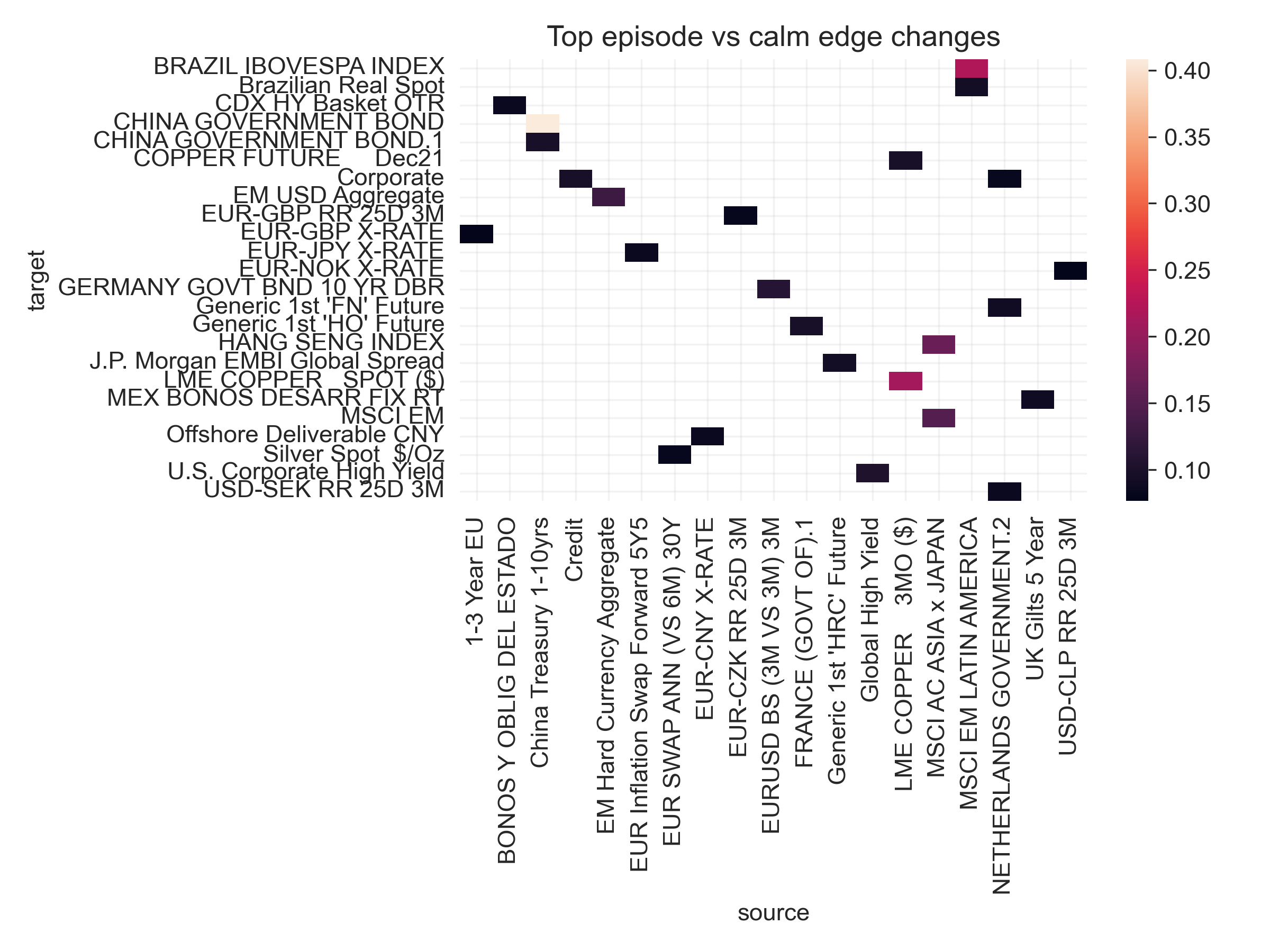}
\caption{Top increases in directed edge strength between episode and calm
periods. Each cell represents the difference in average driver-to-driver
contribution $M_{j,i}(t)$ between statistically significant episodes and
calm windows. The amplification is highly sparse, confirming selective
strengthening of specific transmission channels.}
\label{fig:edge_heatmap}
\end{figure}

Figure~\ref{fig:edge_heatmap} reports the largest increases in directed edge
strength between episodes and calm periods.
The amplification is highly sparse, with only a small subset of edges exhibiting
substantial increases.
The top amplified edges involve cross-asset and cross-regional channels,
confirming that systemic stress is not associated with uniform densification of
the causal network but rather with selective strengthening of specific,
economically interpretable transmission channels.

Table~\ref{tab:edges} reports the top 20 episode-versus-calm directed edge
changes ranked by $\Delta M_{j,i}$.

\begin{table}[!t]
\centering
\caption{Top episode vs.\ calm directed edge changes, ranked by
$\Delta M_{j,i}$. Episode and calm strengths report the average
driver-to-driver contribution within significant and non-significant
windows, respectively.}
\label{tab:edges}
\footnotesize
\begin{tabular}{@{}llrrr@{}}
\toprule
Source & Target & Episode & Calm & $\Delta$ \\
\midrule
China Treasury 1--10yrs & China Govt Bond & 0.560 & 0.152 & 0.408 \\
MSCI EM LATIN AMERICA & Brazil Ibovespa & 0.363 & 0.142 & 0.221 \\
LME COPPER 3MO (\$) & LME COPPER SPOT (\$) & 0.388 & 0.176 & 0.212 \\
MSCI AC ASIA x JAPAN & Hang Seng Index & 0.290 & 0.124 & 0.166 \\
MSCI AC ASIA x JAPAN & MSCI EM & 0.303 & 0.153 & 0.150 \\
EM Hard Currency Agg & EM USD Aggregate & 0.190 & 0.062 & 0.129 \\
EURUSD BS 3M vs 3M & Germany 10Y DBR & 0.131 & 0.021 & 0.110 \\
Global High Yield & U.S.\ Corp High Yield & 0.181 & 0.077 & 0.104 \\
China Treasury 1--10yrs & China Govt Bond.1 & 0.235 & 0.136 & 0.099 \\
Credit & Corporate & 0.141 & 0.043 & 0.098 \\
LME COPPER 3MO (\$) & Copper Future Dec21 & 0.188 & 0.090 & 0.098 \\
France (Govt Of).1 & Generic 1st HO Future & 0.113 & 0.015 & 0.098 \\
MSCI EM LATIN AMERICA & Brazilian Real Spot & 0.157 & 0.064 & 0.093 \\
Generic 1st HRC Future & J.P.\ Morgan EMBI Spread & 0.135 & 0.042 & 0.093 \\
Netherlands Govt.2 & Generic 1st FN Future & 0.114 & 0.023 & 0.091 \\
UK Gilts 5Y & Mex Bonos Desarr Fix Rt & 0.099 & 0.011 & 0.089 \\
Netherlands Govt.2 & USD-SEK RR 25D 3M & 0.101 & 0.014 & 0.087 \\
EUR-CNY X-RATE & Offshore Deliverable CNY & 0.154 & 0.067 & 0.087 \\
Bonos y Oblig del Estado & CDX HY Basket OTR & 0.114 & 0.029 & 0.085 \\
EUR Inflation Swap Fwd 5Y5 & EUR-JPY X-RATE & 0.121 & 0.037 & 0.085 \\
\bottomrule
\end{tabular}
\end{table}

\subsection{Aggregation consistency: full system vs.\ synthetic indexes}
\label{app:emp:aggregation}

To assess robustness to dimensionality reduction, drivers are clustered by the
similarity of their rolling target hub score trajectories using agglomerative
clustering ($n=12$ clusters), and each cluster is aggregated into an
equal-weighted synthetic index.
The full rolling operator analysis is then repeated on the reduced
$12$-dimensional system, and agreement with the original $211$-driver system
is evaluated.

Table~\ref{tab:synthetic_agreement} reports the results.
Directional strength ($\lambda_1$ correlation $0.84$) and episode detection
(agreement rate $0.82$, Jaccard $0.46$) exhibit strong agreement, indicating
that the low-dimensional spectral structure of $C(t)$ is robust to aggregation.
In contrast, hub rank correlations are weak ($0.03$), confirming that
fine-grained directional roles require the full system for resolution.

This result is consistent with the axiomatic uniqueness of
Theorem~\ref{thm:axiomatic-uniqueness}: orthogonally invariant spectral
functionals are by construction insensitive to coordinate reparameterizations
induced by aggregation.
The negative $r_{\mathrm{eff}}$ correlation ($-0.43$) indicates that effective
rank behaves differently under aggregation, reflecting the loss of
high-dimensional spectral detail when the system is compressed to 12 indexes.

\begin{table}[!t]
\centering
\caption{Agreement between the full $211$-driver system and the reduced
$12$-index synthetic representation. Metrics compare rolling operator
statistics, episode detection, and hub rankings.}
\label{tab:synthetic_agreement}
\begin{tabular}{@{}lr@{}}
\toprule
Metric & Value \\
\midrule
$\lambda_1$ correlation & 0.842 \\
$r_{\mathrm{eff}}$ correlation & $-$0.431 \\
Episode mask agreement rate & 0.817 \\
Episode mask Jaccard & 0.458 \\
Mean windowwise hub rank correlation & 0.028 \\
Number of full-system episodes & 2 \\
Number of synthetic episodes & 2 \\
\bottomrule
\end{tabular}
\end{table}

\subsection{Driver-to-driver causal networks and temporal dominance}
\label{app:emp:networks}

Episode-averaged causal energy maps based on $M_{j,i}(t)$ can be dominated by
squared low-rank structure and therefore obscure statistically robust network
organization.
Inference-oriented heatmaps isolate interpretable directional structure by
thresholding against circular-shift nulls that preserve each series' internal
dynamics, directly targeting directional alignment rather than magnitude.

\begin{figure}[!t]
\centering
\includegraphics[width=\linewidth]{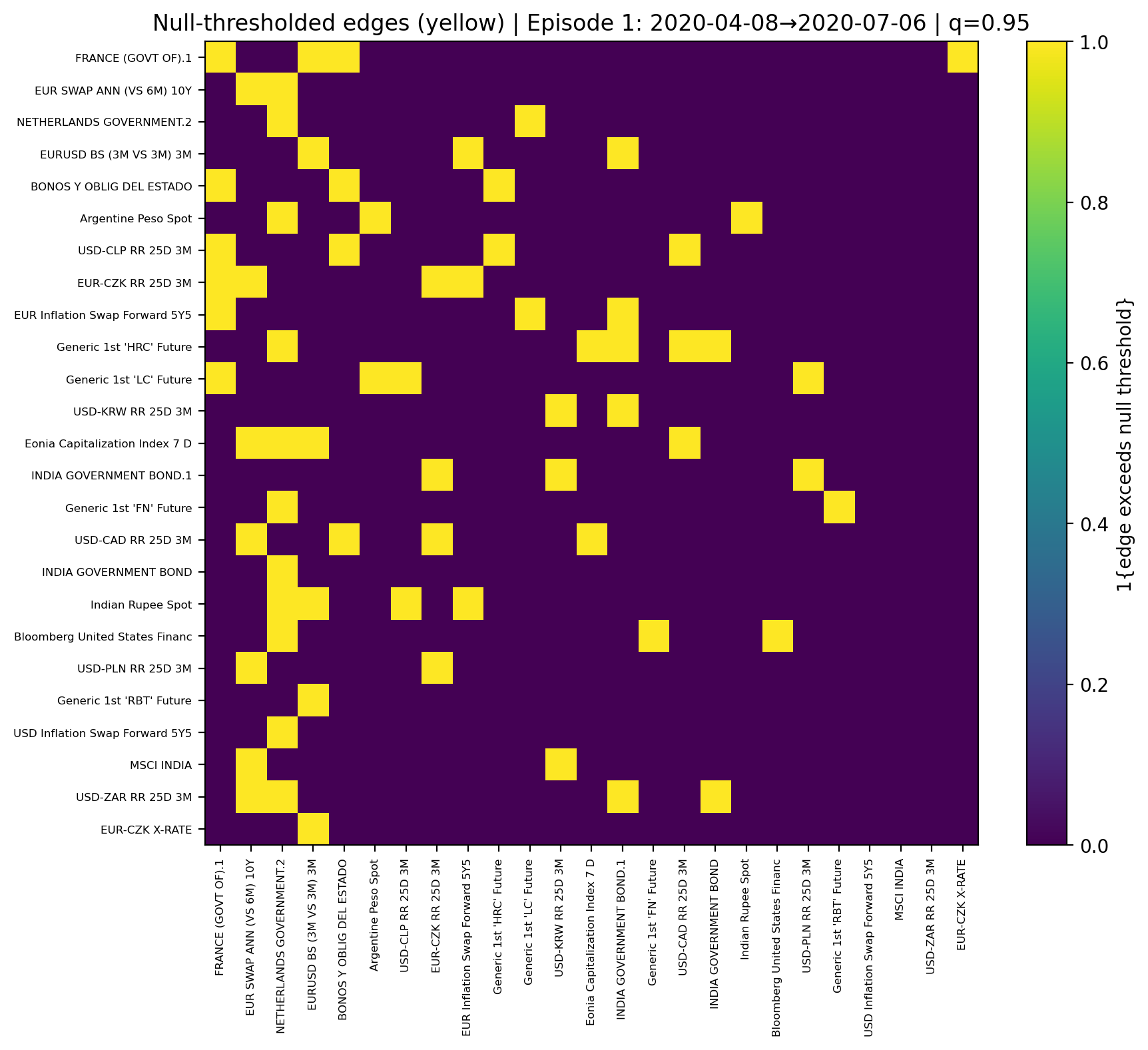}
\caption{Episode~1 (April--July 2020) null-thresholded driver-to-driver network.
Edges represent statistically robust directional transmission channels that
exceed what can be generated by marginal autocorrelation alone.}
\label{fig:heatmap_ep1_yellow}
\end{figure}

\begin{figure}[!t]
\centering
\includegraphics[width=\linewidth]{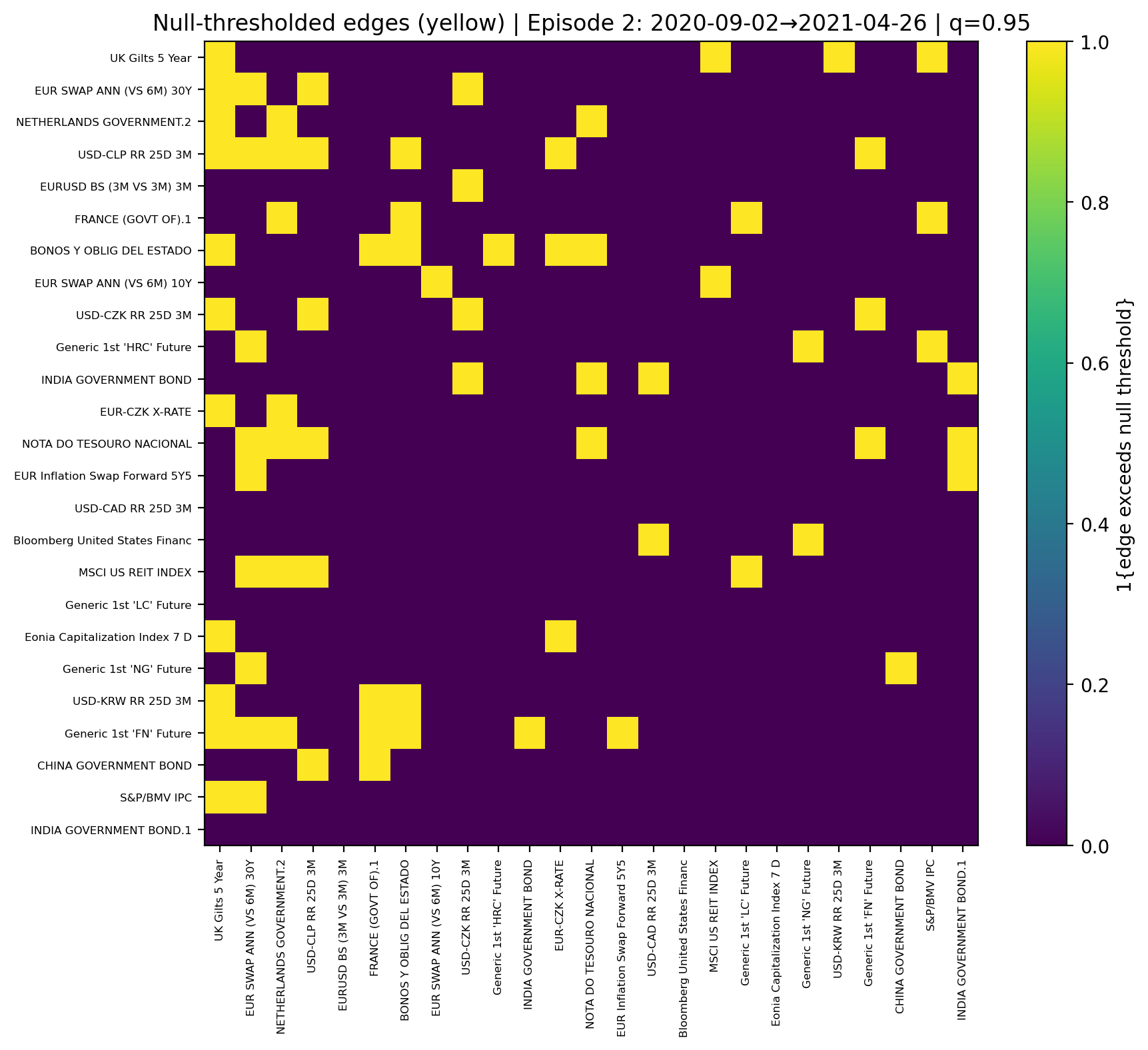}
\caption{Episode~2 (September 2020--February 2021) null-thresholded
driver-to-driver network.
Edges represent statistically robust directional transmission channels.}
\label{fig:heatmap_ep2_yellow}
\end{figure}

Figures~\ref{fig:heatmap_ep1_yellow} and~\ref{fig:heatmap_ep2_yellow} report
null-thresholded driver-to-driver networks for Episodes~1 and~2, respectively.
The resulting networks are sparse by construction and represent statistically
robust transmission channels rather than visually dominant artifacts.

\begin{figure}[!t]
\centering
\includegraphics[width=\linewidth]{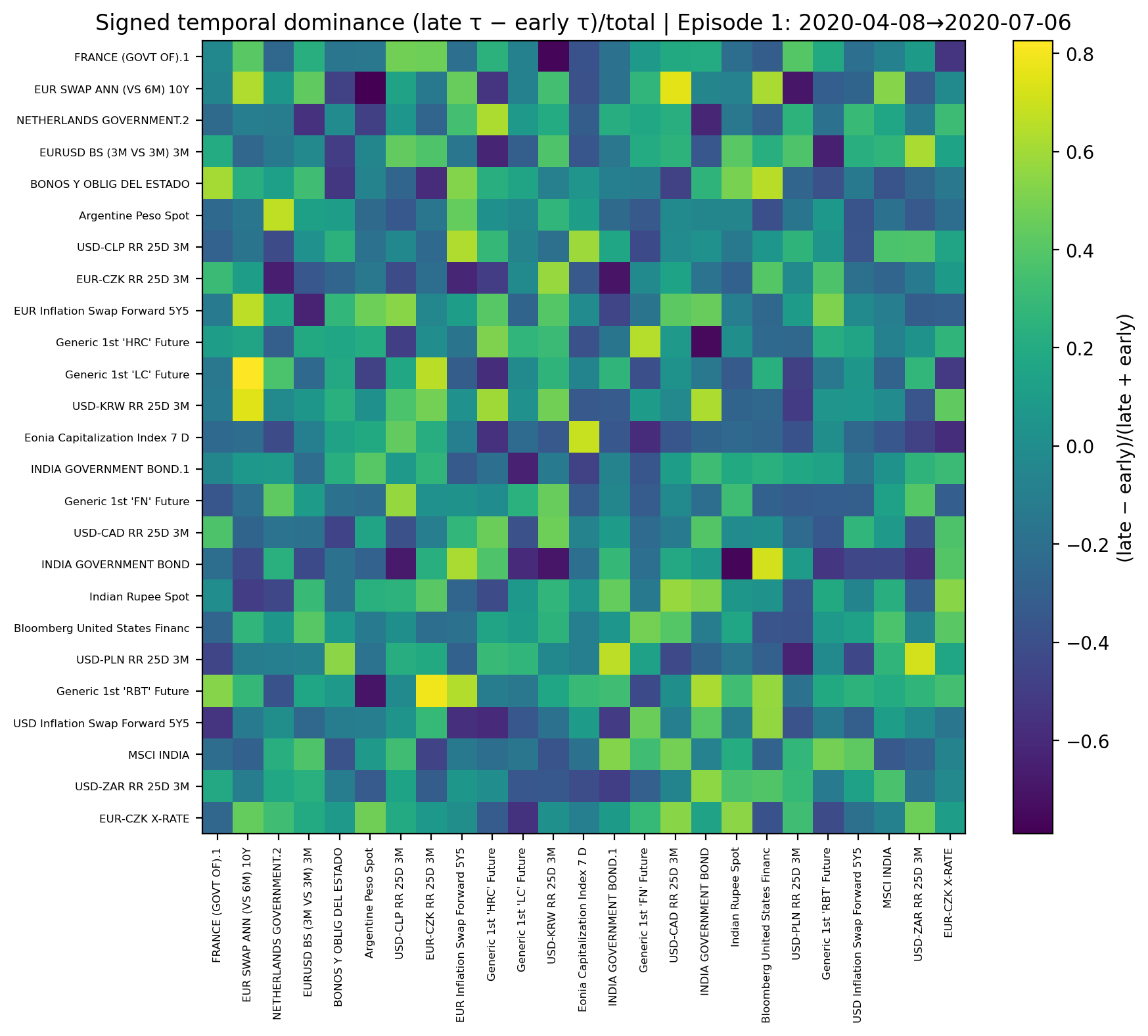}
\caption{Episode~1 signed early--late temporal dominance map.
Positive values indicate dominance of longer-delay transmission ($\tau\in\{3,5\}$)
and negative values indicate dominance of short-delay transmission
($\tau\in\{1,2\}$).}
\label{fig:heatmap_ep1_signed}
\end{figure}

\begin{figure}[!t]
\centering
\includegraphics[width=\linewidth]{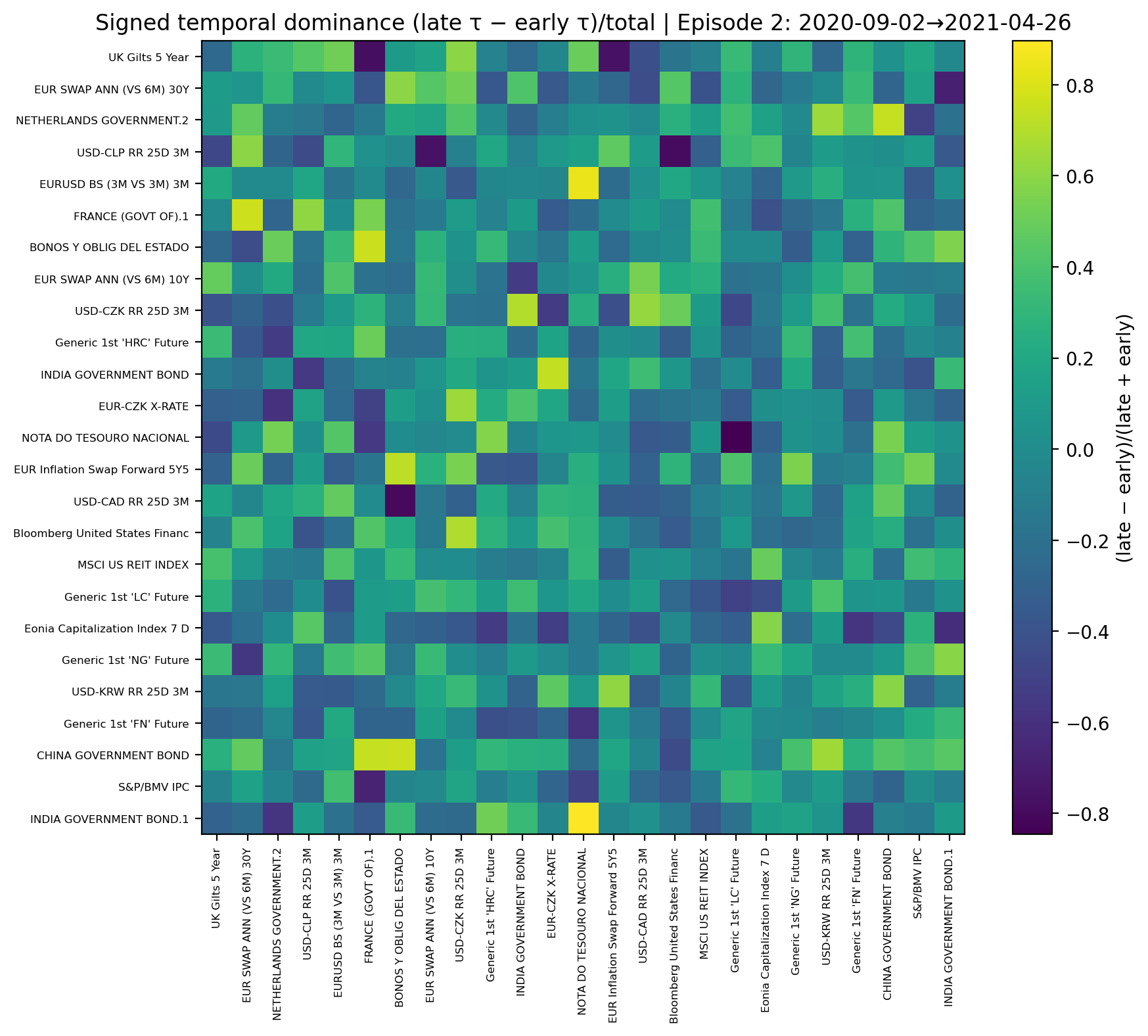}
\caption{Episode~2 signed early--late temporal dominance map.
Positive values indicate dominance of longer-delay transmission and negative
values indicate dominance of short-delay transmission.}
\label{fig:heatmap_ep2_signed}
\end{figure}

Figures~\ref{fig:heatmap_ep1_signed} and~\ref{fig:heatmap_ep2_signed} report
signed early--late temporal dominance maps, decomposing each edge into its
short-delay ($\tau\in\{1,2\}$) and long-delay ($\tau\in\{3,5\}$) components.
Together, these figures reveal heterogeneous propagation speeds across robust
channels: some links transmit information almost immediately, while others
operate with systematic delay.
This heterogeneity is consistent with the lag-level validation in
Table~\ref{tab:lag_validation}, which showed that multiple lags contribute
significant directional content.

\subsection{Macro hub structure and regime interpretation}
\label{app:emp:macrohubs}

To summarize high-dimensional hub dynamics, drivers are grouped according to the
similarity of their rolling target hub score trajectories.
Clustering is performed directly on these operator-derived hub time series using
agglomerative clustering with correlation-based distance, thereby grouping
drivers that exhibit similar temporal patterns of directional causal influence.

\begin{figure}[!t]
\centering
\includegraphics[width=\linewidth]{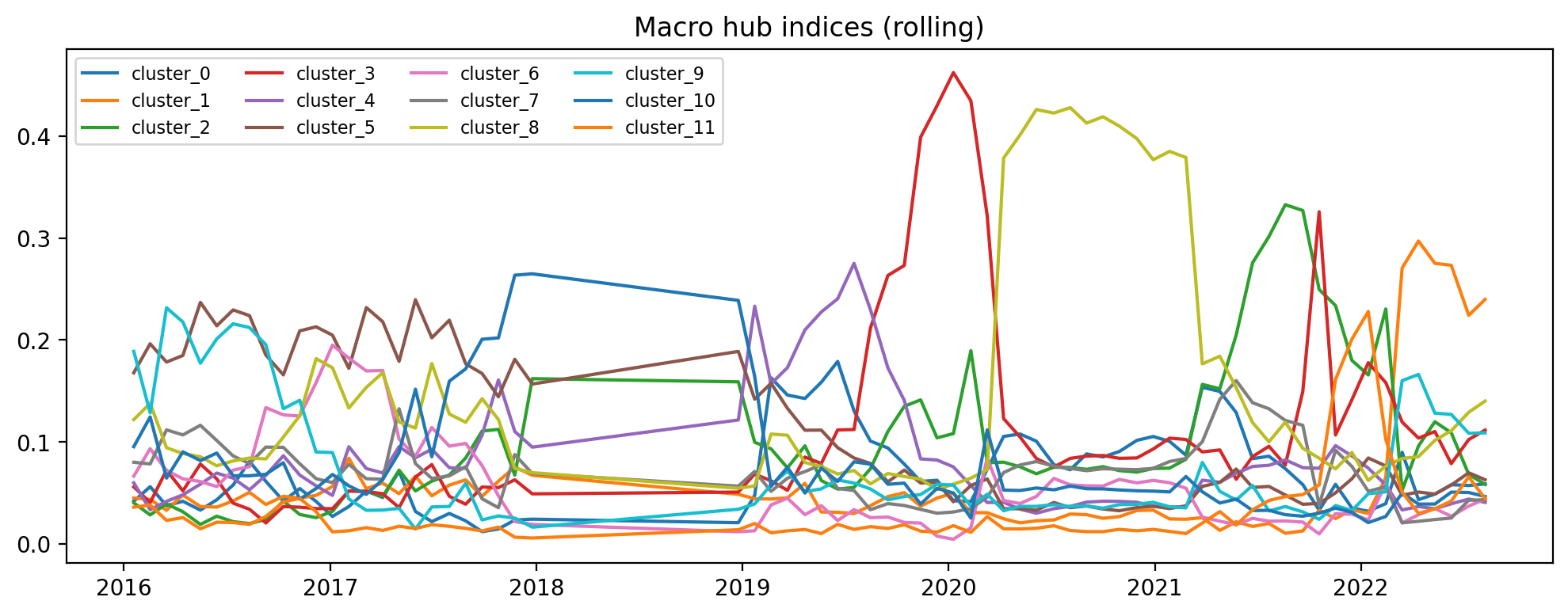}
\caption{Evolution of macro hub indices constructed from clustered rolling target
hub score profiles.
Each curve represents one macro hub cluster.
Periods of single-cluster dominance indicate regime-level concentration of
directional causal influence.}
\label{fig:macro_hub}
\end{figure}

For each group, a macro hub index is defined as the cross-sectional sum of the
target hub scores of its constituent drivers.
Figure~\ref{fig:macro_hub} reports the evolution of these cluster-level hub
indices over time.

The resulting groups should be interpreted as \emph{data-driven collections of
drivers with similar causal roles}, rather than predefined asset classes or
exogenous macroeconomic factors.
Nevertheless, ex post inspection reveals that many groups align with economically
meaningful categories (e.g., global equities, credit, commodities, or rates),
indicating that the operator-based notion of causal influence captures coherent
financial structures.

Periods in which a single cluster-level index dominates correspond to
concentration of directional causal influence within a subset of drivers sharing
similar dynamic roles.
These episodes reflect regime-level organization of the system, in which
system-wide propagation is mediated by a small number of coherent transmission
channels, consistent with the low-rank amplification mechanism identified in the
phase diagram (Figure~\ref{fig:phase_diagram}).

The relationship between cluster dominance and statistically significant episodes
is descriptive rather than inferential, since clustering is performed on the same
sample used for estimation.

Table~\ref{tab:macro_counts} reports the frequency of dominant macro hub regimes
across rolling windows.
Table~\ref{tab:macro_regimes} reports the dominant macro hub regime during each
statistically significant episode.
Table~\ref{tab:cluster_summary} reports qualitative cluster compositions.

\begin{table}[!t]
\centering
\caption{Frequency of dominant macro hub regimes across rolling windows.
Dominance is defined by the maximum macro hub index at each time.}
\label{tab:macro_counts}
\begin{tabular}{@{}lcc@{}}
\toprule
Macro cluster & Dominant windows & Share \\
\midrule
Cluster 8 (EM--Credit--Real Estate) & 14 & 0.27 \\
Cluster 5 (FX--Volatility) & 16 & 0.31 \\
Cluster 3 (Core Rates--Sovereigns) & 8 & 0.15 \\
Cluster 4 (Crypto--Metals) & 7 & 0.13 \\
Cluster 2 (Commodities--Real Assets) & 7 & 0.13 \\
\bottomrule
\end{tabular}
\end{table}

\begin{table}[!t]
\centering
\caption{Dominant macro hub regimes during statistically significant episodes.
Both episodes coincide with the same systemic risk transmission regime.}
\label{tab:macro_regimes}
\begin{tabular}{@{}lcc@{}}
\toprule
Episode & Dominant macro cluster & Regime interpretation \\
\midrule
1 & Cluster 8 & Systemic risk transmission \\
2 & Cluster 8 & Systemic risk transmission \\
\bottomrule
\end{tabular}
\end{table}

\begin{table}[!t]
\centering
\small
\caption{Qualitative macro hub clusters based on economic themes.
Clusters are constructed by grouping drivers according to the similarity of
their rolling target hub score trajectories via agglomerative clustering.
Labels are assigned ex post based on constituent composition and serve as
interpretive summaries rather than estimated latent classes.}
\label{tab:cluster_summary}
\begin{tabular}{@{}lll@{}}
\toprule
Cluster & Regime label & Representative constituents \\
\midrule
Cluster 3 & Core rates and sovereign curves &
USD swaps, US Treasuries, EU sovereign bonds \\
Cluster 5 & FX and volatility &
EUR crosses, CHF-JPY, VIX \\
Cluster 8 & Systemic risk transmission &
EM FX, credit spreads, real estate indices \\
Cluster 2 & Commodities and real assets &
Gold, copper, energy-linked instruments \\
Cluster 4 & Crypto and metals &
Bitcoin, Ethereum, industrial metals \\
\bottomrule
\end{tabular}
\end{table}

\subsection{Practitioner-oriented diagnostic mapping}
\label{app:emp:practitioner}

Table~\ref{tab:practitioner_map} provides a mapping between the operator-based
diagnostics computed in the empirical analysis and their system-level
interpretation.
This mapping is intended as a practical reference for applying the framework
to directional risk monitoring.

\begin{table}[!t]
\centering
\caption{Mapping between operator-based diagnostics and system-level
interpretation. All diagnostics are functionals of $C(t)$ or its spectrum.}
\label{tab:practitioner_map}
\begin{tabular}{@{}ll@{}}
\toprule
Observable & Interpretation \\
\midrule
High $\lambda_1(C(t))$ with low $p$-value &
Emergence of structured directional causality \\
Stable or reduced $r_{\mathrm{eff}}(C(t))$ under high $\lambda_1$ &
Concentration into dominant spectral modes \\
Low hub turnover during significant episodes &
Persistence of dominant transmission channels \\
Positive transmitter-receiver asymmetry &
Net directional propagation role \\
Sparse edge amplification (high $\Delta M_{j,i}$) &
Selective strengthening of transmission channels \\
High $\lambda_1$ correlation under aggregation &
Robustness of spectral dynamics to dim.\ reduction \\
Single macro hub cluster dominance &
Regime-level concentration of causal influence \\
\bottomrule
\end{tabular}
\end{table}

\end{document}